\documentclass[prd,showpacs]{revtex4}
\usepackage{amssymb}
\usepackage{graphicx}
\usepackage{float}

\newcommand{\be}{\begin{equation}}
\newcommand{\ee}{\end{equation}}
\newcommand{\ba}{\begin{eqnarray}}
\newcommand{\ea}{\end{eqnarray}}
\newcommand{\ban}{\begin{eqnarray*}}
\newcommand{\ean}{\end{eqnarray*}}
\newcommand \nn {\nonumber}

\begin{document}

\title{Plasmons in Anisotropic Quark-Gluon Plasma}

\author{Margaret E. Carrington}
\affiliation{Department of Physics, Brandon University,
Brandon, Manitoba, Canada\\
and Winnipeg Institute for Theoretical Physics, Winnipeg, Manitoba, Canada}

\author{Katarzyna Deja}
\affiliation{National Centre for Nuclear Research, Warsaw, Poland}

\author{Stanis\l aw Mr\' owczy\' nski}
\affiliation{Institute of Physics, Jan Kochanowski University, Kielce, Poland \\
and National Centre for Nuclear Research, Warsaw, Poland}

\date{October 14, 2014}

\begin{abstract}

Plasmons of quark-gluon plasma - gluon collective modes - are systematically studied. The plasma is, in general, non-equilibrium but homogeneous. We consider  anisotropic momentum distributions of plasma constituents which are obtained from the isotropic one by stretching or squeezing in one direction. This leads to {\it prolate} or {\it oblate} distributions, respectively. We study all possible degrees of one dimensional deformation from the extremely prolate case, when the momentum distribution is infinitely elongated in one direction, to the extremely oblate distribution, which is infinitely squeezed in the same direction. In between these extremes we discuss arbitrarily prolate, weakly prolate, isotropic, weakly oblate and arbitrarily oblate distributions. For each case, the number of modes is determined using a Nyquist analysis and the complete spectrum of plasmons is found analytically if possible, and numerically when not. Unstable modes are shown to exist in all cases except that of isotropic plasma. We derive conditions on the wave vectors for the existence of these instabilities. We also discuss stable modes which are not limited to small domains of wave vectors and therefore have an important influence on the system's dynamics. 

\end{abstract}



\maketitle

\section{Introduction}

The spectrum of collective excitations is a fundamental characteristic of any many-body system. It carries a great deal of information about the thermodynamic and transport properties of an equilibrium system, and also controls to a large extent the temporal evolution of a non-equilibrium one. In the quark-gluon plasma there are collective modes that correspond to plasma particles, that is, quarks and (transverse) gluons, and there are also collective excitations which are genuine many-body phenomena like longitudinal gluon modes (longitudinal plasmons) and phonons. In this study we focus on longitudinal and transverse gluon collective modes, which we call {\it plasmons}, not limiting the meaning of the term to the equilibrium situation. These modes play a crucial role in the dynamics of quark-gluon plasma. We assume that the plasma is spatially homogeneous, and therefore once the momentum distribution is given, the whole spectrum of plasmons is determined. However, it is only in exceptional cases that the spectrum can be found in closed analytic form. 

The quark-gluon plasma (QGP) occurs as a transient state in relativistic heavy-ion collisions, see {\it e.g.} \cite{Florkowski:2010}. The momenta of the partons produced at the earliest stage of the collisions are mostly along the beam, which means that the characteristic longitudinal momentum is much bigger than the transverse one. The momentum distribution is thus strongly elongated along the beam - it is {\it prolate}. The distribution evolves  - mostly due to the free streaming - and, as discussed in {\it e.g.} \cite{Jas:2007rw}, it becomes squeezed along the beam or {\it oblate} with the characteristic transverse momentum bigger than typical longitudinal momenta. The system moves towards an isotropic local equilibrium state, but does not actually reach it because of viscous effects \cite{Florkowski:2013lya}, see also \cite{Florkowski:2012lba}. Hydrodynamic models of relativistic heavy-ion collisions, which are very successful in describing experimental data,  suggest that the equilibration of quark-gluon plasma is so fast that it occurs in a time interval  shorter than $1~{\rm fm}/c$ \cite{Heinz:2004pj,Bozek:2010aj}. The equilibration process is not fully understood (for recent studies see \cite{Kurkela:2011ti,Kurkela:2011ub,Attems:2012js,Ipp:2010uy}), but it is clear that the spectrum of collective excitations of preequilibrium quark-gluon plasma is a key ingredient of any thermalization scenario. 

Plasmons in anisotropic quark-gluon plasma have been studied by several authors, see the review \cite{Mrowczynski:2005ki} and the papers by Romatschke and Strickland \cite{Romatschke:2003ms,Romatschke:2004jh} and by  Arnold, Lenaghan and Moore \cite{Arnold:2003rq} which are particularly relevant to our work. The authors of \cite{Arnold:2003rq,Romatschke:2003ms,Romatschke:2004jh} focused mainly on unstable modes, which do not exist in isotropic systems, and did not pay much attention to the stable excitations. These stable modes also have an important influence on the system's dynamics in part because, in contrast to the unstable modes, they are not limited to small domains of wave vectors. The stable modes also manifest the interesting property of {\it mode coupling} which, as far as we know, has not been discussed in the context of QGP. 

Our analysis is methodologically very close to the study by Romatschke and Strickland \cite{Romatschke:2003ms,Romatschke:2004jh} who introduced an elegant Ansatz to parameterize an anisotropic momentum distribution by deforming an isotropic one. This distribution, which is appropriate for partons produced in relativistic heavy-ion collisions, has been used to study various aspects of quark-gluon plasma, see {\it e.g.} \cite{Attems:2012js,Florkowski:2012as,Dumitru:2007hy,Martinez:2008di,Schenke:2006yp}. A particularly interesting quantity is the energy loss of a highly energetic parton traversing an unstable QGP, which depends crucially on the spectrum of collective excitations (a preliminary account of our energy-loss study can be found in \cite{Carrington:2012hv,Carrington:2013tz,Carrington:2014yra}). Complete information about the collective modes of an anisotropic system is therefore important and useful. 

In this paper we systematically study gluon collective modes in quark-gluon plasma. We consider a series of momentum distributions $f({\bf p})$ which vary from an extremely prolate one, when the momentum distribution is infinitely elongated in the beam direction and $f({\bf p}) \sim \delta (p_T)$, through arbitrarily prolate, weakly prolate, isotropic, weakly oblate, arbitrarily oblate, to an extremely oblate distribution, which is infinitely squeezed in the beam direction and $f({\bf p}) \sim \delta (p_L)$. Except for the case of the extremely prolate distribution, the complete and exact mode spectrum cannot be obtained analytically. We solve the dispersion equations analytically by looking at certain special cases such as large or small anisotropy, small wave vector, or wave vector (almost) parallel or perpendicular to the anisotropy direction. In more general situations, where the equations cannot be solved analytically, we obtain the dispersion relations numerically. We show that there are unstable modes in all cases except that of isotropic plasma, and we obtain conditions on the wave vectors for which instabilities exist. 

Our paper is organized as follows. In Sec.~\ref{section:form} we formulate the problem. First we define a general dispersion equation, the solutions of which are the gluon collective modes that we want to study. Then, we discuss the momentum distributions that we are interested in, and define the tensor basis we will use to decompose the chromodielectric tensor, or equivalently the gluon polarization tensor. Our analysis of the collective modes begins with the isotropic plasma in Sec.~\ref{sec-iso}, which provides a frame of reference for all other cases. Section \ref{sec-weakly-aniso} is devoted to the weakly anisotropic plasma where the spectrum of collective modes changes qualitatively. Instabilities occur for any anisotropy, and calculations are almost completely analytical when the anisotropy is weak, which makes this situation important and interesting. In Sec.~\ref{sec-finite-aniso} we present numerically obtained dispersion relations for finite anisotropy. The next two sections \ref{sec-ex-prolate} and \ref{sec-ex-oblate} deal with the extremely prolate and oblate systems, respectively, which again can be treated analytically to some extent. In Sec.~\ref{sec-Nyquist} we first introduce and then use a Nyquist analysis to determine the numbers of solutions of each dispersion equation under consideration. When an approximation is used, it is possible to miss a solution, or to produce a spurious solution that is an artifact of the approximation. Similarly, any numerical method searches for solutions within a given range, and will miss solutions which are outside of this range. The Nyquist analysis verifies that we have obtained the correct number of modes in each case. Although we refer to results of the Nyquist method already in Secs.~\ref{sec-iso}-\ref{sec-ex-oblate}, the complete Nyquist analysis is given in Sec.~\ref{sec-Nyquist} to facilitate its presentation.  In Sec.~\ref{section:conclusions} we summarize our study and make some final remarks. In Appendix \ref{app-dis-eq} we remind the reader how the dispersion equation of plasma waves is derived in classical electrodynamics, and in  Appendix \ref{app-self-energy-components} we give some useful results for the components of the anisotropic polarization tensor. 

Throughout the paper we use natural units where $\hbar = c =1$. The indices $i,j,k = 1, 2, 3$ and $\mu, \nu = 0, 1, 2, 3$ label, respectively, the Cartesian spatial coordinates and those of Minkowski space.

\section{Formulation of the problem}
\label{section:form}

In this section we present the general dispersion equation and discuss the parameterization of the momentum distributions that we will use. We then derive the dispersion equations which will be solved in the subsequent sections.  

\subsection{General dispersion equation}

Dispersion equations for plasma collective excitations can be obtained in two ways, which are rather different at first glance. These equations are the conditions for existence of solutions of the homogeneous equation of motion. In the case of QGP the equations of motion are the Yang-Mills equations of the chromodynamic field.  In classical theory, the equations of motion depend on the chromodynamic permeability, or chromodielectric tensor, which contains the effect of the plasma medium. In a quantum field theory, dynamical information about the medium is contained in the polarization tensor which enters the gluon propagator, and the dispersion equation is just the equation that determines the poles of the gluon propagator. The actual character of the approach - whether it is classical or quantum mechanical - depends on how the polarization tensor is  calculated. When using kinetic theory, one typically applies a linear response analysis of classical (or semiclassical) transport equations. Within the quantum field theory formulation, the standard calculation is to use a perturbative method within the hard loop approximation. These two approaches are fully equivalent (the latter is essentially classical in spite of its quantum-field-theory formulation) and the chromodielectric tensor can be expressed directly in terms of the polarization tensor and {\it vice versa}. The only quantum effects that are taken into account are those that are due to the quantum statistics of the plasma constituents. The equivalence of the two approaches was first discovered for the case of equilibrium plasma, see the reviews \cite{Blaizot:2001nr,Litim:2001db}, and the result was later generalized to non-equilibrium systems \cite{Mrowczynski:2000ed,Mrowczynski:2004kv}.

Linearized Yang-Mills equations, or equivalently Maxwell's equations, of a system where external charges and currents are absent, tell us that the Fourier transformed chromoelectric field satisfies the equation
\be
\label{maxwell-1}
\Sigma^{ij}(\omega,{\bf k})E^j(\omega,{\bf k})=0\,,
\ee
with the matrix $\Sigma$ defined as
\be
\label{matrix-Sigma}
\Sigma^{ij}(\omega,{\bf k}) \equiv
- {\bf k}^2 \delta^{ij} + k^ik^j 
+ \omega^2 \varepsilon^{ij}(\omega,{\bf k}) ,
\ee
where $\omega$ is the frequency, ${\bf k}$ denotes the wave vector and $\varepsilon^{ij}(\omega,{\bf k})$ is the chromodielectric tensor. Color indices are suppressed. A solution of Eq.~(\ref{maxwell-1}) exists if
\be
\label{general-dis-eq-det}
{\rm det}[ \Sigma(\omega,{\bf k}) ] = 0 ,
\ee
which is the dispersion equation of plasmons - gluon collective modes. The frequency $\omega$ is, in general, complex but the wave vector ${\bf k}$  is assumed in our study to be real. For a locally colorless anisotropic plasma in the collisionless limit the dielectric tensor equals 
\be
\label{eij}
\varepsilon^{ij}(\omega,{\bf k}) = \delta^{ij} +
\frac{g^2}{2\omega} \int {d^3p \over (2\pi)^3} \,
\frac{v^i}{\omega - {\bf v}\cdot {\bf k}+i0^+}
\Big(\big(1-\frac{{\bf k}\cdot {\bf v}}{\omega}\big) \delta^{jk}
+ \frac{v^jk^k}{\omega} \Big) \nabla_p^k f({\bf p}) ,
\ee
where ${\bf p}$ and ${\bf v} \equiv {\bf p}/|{\bf p}|$ are the momentum and velocity of a {\em massless} parton, and $f({\bf p})$ is the effective parton distribution function. For the ${\rm SU}(N_c)$ gauge group $f({\bf p})= n({\bf p})+ \bar n({\bf p}) +2N_c n_g({\bf p})$, where $n({\bf p})$, $\bar n({\bf p})$, $n_g({\bf p})$ are the distribution functions of quarks, antiquarks and gluons of a single color component. We note that the chromodielectric tensor does not carry any color indices, as the state corresponding to the momentum distribution $f({\bf p})$ is assumed to be colorless. The $i0^+$ prescription makes the Fourier transformed dielectric tensor $\varepsilon^{ij}(t,{\bf r})$ vanish for $t<0$, which is required by causality. In kinetic theory, the infinitesimal quantity  $i0^+$ can be treated as a remnant of inter-particle collisions. Integrating by parts the chromodielectric tensor (\ref{eij}) can be rewritten in the form
\be
\label{eij-1}
\varepsilon^{ij}(\omega,{\bf k}) = \delta^{ij} -
\frac{g^2}{2\omega^2} \int {d^3p \over (2\pi)^3} \,
\frac{f({\bf p})}{|{\bf p}|}
\bigg[\delta^{ij} +
\frac{k^i v^j + v^i k^j}
{\omega - {\bf v}\cdot {\bf k}+i0^+}
+ \frac{({\bf k}^2 - \omega^2) v^i v^j}
{(\omega - {\bf v}\cdot {\bf k}+i0^+)^2} \bigg] ,
\ee
which is often more convenient than the expression (\ref{eij}).  In Appendix~\ref{app-dis-eq} we remind the reader how the dielectric tensor is defined and how Eqs. (\ref{maxwell-1}, \ref{matrix-Sigma}, \ref{general-dis-eq-det}) are obtained in classical electrodynamics. 

In the field theory formulation, where collective modes are determined by the location of the poles of the propagators, the matrix $\Sigma^{ij}(\omega, {\bf k})$ defined by Eq.~(\ref{matrix-Sigma}) equals the inverse {\em retarded} gluon propagator in the temporal axial gauge ($A^0=0$). The dielectric tensor $\varepsilon^{ij}(\omega,{\bf k})$ is related to the retarded gluon polarization tensor $\Pi^{i j}(\omega,{\bf k})$ as
\be
\label{diel-tensor}
\varepsilon^{ij} (\omega,{\bf k}) = \delta^{ij} - \frac{1}{\omega^2} \, \Pi^{i j}(\omega,{\bf k}) .
\ee
The polarization tensor carries Lorentz indices $(\mu, \nu = 0, 1, 2, 3)$, which label coordinates in Minkowski space, and not Cartesian indices $(i,j = 1, 2, 3)$. The components of the polarization tensor, which are not determined by Eq.~(\ref{diel-tensor}), can be reconstructed from the transversality condition $k_\mu \Pi^{\mu \nu}(k) = 0$ with $k^\mu =(\omega,{\bf k})$, which is required by gauge invariance.

Using kinetic theory in the linear response regime, or equivalently working in the hard loop approximation, the dielectric and polarization tensors have the same form for chromodynamic and electrodynamic plasmas of massless constituents, see {\it e.g.} \cite{Mrowczynski:2007hb}. The spectrum of plasmons is also qualitatively the same in chromodynamic and electrodynamic plasmas. Therefore, we often use the more familiar electromagnetic terminology to discuss our results.  

\subsection{Momentum distributions}

The dielectric tensor given by Eq.~(\ref{eij}) or (\ref{eij-1}) is fully determined by the momentum distribution of plasma constituents. Romatschke and Strickland \cite{Romatschke:2003ms} introduced an Ansatz to model anisotropic distributions by deforming isotropic ones. They considered a momentum distribution of the form 
\be
\label{R-S-ansatz}
f_\xi({\bf p}) = C_\xi
f_{\rm iso}\big(\sqrt{{\bf p}^2 +\xi ({\bf p}\cdot{\bf n})^2}\big) ,
\ee
where $ f_{\rm iso}(|{\bf p}|)$ is an isotropic distribution,  $C_\xi$ is a normalization constant, ${\bf n}$ is a unit vector, and the parameter $\xi \in (-1, \infty)$ controls the shape of the distribution. The vector ${\bf n}$ is usually chosen along the beam direction, so that $p_L \equiv {\bf p}\cdot{\bf n}$ and $p_T \equiv |{\bf p} - ({\bf p}\cdot{\bf n}){\bf n}|$. When $\xi =0$ the distribution is isotropic. For $-1<\xi < 0$ the distribution is elongated in the direction of ${\bf n}$ - it is  {\em prolate}. For $\xi > 0$ the distribution is squeezed in the direction of the vector ${\bf n}$ - it is  {\em oblate} - becoming more and more oblate as the parameter $\xi$ increases. 

There is some freedom in choosing the normalization constant $C_\xi$ of the distribution (\ref{R-S-ansatz}).  Initially Romatschke and Strickland put $C_\xi =1$ \cite{Romatschke:2003ms} but in a later publication \cite{Romatschke:2004jh} they used $C_\xi = \sqrt{1 + \xi}$, which is equivalent to normalizing the anisotropic number density to the isotropic one, so that  
\be
\label{norm-xi-1}
\int \frac{d^3p}{(2\pi)^3} \, f_\xi ({\bf p}) = 
\int \frac{d^3p}{(2\pi)^3} \, f_{\rm iso}(|{\bf p}|) .
\ee
In this paper we adopt a different normalization. In case of massless partons, the whole spectrum of collective excitations depends on a single mass parameter which is usually chosen to be
\be
\label{mass2}
m^2 \equiv g^2 \int {d^3p \over (2\pi)^3} \, 
\frac{f_\xi ({\bf p})}{|{\bf p}|} \,.
\ee
When $\xi = 0$ (and the momentum distribution is isotropic), the parameter $m$ reduces to the usual Debye mass. To compare collective modes at different anisotropies it is natural to use a mass parameter that is independent of $\xi$. To accomplish this we require the momentum distribution (\ref{R-S-ansatz}) to be normalized so that
\be
\label{mass2-iso}
\int {d^3p \over (2\pi)^3} \, 
\frac{f_\xi ({\bf p})}{|{\bf p}|} =\int {d^3p \over (2\pi)^3} \, 
\frac{f_{\rm iso}(|{\bf p}|)}{|{\bf p}|} ,
\ee
which determines the normalization constant as
\ba
\label{norm-constant-xi}
C_\xi  =\left\{ \begin{array}{lll}  
\frac{\sqrt{|\xi|} }{ {\rm Arctanh}\sqrt{|\xi|}}
&& \textrm {for} ~~   -1 \le \xi < 0 ,
\\[3mm] 
\frac{\sqrt{\xi} }{ {\rm Arctan}\sqrt{\xi}}
&& \textrm{for}  ~~~~~~ 0 \le \xi . 
\end{array} \right.
\ea

In addition to Eq.~(\ref{R-S-ansatz}), which we refer to as the $\xi$-distribution, we also consider a distribution of the form
\be
\label{alter-ansatz}
f_\sigma({\bf p}) \equiv 
C_\sigma f_{\rm iso} \big(\sqrt{(\sigma +1) {\bf p}^2 - \sigma ({\bf p}\cdot{\bf n})^2 }\:\big) ,
\ee
where $\sigma\ge - 1$, which we call the $\sigma$-distribution.  For $0 > \sigma \ge - 1$ the distribution (\ref{alter-ansatz}) is oblate, for $\sigma=0$ it is isotropic, for $\sigma > 0$ it is prolate, increasing in prolateness as the parameter $\sigma$ grows. If the normalization constant $C_\sigma$ is determined by requiring that the distributions $f_\sigma ({\bf p})$ and $f_{\rm iso}(|{\bf p}|)$ satisfy the condition analogous to Eq.~(\ref{norm-xi-1}), one finds $C_\sigma = \sigma + 1$.  We will require the condition analogous to (\ref{mass2-iso}), so that the mass parameter (\ref{mass2}) is independent of $\sigma$, which gives 
\ba
\label{norm-constant-eta}
C_\sigma  =\left\{ \begin{array}{lll} 
\frac{\sqrt{|\sigma(\sigma + 1)|}}{{\rm Arctan}\sqrt{|\frac{\sigma}{\sigma + 1}|}} 
&& \textrm {for} ~~   -1 \le \sigma < 0 ,
\\[3mm] 
\frac{\sqrt{\sigma(\sigma + 1)} }{{\rm Arctanh}\sqrt{\frac{\sigma}{\sigma + 1}}}
&& \textrm{for}  ~~~~~~ 0 \le \sigma . 
\end{array} \right.
\ea

We are particularly interested in two special cases which are easier to deal with analytically: the extremely prolate and extremely oblate distributions. The latter is proportional to $\delta({\bf n} \cdot {\bf p})=\delta(p_L)$ and can be obtained from the $\xi$-distribution (\ref{R-S-ansatz}) by taking the limit $\xi\to \infty$ (it does not correspond to the limit $\sigma\to -1$ of the $\sigma$-distribution (\ref{alter-ansatz})). The extremely prolate distribution is proportional to $\delta({\bf p}^2-({\bf n} \cdot {\bf p})^2) \sim \delta(p_T)$ and corresponds to the limit $\sigma\to \infty$ of the $\sigma$-distribution (but not the limit $\xi\to -1$ of the $\xi$-distribution).

In practice, the simplest way to obtain the extremely oblate and extremely prolate distributions is not to take the limits described above, but to start from the forms
\ba
\label{extreme-oblate} 
f_{\rm ex-oblate}({\bf p}) &=& \delta(p_L) \, h (p_T), 
\\[2mm]
\label{extreme-prolate} 
f_{\rm ex-prolate}({\bf p}) &=& \delta(p_T)\, \frac{|p_L|}{p_T} \, g(p_L) \,,
\ea
and determine the functions $h(p_T)$ and $g(p_L)$ from the normalization conditions analogous to Eq.~(\ref{mass2-iso}): 
\be
m^2 = {g^2 \over 4\pi^2} \int_0^\infty dp_T\, h(p_T) 
= {g^2 \over 4\pi^2} \int_{-\infty}^\infty dp_L \, g(p_L) .
\ee

Using any one of the momentum distributions (\ref{R-S-ansatz}, \ref{alter-ansatz}, \ref{extreme-oblate}, \ref{extreme-prolate}), the dielectric tensor  (\ref{eij}) or (\ref{eij-1}) is uniquely defined and the mass (\ref{mass2}) is the only dimensional parameter which enters the problem. We define our system of units by rescaling all dimensional quantities by the appropriate power of the mass $m$, which is numerically equivalent to setting $m=1$.

\subsection{Decomposition of $\Sigma$}
\label{sec-decom-ABCD}

To solve the general dispersion equation (\ref{general-dis-eq-det}), one must either find the zeros of the determinant of the matrix $\Sigma$ (\ref{matrix-Sigma}), or invert $\Sigma$  and find the poles of the inverted matrix. We will follow the second strategy.

The first step is to decompose the matrix using a complete set of projection operators. In isotropic plasmas, an arbitrary tensor depends only on the wave vector ${\bf k}$, and can be decomposed into two components, which are transverse and longitudinal with respect to ${\bf k}$. In anisotropic plasmas, the number of projection operators that is needed is larger. An important simplifying feature of the distributions (\ref{R-S-ansatz}) and (\ref{alter-ansatz}) is that the momentum distribution is deformed in only one direction, which is given by the vector ${\bf n}$. An arbitrary symmetric tensor which depends on two vectors can be decomposed in terms of four projection operators. Following \cite{Romatschke:2003ms,Kobes:1990dc}, we introduce the vector ${\bf n}_T$ transverse to ${\bf k}$, which equals
\be
\label{nT-def}
n_T^i = \big(\delta^{ij} - \frac{k^i k^j}{{\bf k}^2}\big) \, n^j ,
\ee 
and define four projectors 
\ba
\begin{array}{ccc}
A^{ij}({\bf k}) = \delta^{ij} - \frac{k^i k^j}{{\bf k}^2}, & &
B^{ij}({\bf k}) = \frac{k^i k^j}{{\bf k}^2} ,
\\[2mm] 
C^{ij}({\bf k},{\bf n}) = \frac{n_T^i n_T^j}{{\bf n}_T^2}, & &
D^{ij}({\bf k},{\bf n}) = k^i n_T^j + k^j n_T^i ,
\end{array}
\ea
which obey the following relations 
\ba
\label{ortho}
\begin{array}{cccc}
AA=A, & AB = 0, & AC = C, & (AD)^{ij} = n_T^i k^j,  \\[2mm]
BA =0, & BB = B, & BC = 0, & (BD)^{ij} = k^i n_T^j,  \\[2mm]
CA= C, & CB = 0, & CC =C, & (CD)^{ij} = n_T^i k^j,  \\[2mm]
(DA)^{ij} = k^i n_T^j, & (DB)^{ij} = n_T^i k^j, & (DC)^{ij} = k^i n_T^j, & DD = n_T^2{\bf k}^2 (B+C).
\end{array}
\ea
Using this projector basis, the inverse propagator $\Sigma$ can be decomposed as 
\be
\label{Sigma-A-B-C-D}
\Sigma^{ij}(\omega, {\bf k}) = a(\omega, {\bf k})\,A^{ij} +b(\omega, {\bf k})\,B^{ij} +c(\omega, {\bf k})\,C^{ij} +d(\omega, {\bf k})\,D^{ij}\, ,
\ee
and the coefficients $a$, $b$, $c$ and $d$ can be found from the equations
\ba
\label{a-b-c-d}
k^i \Sigma^{ij} k^j = {\bf k}^2 b , \;\;\;\;\;
n_T^i \Sigma^{ij} n_T^j = {\bf n}_T^2 (a + c) , \;\;\;\;\;
n_T^i \Sigma^{ij} k^j = {\bf n}_T^2 {\bf k}^2 d , \;\;\;\;\;
{\rm Tr}\Sigma = 2a + b + c .
\ea
With the help of the relations (\ref{ortho}), we invert the matrix (\ref{Sigma-A-B-C-D}) and obtain 
\ba
\label{dispXX}
(\Sigma^{-1})^{ij}&& =
\frac{1}{a} \,A^{ij} 
+ \frac{-a(a+c)\,B^{ij} 
+ (- d^2{\bf k}^2{\bf n}_T^2 +bc)\,C^{ij}
+ad \,D^{ij}}
{a(d^2{\bf k}^2{\bf n}_T^2 - b(a+c))} .
\ea
The Lorentz covariant version of the decomposition (\ref{Sigma-A-B-C-D}) can be found in \cite{Dumitru:2007hy}. 

The inverse propagator $\Sigma$ can be written in terms of the polarization tensor as 
\ba
\label{inv-prop}
(\Delta^{-1})^{ij}(\omega,{\bf k}) =  \Sigma^{ij}(\omega,{\bf k}) 
=  \delta^{ij}(\omega^2-{\bf k}^2) + k^i k^j - \Pi^{ij}(\omega,{\bf k}) ,
\ea
and the polarization tensor is decomposed as
\ba
\label{pi-decomposition}
\Pi^{ij}(\omega,{\bf k}) = \alpha(\omega,{\bf k}) A^{ij}+\beta(\omega,{\bf k}) B^{ij} + \gamma(\omega,{\bf k}) C^{ij} + \delta(\omega,{\bf k}) D^{ij}\,.
\ea
The coefficients $\alpha, \beta, \gamma, \delta$ are related to the functions $a, b, c, d$ from Eq.~(\ref{Sigma-A-B-C-D}) as
\ba
\label{a-alpha}
a(\omega,{\bf k}) &=& \omega^2 - {\bf k}^2- \alpha(\omega,{\bf k}), 
\\
\label{b-beta}
b(\omega,{\bf k}) &=& \omega^2- \beta(\omega,{\bf k}), 
\\
\label{c-gamma}
c(\omega,{\bf k}) &=& - \gamma(\omega,{\bf k}) ,
\\
\label{d-delta}
d(\omega,{\bf k}) &=& - \delta(\omega,{\bf k}) .
\ea
Inverting the matrix (\ref{inv-prop}), the propagator is written
\ba 
\label{propagator-meg-notation}
\Delta^{ij} = (A^{ij}-C^{ij})\,\Delta_A +  \big((\omega^2 - {\bf k}^2 -\alpha - \gamma)B^{ij} 
- (\beta - \omega^2 ) C^{ij} + \delta D^{ij}\big)\,\Delta_G ,
\ea
where the functions $\Delta_A$ and $\Delta_G$ are defined below. 

The dispersion equations are obtained from the poles of the propagator (\ref{dispXX}) or (\ref{propagator-meg-notation}) and are
\ba
\label{dis-eq-A}
 \Delta_A^{-1}(\omega,{\bf k}) &=& a(\omega,{\bf k}) =   \omega^2 - {\bf k}^2 - \alpha(\omega,{\bf k}) = 0 ,
\\[2mm]
\label{dis-eq-G}
\frac{1}{\omega^2}\Delta^{-1}_G(\omega,{\bf k})&=& 0,
\\[2mm]
 \Delta^{-1}_G(\omega,{\bf k}) &=& b(\omega,{\bf k}) \big( a(\omega,{\bf k}) + c(\omega,{\bf k}) \big)-{\bf k}^2 {\bf n}_T^2 d^2(\omega,{\bf k})  \nonumber
\\
&=&  \big(\omega^2 - \beta(\omega,{\bf k})\big)
\big(\omega^2 - {\bf k}^2 - \alpha(\omega,{\bf k}) - \gamma(\omega,{\bf k})\big) 
- {\bf k}^2  {\bf n}_T^2 \delta^2(\omega,{\bf k}).
\ea
We will refer to solutions of the dispersion equation $\Delta^{-1}_A=0$ as $A$-modes, and solutions of the equation $\Delta^{-1}_G/\omega^2=0$ will be called $G$-modes. In the $G$-mode dispersion equation, the factor $1/\omega^2$ is introduced to remove two trivial zero solutions that are of no physical interest, see the discussion below Eq.~(\ref{dis-eq-iso-T-L-eps}). Removing the zero solutions is important in the context of the Nyquist analysis discussed in Sec.~\ref{sec-Nyquist} which then provides the number of {\it physical} solutions of a given dispersion equation.

When the anisotropy is weak, the coefficient $\delta(\omega,{\bf k})$ can be neglected, as shown in Sec.~\ref{sec-weakly-aniso}, and the second dispersion equation (\ref{dis-eq-G}) factors into two simpler equations which are
\ba
\label{b-mode}
&& \frac{1}{\omega^2}\Delta_B^{-1}(\omega, {\bf k})=0,~~ \Delta_B^{-1}(\omega, {\bf k}) = b(\omega,{\bf k})=\omega^2 - \beta(\omega,{\bf k}),
\\[2mm]
\label{c-mode}
&& \Delta_C^{-1}(\omega, {\bf k}) = a(\omega,{\bf k})+c(\omega,{\bf k})=\omega^2 - {\bf k}^2 - \alpha(\omega,{\bf k}) - \gamma(\omega,{\bf k}) = 0 .
\ea
We will refer to the solutions of these equations as $B$-modes and $C$-modes, respectively. In the $B$-mode equation we have again removed two zero solutions. 

In isotropic plasmas, the vector ${\bf n}$ drops out, and the propagator and its inverse can be written in terms of the two projection operators $A$ and $B$
\be
\label{Sigma-A-B}
\Sigma^{ij} = a\,A^{ij} +b\,B^{ij} \,,~~~
(\Sigma^{-1})^{ij} = \Delta^{ij} =  \frac{1}{a}\,A^{ij} + \frac{1}{b}\,B^{ij}   . 
\ee
The dispersion relations are $a(\omega,{\bf k})=0$ and $\omega^{-2}b(\omega,{\bf k})=0$, where, as previously, we have introduced the factor $\omega^{-2}$ to remove trivial zero solutions. In the anisotropic case when the wave vector ${\bf k}$ is parallel to the direction of anisotropy ${\bf n}$, the vector ${\bf n}_T$ vanishes. In this situation, as in the isotropic case, the propagator depends on only one vector. It can be decomposed in terms of the two projectors $A$ and $B$ as in equation (\ref{Sigma-A-B}) and the dispersion relations are again $a(\omega,{\bf k})=0$ and $\omega^{-2}b(\omega,{\bf k})=0$.

\subsection{Coefficients $\alpha, \; \beta, \; \gamma, \; \delta$}
\label{sec-coefficients}

Starting with the decomposition (\ref{pi-decomposition}) and solving the set of equations analogous to (\ref{a-b-c-d}), one finds the coefficients $\alpha, \; \beta, \; \gamma, \; \delta$:
\ba
\label{alpha-gen}
\alpha(\omega,{\bf k}) &=& 
\frac{g^2}{2} \int {d^3p \over (2\pi)^3} \,
\frac{f({\bf p})}{|{\bf p}|}
\bigg[1 + \frac{{\bf k}^2 - \omega^2}{(\omega - {\bf k}\cdot {\bf v} +i0^+)^2}
\Big(1 - \frac{({\bf n}_T \cdot {\bf v})^2}{{\bf n}_T^2} - \frac{({\bf k} \cdot {\bf v})^2}{{\bf k}^2} \Big)\bigg] ,
\\[2mm]
\label{beta-gen}
\beta(\omega,{\bf k}) &=& 
\frac{g^2}{2} \int {d^3p \over (2\pi)^3} \,
\frac{f({\bf p})}{|{\bf p}|}
\bigg[1 + \frac{2({\bf k}\cdot {\bf v})}{\omega - {\bf k}\cdot {\bf v} +i0^+}
 + \frac{({\bf k}^2 - \omega^2)({\bf k}\cdot {\bf v})^2}{{\bf k}^2 (\omega - {\bf k}\cdot {\bf v} +i0^+)^2} \bigg],
\\[2mm]
\label{gamma-gen}
\gamma(\omega,{\bf k}) &=&  
\frac{g^2}{2} \int {d^3p \over (2\pi)^3} \,
\frac{f({\bf p})}{|{\bf p}|}
\bigg[\frac{{\bf k}^2 - \omega^2}{(\omega - {\bf k}\cdot {\bf v} +i0^+)^2}
\Big(-1 +2 \frac{({\bf n}_T \cdot {\bf v})^2}{{\bf n}_T^2} + \frac{({\bf k} \cdot {\bf v})^2}{{\bf k}^2} \Big) \bigg],
\\[2mm]
\label{delta-gen}
\delta(\omega,{\bf k}) &=&  
\frac{g^2}{2} \int {d^3p \over (2\pi)^3} \,
\frac{f({\bf p})}{|{\bf p}|}
\bigg[\frac{1}{\omega - {\bf k}\cdot {\bf v} +i0^+} \frac{{\bf n}_T \cdot {\bf v}}{{\bf n}_T^2}  + \frac{{\bf k}^2 - \omega^2}{(\omega - {\bf k}\cdot {\bf v} +i0^+)^2}
\frac{({\bf n}_T \cdot {\bf v})({\bf k} \cdot {\bf v})}{{\bf n}_T^2 {\bf k}^2} \bigg].
\ea
An important advantage of a momentum distribution in the form (\ref{R-S-ansatz}) or (\ref{alter-ansatz}) is that, for massless plasma constituents, the integral over the magnitude of the momentum and the angular integrals factorize. The momentum distributions (\ref{R-S-ansatz}, \ref{alter-ansatz}) can be written as
\ba
\label{M-defns}
&& f_\xi({\bf p}) = C_\xi f_{\rm iso}(M_\xi |{\bf p}|),
~~~~~M_\xi \equiv \sqrt{ 1+\xi({\bf n}\cdot{\bf v})^2} , 
\\[2mm]
&& f_\sigma({\bf p}) = C_\sigma f_{\rm iso}(M_\sigma |{\bf p}|),
~~~~M_\sigma \equiv \sqrt{1+\sigma + \sigma({\bf n}\cdot{\bf v})^2} ,
\ea
where the functions $M_\xi$ and $M_\sigma$ do not depend on the magnitude $p=|{\bf p}|$. Introducing the variable $\tilde p = M_{\xi/\sigma}|{\bf p}|$, the integrals over $\tilde p$ can be done analytically, and the formulas (\ref{alpha-gen}, \ref{beta-gen}, \ref{gamma-gen}, \ref{delta-gen}) are written as 
\ba
\label{intM}
X_{\xi/\sigma} = \frac{m^2}{2} \int \frac{d\Omega}{4\pi}  \, \frac{F_X}{M_{\xi/\sigma}^2} ,
\ea
where $X$ stands for $\alpha, \,\beta,\, \gamma$ or $\delta$, and the functions $F_X$ equal the expressions in the square brackets in Eqs.~(\ref{alpha-gen}, \ref{beta-gen}, \ref{gamma-gen}, \ref{delta-gen}). These functions do not depend on the magnitude $\tilde p$, and the $\tilde p$-integral just produces the factor $m^2/g^2$ (see Eq. (\ref{mass2})). After performing the angular integration, $\alpha, \,\beta,\, \gamma$ and $\delta$ depend only on the ratio $\omega/k$, and the angle between the wave vector $\bf k$ and the direction of the anisotropy $\bf n$.

The azimuthal integrals can be done analytically in a straightforward manner. The polar integration can also be done analytically, but the resulting expressions are complicated and not very enlightening. In Appendix \ref{app-self-energy-components} we present analytic expressions for the coefficients $\alpha, \; \beta, \; \gamma, \; \delta$ for the $\xi$-distribution (\ref{R-S-ansatz}) and the $\sigma$-distribution (\ref{alter-ansatz}) in which only the polar integration has been done. In the same appendix, we also show some numerical results for the four components of the polarization tensor, after the azimuthal integration is done. In the sections below, we give analytic expressions (after performing both angular integrations) for $\alpha, \; \beta, \; \gamma, \; \delta$  for some special cases where the results are relatively simple. 

The analytic structure of the coefficients $\alpha$ and $\beta$ for finite $\xi$ or $\sigma$ is the same as in the isotropic case. For real valued $\omega$ all four coefficients are complex for $\omega^2 < k^2$ and real for $\omega^2 >k^2$, and for imaginary valued $\omega$ all four coefficients are real. This can be understood as follows. From the formulas (\ref{alpha-gen}, \ref{beta-gen}, \ref{gamma-gen}, \ref{delta-gen}) we see that for real $\omega$, an imaginary contribution to any component of the polarization tensor comes from the denominators $\omega-{\bf k}\cdot{\bf v} +i0^+$ or $(\omega-{\bf k}\cdot{\bf v} +i0^+)^2$ where the Landau infinitesimal elements $i0^+$ are needed to define the integrands when $\omega={\bf k}\cdot {\bf v}$. If $\omega^2 > k^2$ the denominators are always positive and the polarization tensor is pure real. If $\omega^2 < k^2$, the denominators produce an imaginary part due to the $i0^+$ prescription. When $\omega$ is imaginary, it is easy to see that the complex conjugate of each integrand in the formulas (\ref{alpha-gen}, \ref{beta-gen}, \ref{gamma-gen}, \ref{delta-gen}) equals the original integrand with the change ${\bf p} \to - {\bf p}$. Changing the sign of the integration variable and using the fact that the momentum distributions under consideration are even functions of ${\bf p}$, one finds that the polarization tensor is pure real for imaginary $\omega$. We summarize this information as 
\begin{eqnarray}
\label{ana-info}
\begin{array} {lllll}
\omega \in \mathbb{R}  &~~\&~~& \omega^2>k^2            &~~\Rightarrow ~~&     \{\alpha,\beta,\gamma,\delta\} \in \mathbb{R} ,
\\
\omega \in \mathbb{R}  &~~\& ~~& \omega^2<k^2           &~~ \Rightarrow ~~&   \{\alpha,\beta,\gamma,\delta\} \in \mathbb{C} ,
\\
\omega = i\gamma        &~~\&~~& ~\gamma \in \mathbb{R} &~~ \Rightarrow ~~&   \{\alpha,\beta,\gamma,\delta\} \in \mathbb{R} .
\end{array}
\end{eqnarray}

\subsection{Collective modes}
\label{sec-sol-dis-eq}

Solutions $\omega ({\bf k})$ of the dispersion equations (\ref{dis-eq-A}) and (\ref{dis-eq-G}) represent plasmons that are gluon collective modes. There are {\it transverse} plasmons, for which the chromoelectric field is transverse to the wave vector ${\bf k}$, and {\it longitudinal} plasmons with chromoelectric field parallel to ${\bf k}$. The transverse modes correspond to oscillations of current, and the longitudinal ones to oscillations of charge density.  

A mode is called {\it unstable} if $\Im \omega ({\bf k}) > 0$, because the amplitude $\sim \! e^{\Im\omega ({\bf k}) \, t}$ grows exponentially in time. When $\Im \omega ({\bf k}) \le 0$, the mode is {\it stable}. The mode is {\it damped} whenever $\Im \omega ({\bf k}) < 0$  and it is {\it over-damped} when additionally $\Re \omega ({\bf k}) = 0$.  We will show that there are no complex solutions of the dispersion equations (\ref{dis-eq-A}, \ref{dis-eq-G}), only pure real and pure imaginary ones. The real solutions correspond to undamped propagating modes, and the imaginary ones to unstable or over-damped modes (depending on the sign of the solution). These modes are sometimes called, respectively, stable and unstable solutions, but this terminology is confusing since both the over-damped and propagating modes are stable. We will refer to them as real and imaginary solutions. Every solution has a partner with opposite sign. In the case of imaginary solutions, every unstable mode has a partner over-damped mode. In the case of real solutions, the change of sign corresponds to a phase shift of the plasma wave and is physically unimportant.  

When the system is isotropic, the components of the polarization tensor depend only on the magnitude of the vector ${\bf k}$, but this is no longer true for anisotropic systems. For anisotropic systems we choose, without loss of generality, the direction of the anisotropy to be along the $z$-axis: ${\bf n}=(0,0,1)$, and the vector ${\bf k}$ to lie in the $xz$-plane: ${\bf k} = (k_x,0,k_z) = k(\sin\theta, 0,\cos\theta)$. Using this notation, the components of the polarization tensor can be written as functions of the three variables ($\omega,k,\theta$). Actually, the tensor depends on ($\omega/k,\theta$).

In subsequent sections we present spectra of plasmons for various momentum distributions. The extremely prolate system is special in several ways and the notation we use in this case is explained in section \ref{sec-ex-prolate}. In all other cases, we will use the following notation for the dispersion curves: 
\begin{itemize}

\item
red (solid) - real $A$-modes denoted $\omega_\alpha$,

\item
green (dashed) - real $G$-modes which stay above the light cone, denoted $\omega_+$,

\item
blue (dotted) - real $G$-modes which cross the light cone, denoted $\omega_-$,

\item
orange (dashed) - imaginary $A$-modes denoted $\omega_{\alpha i}=i\gamma_\alpha$,

\item
pink (solid) - imaginary $G$-modes  denoted $\omega_{-i}=i\gamma_-$,

\end{itemize}

When plotting real solutions we show only the positive partner, and for imaginary solutions we show the positive imaginary part of the frequency. The curves for the real and imaginary modes are not similar, and therefore there is no ambiguity in plots with two dashed or solid lines. The light cone is always represented as a thin light gray (solid) line. 

We will show that imaginary solutions exist in anisotropic plasmas, when certain conditions on the wave vector are satisfied. These conditions are even functions of $\cos\theta$, and therefore when we discuss them we will consider only $0<\theta<90^\circ$.

\section{Isotropic Plasma}
\label{sec-iso}

In this section we discuss the dispersion relations of a plasma system which is isotropic but not necessarily in equilibrium. In the case of massless plasma constituents, the actual shape of an isotropic momentum distribution enters only through the mass parameter (\ref{mass2}). Plasmons in isotropic plasmas are discussed in textbooks, see {\it e.g.} \cite{Landau-Lifshitz-1981,lebellac}, and we include them in this article for the sake of completeness, and as a reference for  our analysis of anisotropic plasmas. 

When the momentum distribution is isotropic, one uses the decomposition (\ref{Sigma-A-B}) and derives the coefficients $\alpha_{\rm iso}$ and $\beta_{\rm iso}$ which have the form 
\ba
\label{alpha-iso}
\alpha_{\rm iso}(\omega,{\bf k})  &=& \frac{m^2 \omega^2}{2 k^2}
 \bigg[ 1- \Big(\frac{\omega}{2 k} -\frac{k }{2\omega}\Big) 
\ln{\Big(\frac{\omega + k +i0^+}{\omega - k +i0^+}\Big)}  \bigg]  ,
\\[2mm] 
\label{beta-iso}
\beta_{\rm iso}(\omega,{\bf k}) &=& -\frac{m^2 \omega^2}{k^2} 
\bigg[ 1- \frac{\omega}{2 k} \ln{\bigg(\frac{\omega + k +i0^+}{\omega - k +i0^+}\bigg)} \bigg] .
\ea
If one uses the general decomposition (\ref{pi-decomposition}), one finds the same result for $\alpha_{\rm iso}$ and $\beta_{\rm iso}$, and $\gamma_{\rm iso}=\delta_{\rm iso}$=0. The $i0^+$ prescription is needed only if $\omega, k \in \mathbb{R}$ and $\omega^2 \le k^2$.  When  $\omega$ and $k$ are both real, the coefficients can be written as
\ba
\label{alpha-iso-real}
\alpha_{\rm iso}(\omega,{\bf k})  &=& \frac{m^2 \omega^2}{2 k^2}
 \bigg[1 - \Big(\frac{\omega}{2k} -\frac{k}{2\omega}\Big)\Big( \ln{\Big|\frac{k+\omega}{k-\omega}\Big|} -i\pi \Theta(k-\omega) \Big) \bigg]  ,
\\[2mm] 
\label{beta-iso-real}
\beta_{\rm iso}(\omega,{\bf k}) &=& -\frac{m^2 \omega^2}{k^2} 
\bigg[ 1 - \frac{\omega}{2k} \Big( \ln{\Big|\frac{k+\omega}{k-\omega}\Big|} -i\pi \Theta(k-\omega) \Big) \bigg] \,.
\ea
For $k^2 \ll \omega^2$, the logarithm in Eqs.~(\ref{alpha-iso}, \ref{beta-iso}) can be expanded in powers of $k/\omega$ and the
functions $\alpha_{\rm iso}(\omega,{\bf k})$ and $\beta_{\rm iso}(\omega,{\bf k})$ are approximated as 
\ba
\label{alpha-iso-small-k}
\alpha_{\rm iso}(\omega,{\bf k})  &=& \frac{m^2}{3} \Big[ 1 + \frac{k^2}{ 5\omega^2}
+ {\cal O}\Big( \frac{k^4}{ \omega^4}\Big) \Big] ,
\\[2mm] 
\label{beta-iso-small-k}
\beta_{\rm iso}(\omega,{\bf k}) &=& \frac{m^2}{3} \Big[ 1 + \frac{3 k^2}{5 \omega^2}
+ {\cal O}\Big( \frac{k^4}{ \omega^4}\Big) \Big] .
\ea

The dispersion equations for isotropic plasma are given by Eqs.~(\ref{dis-eq-A}, \ref{b-mode}) together with the formulas (\ref{alpha-iso}, \ref{beta-iso}) and read
\ba
\label{dis-eq-iso-T}
\omega^2 - {\bf k}^2 - \alpha_{\rm iso}(\omega,{\bf k}) =  0 ,
\\[2mm]
\label{dis-eq-iso-L}
\frac{1}{\omega^2}\big(\omega^2 - \beta_{\rm iso}(\omega,{\bf k})\big) =  0 .
\ea
These equations describe transverse and longitudinal plasmons, respectively. Expressing the coefficients $\alpha_{\rm iso}$, $\beta_{\rm iso}$ through the transverse and longitudinal components of the dielectric tensor as
\be
\label{alpha-beta-eT-eL}
\alpha_{\rm iso}(\omega,{\bf k})  = \omega^2 \big(1- \varepsilon_T (\omega,{\bf k}) \big) , 
~~~~~~~~~~
\beta_{\rm iso}(\omega,{\bf k})  = \omega^2 \big(1- \varepsilon_L (\omega,{\bf k}) \big) ,
\ee
the dispersion equations (\ref{dis-eq-iso-T}, \ref{dis-eq-iso-L}) can be written in the form 
\ba
\label{dis-eq-iso-T-L-eps}
\omega^2\varepsilon_T(\omega,{\bf k}) - {\bf k}^2 =  0 ,
~~~~~~~~~~~
\varepsilon_L(\omega,{\bf k}) =  0 ,
\ea
which is well known in classical electrodynamics. We note that in the vacuum, where $\varepsilon_{T}=\varepsilon_{L} = 1$, Eqs.~(\ref{dis-eq-iso-T-L-eps}) give two transverse modes $\omega = \pm |{\bf k}|$ and no longitudinal one. We also note that without the multiplier $1/\omega^{2}$ in dispersion equation (\ref{dis-eq-G}) or (\ref{b-mode}), the equation for longitudinal modes would become $\omega^2 \varepsilon_L(\omega,{\bf k}) =  0$, which has a doubled trivial solution $\omega =0$ even in the vacuum. 

Eqs.~(\ref{dis-eq-iso-T}, \ref{dis-eq-iso-L}) cannot be solved analytically but using a Nyquist analysis, which is described in detail in Sec.~\ref{sec-Nyquist}, one shows that each equation has two solutions (a pair of solutions of opposite sign). We note that when counting the number of solutions, one should be careful to specify the form of the dispersion equation under consideration. If one looks at the determinant of the inverse propagator, as in Eq.~(\ref{general-dis-eq-det}), the number of solutions is 8 and not 4. This happens because 
there are two trivial $\omega=0$ solutions and there are two possible orientations of the chromoelectric vector ${\bf E}(\omega, {\bf k})$ that are transverse to ${\bf k}$, which means that the transverse mode appears twice. For an isotropic system one can see this directly from the matrix $\Sigma(\omega, {\bf k})$. Choosing  ${\bf k}=(k,0,0)$ we have
\ba
\label{sigma-iso}
\Sigma (\omega, {\bf k}) = \left[
\begin{array}{ccc}
\omega^2 - \beta_{\rm iso}(\omega,{\bf k}) & 0 & 0 
\\[2mm]
0 & \omega^2 - k^2 - \alpha_{\rm iso}(\omega,{\bf k})  & 0
\\[2mm] 
0 & 0 & 
\omega^2 - k^2 - \alpha_{\rm iso}(\omega,{\bf k})
\end{array}
\right] .
\ea
From Eq.~(\ref{maxwell-1}),  the component $\Sigma^{xx}=\omega^2  - \beta_{\rm iso}$ acts on $E^x $ and thus the solution of Eq.~(\ref{dis-eq-iso-L}) represents, as expected, the longitudinal mode. The components  $\Sigma^{yy}$, $\Sigma^{zz}$ act on $E^y$, $E^z$, and thus the solutions of Eq.~(\ref{dis-eq-iso-T}) correspond to two transverse modes.

Using the approximations (\ref{alpha-iso-small-k}, \ref{beta-iso-small-k}), the dispersion equations can be solved analytically in the long wavelength limit ($\omega^2\gg k^2$) and one obtains
\ba
\label{omega-trans}
\omega^2_T({\bf k}) = \frac{m^2}{3} + \frac{6}{5} k^2  + {\cal O}\Big(\frac{k^4}{m^2} \Big) ,
\\[2mm]
\label{omega-long}
\omega^2_L({\bf k}) = \frac{m^2}{3} + \frac{3}{5} k^2  + {\cal O}\Big(\frac{k^4}{m^2} \Big) .
\ea
The frequency at $k=0$ is the lowest possible frequency of the plasma wave and is known as the {\it plasma frequency} (usually denoted $\omega_p$). For both transverse and longitudinal modes in isotropic plasma we have $\omega_p = m/\!\sqrt{3}$. The equality of the frequency for transverse and longitudinal modes results from the fact that no direction can be distinguished in an isotropic medium when ${\bf k}=0$. 

The dispersion equations can also be solved analytically in the short wavelength limit ($k^2 \gg m^2$) and the dispersion relations are
\ba
\omega^2_T({\bf k}) &\approx& \frac{m^2}{2} + k^2 ,
\\[2mm]
\label{omega-long-large-k}
\omega^2_L({\bf k}) &\approx& k^2  \Big( 1 + 4 e^{-\frac{2k^2}{m^2} -2} \Big).
\ea

\begin{figure}[t]
\begin{minipage}{8.6cm}
\center
\includegraphics[width=1.0\textwidth]{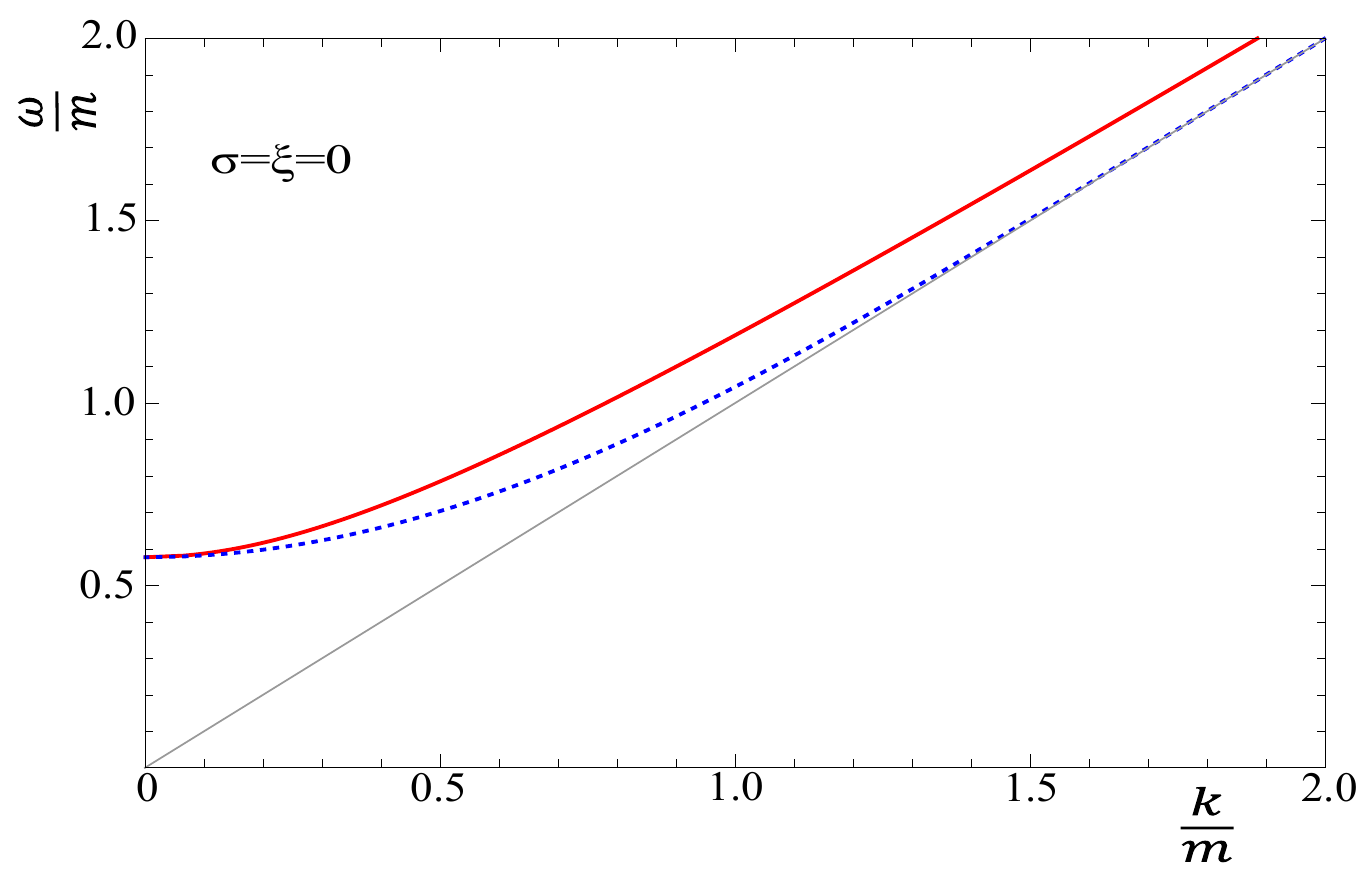}
\caption{(Color online) Dispersion curves of transverse and longitudinal plasmons in isotropic plasma.}
\label{fig-iso}
\end{minipage}
\hspace{5mm}
\begin{minipage}{8.6cm}
\center
\includegraphics[width=1.0\textwidth]{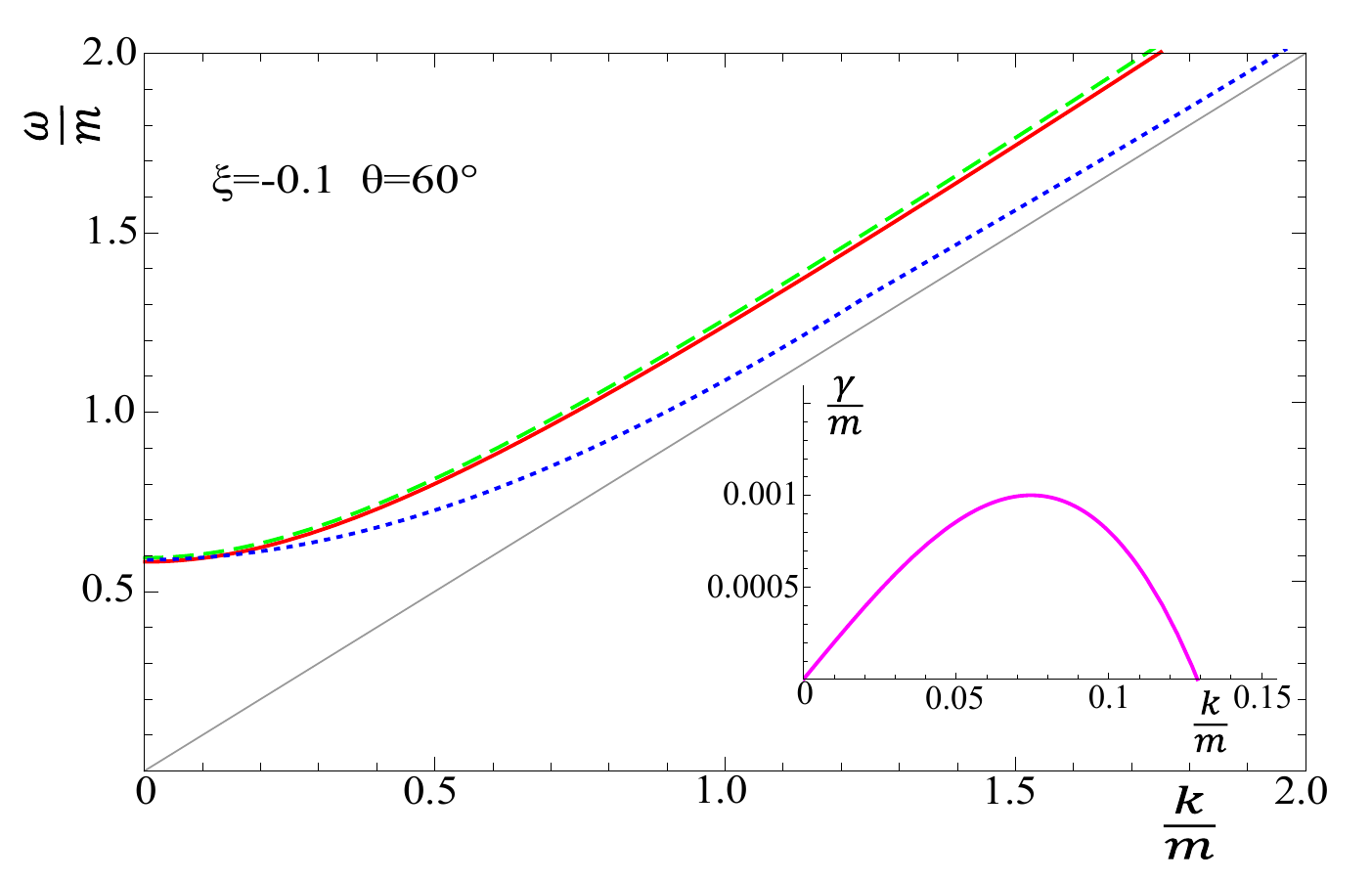}
\caption{(Color online) Dispersion curves of plasmons in weakly prolate plasma with $\xi = -0.1$ for $\theta = 60^\circ$.}
\label{fig-weak-pro-60}
\end{minipage}
\end{figure}

Numerical results for the transverse and longitudinal dispersion relations for arbitrary $k$ are shown in Fig.~\ref{fig-iso}.  The curves stay above the light cone and consequently there is no Landau damping, as the phase velocity of the plasma waves exceeds the speed of light. The longitudinal mode approaches the light cone as $k \rightarrow \infty$ in agreement with the formula (\ref{omega-long-large-k}). 

\section{Weakly Anisotropic Plasma}
\label{sec-weakly-aniso}

It is interesting to study a weakly anisotropic system because it can be treated analytically to a large extent. We will show that the spectrum of plasmons changes qualitatively when an infinitesimal anisotropy is introduced. In Sec.~\ref{sec-finite-aniso} we will demonstrate that all qualitative features of the weakly anisotropic plasma survive in case of strong anisotropy. 

To derive the spectrum of collective modes in a weakly anisotropic plasma, we use the $\xi$-distribution (\ref{R-S-ansatz}) with the assumption $\xi\ll 1$ which gives 
\be
\label{R-S-ansatz-weak-aniso}
f_\xi({\bf p}) = \Big(1+\frac{\xi}{3} \Big) f_{\rm iso}(p) 
+ \frac{\xi}{2} \frac{d f_{\rm iso}(p)}{dp} \, p \, ({\bf v}\cdot{\bf n})^2 ,
\ee 
where we have taken into account that the normalization constant (\ref{norm-constant-xi}) equals 
\be
C_\xi = 1 + \frac{\xi}{3} + {\cal O}(\xi^2) .
\ee
The distribution (\ref{R-S-ansatz-weak-aniso}) is weakly prolate for $\xi<0$ and weakly oblate for $\xi>0$. 

Using the formula (\ref{R-S-ansatz-weak-aniso}), the coefficients  $\alpha, \; \beta, \; \gamma, \; \delta$ given by Eqs.~(\ref{alpha-gen}-\ref{delta-gen}) can be computed analytically. For $\alpha$ and $\beta$ there are contributions of order $\xi^0$ which are just the isotropic results of the previous section. All four functions  $\alpha\,,~\beta\,,~\gamma\,,~\delta$ have contributions of order $\xi$. Since the coefficient $\delta$ enters the dispersion equation (\ref{dis-eq-G}) quadratically, it does not contribute to linear order in $\xi$ and the dispersion equation factorizes into two pieces, so that we have the three dispersion equations of $A$-modes (\ref{dis-eq-A}), $B$-modes (\ref{b-mode}) and $C$-modes (\ref{c-mode}). 

The  coefficients  $\alpha, \; \beta, \; \gamma$ are computed as
\ba
\nn
\alpha (\omega, {\bf k}) &=&  \Big(1 + \frac{\xi}{3}\Big) \alpha_{\rm iso} (\omega, {\bf k}) 
-\xi  \frac{m^2}{8}
\bigg\{
\frac{8}{3} \cos^2\theta  + \frac{2}{3} \big( 5 - 19 \cos^2\theta \big)\frac{\omega^2}{k^2}
- 2 \big( 1 - 5 \cos^2\theta \big) \frac{\omega^4}{k^4} 
\\[2mm] 
\label{alpha-final}
&& 
+ \bigg[ 1 - 3\cos^2\theta 
-  \Big(2 - 8 \cos^2\theta \Big)  \frac{\omega^2}{k^2} 
+ \Big(1  - 5  \cos^2\theta \Big) \frac{\omega^4 }{k^4}  \bigg] 
\frac{\omega}{k}  \ln\Big( \frac{\omega + k + i0^+}{\omega -k + i0^+} \Big) \bigg\} ,
\\[4mm] 
\nn
\beta (\omega, {\bf k}) &=& \Big(1 + \frac{\xi}{3}\Big) \beta_{\rm iso} (\omega, {\bf k}) 
-\xi  m^2 \bigg\{ \Big(- \frac{2}{3} + \cos^2\theta  \Big) \frac{\omega^2}{k^2}
+  (1 - 3 \cos^2\theta )  \frac{\omega^4 }{k^4}
\\[2mm] 
\label{beta-final}
&& 
+ \frac{1}{2} \bigg[  ( 1 - 2 \cos^2\theta ) \frac{\omega^2}{k^2} 
- (1 - 3 \cos^2\theta ) \frac{\omega^4 }{k^4}  \bigg] 
\frac{\omega}{k } \ln\Big( \frac{\omega + k+ i0^+}{\omega -k + i0^+} \Big) \bigg\} ,
\\[4mm]
\label{gamma-final}
\gamma (\omega, {\bf k}) &=& -\xi  \frac{m^2}{4}  
 \, \sin^2\theta  \bigg[ - \frac{4}{3}
 +  \frac{10}{3} \frac{\omega^2}{k^2}
- 2 \frac{\omega^4}{k^4}
+ \Big( 1-  2\frac{\omega^2}{k^2} + \frac{\omega^4}{k^4} \Big) 
\frac{\omega}{k}
\ln\Big( \frac{\omega + k + i0^+}{\omega -k + i0^+} \Big) \bigg] ,
\ea
where  $\alpha_{\rm iso}, ~\beta_{\rm iso}$ are given by Eqs.~(\ref{alpha-iso},  \ref{beta-iso}).  These results appeared previously in \cite{Romatschke:2003ms} with a different normalization.

As in the case of the isotropic plasma, the dispersion relations cannot be solved analytically for arbitrary $k$. When $k^2 \ll \omega^2$, the functions $\alpha (\omega,{\bf k})$, $\beta (\omega,{\bf k})$, $\gamma (\omega,{\bf k})$ are approximated as 
\ba
\label{alpha-small-k}
\alpha(\omega,{\bf k})  
&=& 
m^2 \bigg\{ \frac{1}{3} \Big(1 - \frac{\xi}{15}\Big)  
+ \frac{1}{5} \Big[ \frac{1}{3} + \frac{\xi}{7}  \Big(\frac{1}{9} + \cos^2\theta \Big) \Big]
\frac{k^2}{\omega^2}
+ {\cal O}\Big( \frac{k^4}{ \omega^4}\Big) \bigg\} ,
\\[4mm]
\label{beta-small-k}
\beta(\omega,{\bf k})  
&=& 
m^2 \bigg\{ \frac{1}{3} \Big[1 + \frac{\xi}{5}\Big(-\frac{1}{3} + \cos^2\theta \Big) \Big]  
+ \frac{1}{5} \Big[ 1 + \frac{\xi}{7}  \Big(\frac{1}{3} - \cos^2\theta \Big) \Big]
\frac{k^2}{\omega^2}
+ {\cal O}\Big( \frac{k^4}{ \omega^4}\Big) \bigg\} ,
\\[4mm]
\label{gamma-small-k}
\gamma(\omega,{\bf k})  
&=& \xi  \,m^2 \sin^2\theta \Big[ \frac{1}{15} - \frac{4 \, k^2}{105 \, \omega^2}
+ {\cal O}\Big( \frac{k^4}{ \omega^4}\Big) \Big] .
\ea

In the next three subsections we discuss solutions of the dispersion equations (\ref{dis-eq-A}, \ref{b-mode}, \ref{c-mode}) using the coefficients (\ref{alpha-final}, \ref{beta-final}, \ref{gamma-final}) or (\ref{alpha-small-k}, \ref{beta-small-k}, \ref{gamma-small-k}). In every case we begin with a discussion of the number of solutions which can be found using a Nyquist analysis, as described in Sec.~\ref{app-weakly-aniso}. 

\subsection{$B$-modes}

We start by looking at the $B$-mode dispersion equation (\ref{b-mode}) which describes longitudinal modes. In Sec.~\ref{app-weakly-aniso} we show that there are always two solutions. In the limit $\omega^2 \gg k^2$ we can find these solutions analytically. The coefficient $\beta (\omega,{\bf k})$ is approximated by the formula (\ref{beta-small-k}) and the dispersion equation (\ref{b-mode}) is solved by
\be
\label{sol-weak-aniso-L}
\omega^2({\bf k}) = \frac{m^2}{3} \Big[1 + \frac{\xi}{5}\Big(-\frac{1}{3} + \cos^2\theta \Big) \Big]  +
\frac{3}{5} \Big[ 1 + \frac{4\xi}{35}  \big(1 - 3 \cos^2\theta \big) \Big] k^2 
+  {\cal O}\Big(\frac{k^4}{m^2} \Big) ,
\ee
which reduces to the well-known result for the longitudinal plasmon (\ref{omega-long}) when $\xi=0$. The first term on the right side gives the plasmon mass which depends on the anisotropy parameter $\xi$ and the orientation of wave vector ${\bf k}$.  The formula analogous to (\ref{omega-long-large-k}) shows that the longitudinal mode approaches the light cone as $k \rightarrow \infty$.

\subsection{$A$-modes}

In Sec.~\ref{app-weakly-aniso} we prove that the $A$-mode dispersion equation (\ref{dis-eq-A}) has four solutions when 
\be
\label{cond-insta-alpha}
k^2 - \xi \, \frac{m^2}{3} \,\cos^2\theta < 0
\ee 
and two solutions otherwise. The condition (\ref{cond-insta-alpha}) is never fulfilled for the prolate plasma ($\xi <0$) and it is fulfilled for any oblate momentum distribution ($\xi >0$) when 
\ba
\label{k-crit-A}
k < k_{\rm A} \equiv \Re\sqrt{\frac{\xi}{3}}\, m|\cos\theta|.
\ea 
Because of the real value in the definition of $k_A$, it vanishes for $\xi < 0$.

We solve here the $A$-mode dispersion equation analytically, by looking at certain simplifying limits, and show that the results agree with the those of the Nyquist analysis mentioned above. We first look for real $A$-modes which satisfy $\omega^2 \gg k^2$. In this limit $\alpha (\omega,{\bf k})$ is approximated by the formula (\ref{alpha-small-k}) and Eq.~(\ref{dis-eq-A}) is solved by
\be
\label{sol-weak-aniso-T1}
\omega^2({\bf k}) = \frac{m^2}{3} \Big(1 - \frac{\xi}{15}\Big)  +
\frac{6}{5} \Big[ 1 + \frac{\xi }{14} \Big(\frac{4}{15} + \cos^2\theta \Big) \Big] k^2 
+  {\cal O}\Big(\frac{k^4}{m^2} \Big) ,
\ee
which reduces to the well-known result for the transverse plasmon (\ref{omega-trans}) when $\xi=0$. The plasmon mass, which is given by the first term on the right side of Eq.~(\ref{sol-weak-aniso-T1}), depends on the anisotropy parameter $\xi$ but is independent of the orientation of the wave vector ${\bf k}$. When compared to isotropic plasma, the plasmon mass is smaller for oblate momentum distributions ($\xi > 0$) and bigger for prolate ones ($\xi < 0$). 

We can also look for pure imaginary solutions by substituting $\omega = i \gamma$ with $\gamma \in \mathbb{R}$ and assuming $\gamma^2 \ll k^2$. Using the approximate formula
\be
\frac{\omega + k}{\omega - k} = \frac{\gamma^2 - k^2} {\gamma^2 + k^2} -i \frac{2\gamma k} {\gamma^2 + k^2} 
\buildrel{\gamma^2 \ll k^2}\over{\approx} \exp\Big( - i \pi \frac{\gamma}{|\gamma |} \Big) ,
\ee
the coefficient $\alpha (\omega, {\bf k})$ becomes
\be
\alpha (\omega, {\bf k}) =  - \frac{1}{3}  \xi m^2
 \cos^2\theta + \frac{\pi}{4}\Big[1 -\frac{\xi}{2}\Big(\frac{1}{3} - 3\cos^2\theta \Big)\Big] m^2 \frac{|\gamma |}{k} 
+ {\cal O}\Big(\frac{\gamma^2}{k^2} \Big) ,
\ee
and the dispersion equation (\ref{dis-eq-A}) is written in the form
\be
\label{dis-eq-77}
\gamma^2 + \frac{\lambda}{k} |\gamma | - k_{\rm A}^2 + k^2 = 0,
\ee
where $ k_{\rm A}$ is defined by the formula (\ref{k-crit-A}) and 
\be
\label{wqwq}
\lambda \equiv \frac{\pi}{4}\Big[1 -\frac{\xi}{2}\Big(\frac{1}{3} - 3\cos^2\theta \Big)\Big] m^2 .
\ee
Eq. (\ref{dis-eq-77}) has no roots for an isotropic or prolate system, since $k_A^2 \le 0$ when $\xi\le 0$. For oblate systems, $\xi$ and $k_A^2$ are positive and there are two solutions which read
\be
\label{solution-77}
\gamma ({\bf k}) = \pm \frac{1}{2} \Big(\sqrt{\frac{\lambda^2}{k^2} + 4 (k_{\rm A}^2 -k^2)} - \frac{\lambda}{k} \Big) .
\ee
Equations (\ref{k-crit-A}, \ref{wqwq}) show that in the limit of weak anisotropy $\lambda \gg k_{\rm A}^2$, and therefore the expression (\ref{solution-77}) can be approximated as
\be
\label{solution-78}
\gamma ({\bf k}) \approx \pm \frac{1}{\lambda} \, k(k_{\rm A}^2 -k^2).
\ee
The solutions  (\ref{solution-77}) or  (\ref{solution-78}) represent the unstable and overdamped transverse modes which exist only for oblate plasmas ($\xi >0$) provided the condition (\ref{k-crit-A}) is satisfied.

\begin{figure}[t]
\begin{minipage}{8.5cm}
\center
\includegraphics[width=1.04\textwidth]{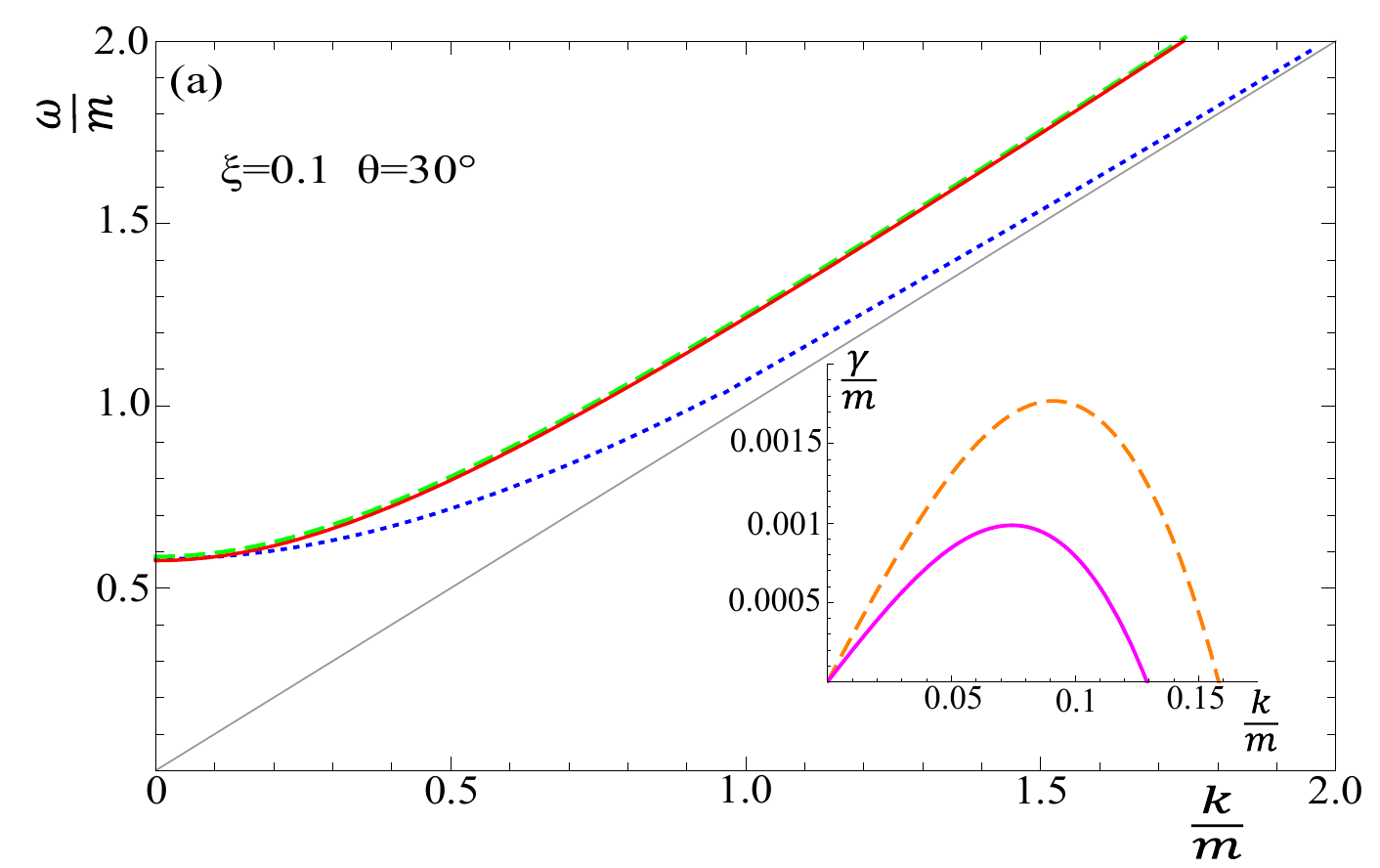}
\end{minipage}
\hspace{2mm}
\begin{minipage}{8.5cm}
\center
\includegraphics[width=1.03\textwidth]{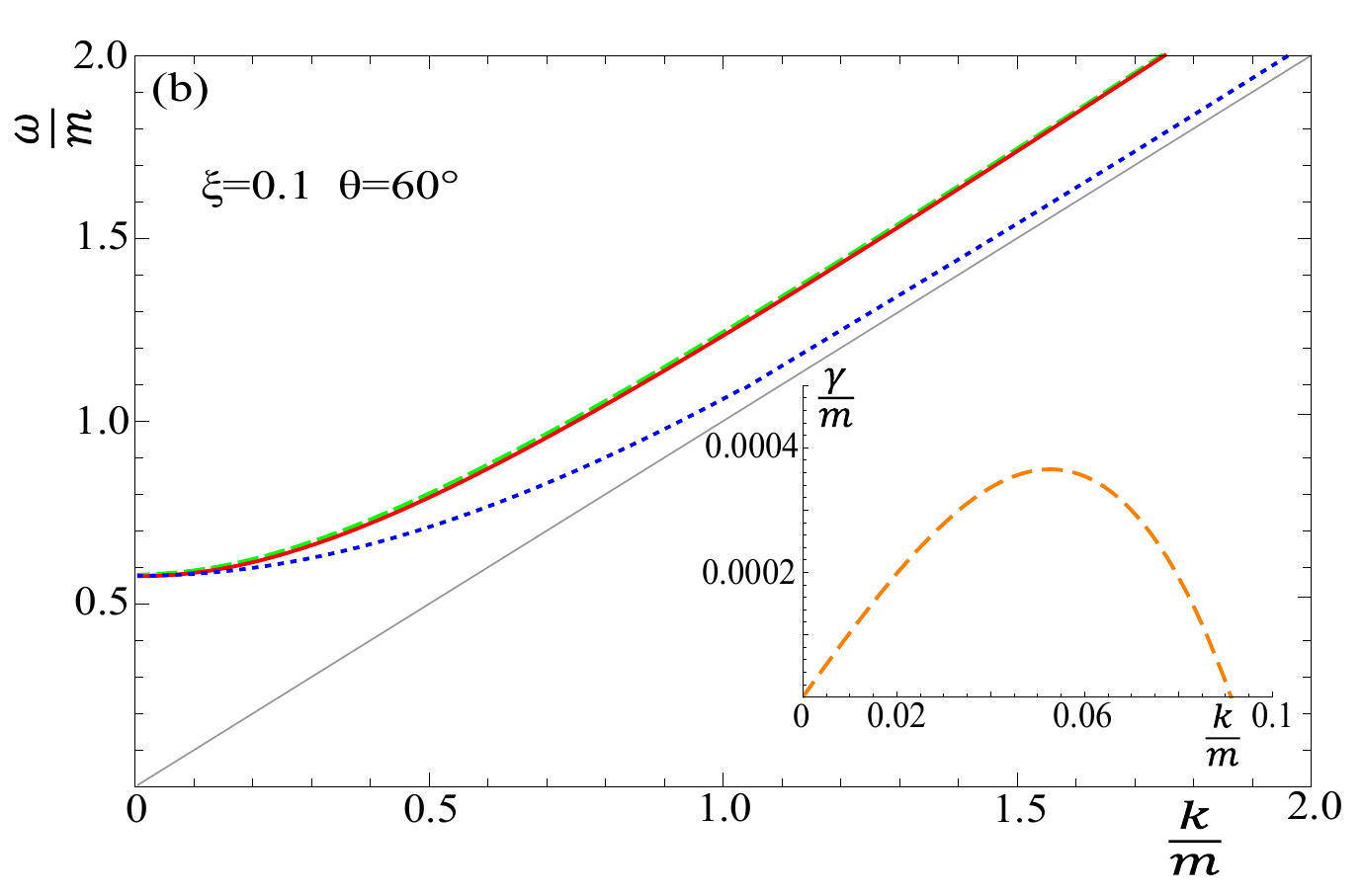}
\end{minipage}
\caption{(Color online) Dispersion curves of plasmons in weakly oblate plasma with $\xi=0.1$ for $\theta = 30^\circ$ (a) and $\theta = 60^\circ$ (b).}
\label{fig-weak-oble-30-60}
\end{figure}

\subsection{$C$-modes}

The $C$-mode dispersion equation  (\ref{c-mode}) has the richest structure. In Sec.~\ref{app-weakly-aniso} we show that the $C$-mode dispersion equation has four solutions when 
\be
\label{cond-insta-alpha-gamma}
k^2 + \xi \, \frac{m^2}{3} \,\big( 1 - 2 \cos^2\theta \big) < 0
\ee
and two solutions otherwise. The condition (\ref{cond-insta-alpha-gamma}) can be fulfilled for oblate plasma ($\xi >0$) when $1/2 < \cos^2\theta$ and for prolate plasma ($\xi <0$) when $1/2> \cos^2\theta$. In both cases the wave vector must satisfy
\be
\label{k-crit-C}
k < k_{\rm C} \equiv m \, \Re\sqrt{\frac{\xi}{3}\,\big(2 \cos^2\theta - 1 \big)} .
\ee
When the argument of the square root is negative, the real part of the root is zero and the critical wave vector $k_{\rm C}$ vanishes. 

In the rest of this subsection we look at specific limits and solve the $C$-mode dispersion equation analytically. The results agree with the those of the Nyquist analysis. We first look for real solutions in the long wavelength limit ($\omega^2 \gg k^2$) when the coefficients $\alpha (\omega,{\bf k})$  and $\gamma (\omega,{\bf k})$ are approximated by the formulas (\ref{alpha-small-k}, \ref{gamma-small-k}). The dispersion equation (\ref{c-mode}) is solved by 
\be
\label{sol-weak-aniso-T2}
\omega^2({\bf k}) = \frac{m^2}{3} \Big[ 1 + \frac{\xi}{5} \Big( \frac{2}{3}  - \cos^2\theta  \Big) \Big]
 +
\frac{6}{5} \Big[ 1 - \frac{\xi }{5} \Big(\frac{23}{42} - \cos^2\theta \Big) \Big] k^2 
+  {\cal O}\Big(\frac{k^4}{m^2} \Big) ,
\ee
which reduces to the well-known transverse plasmon (\ref{omega-trans}) when $\xi=0$. The plasmon mass, which is given by the first term on the right side, depends on the anisotropy parameter $\xi$ and on the orientation of ${\bf k}$. 

One  also finds pure imaginary solutions by substituting $\omega = i \gamma$ with $\gamma \in \mathbb{R}$ and assuming $ \gamma^2 \ll k^2$. The dispersion equation and its solutions have the same form as in the previous section, see Eqs.~(\ref{dis-eq-77}, \ref{solution-77}, \ref{solution-78}), but the coefficient $\lambda$ is now defined as
\be
\lambda \equiv \frac{\pi}{4}\Big[1 - \frac{\xi}{2} \Big( \frac{7}{3} -  5 \cos^2\theta \Big) \Big] m^2 ,
\ee
and $k_{\rm A}$ is replaced by $k_{\rm C}$ given in Eq.~(\ref{k-crit-C}).

\subsection{Discussion}

We have found a complete spectrum of plasmons in weakly anisotropic QGP solving numerically the dispersion equations  (\ref{dis-eq-A}, \ref{b-mode}, \ref{c-mode}). The numerical solutions agree very well with the approximated analytical ones (\ref{sol-weak-aniso-L}, \ref{sol-weak-aniso-T1}, \ref{solution-77}, \ref{sol-weak-aniso-T2}) in the domains of their applicability. Fig.~\ref{fig-weak-pro-60} shows the spectrum for weakly prolate plasma ($\xi = - 0.1$) at $\theta=60^\circ$  and  Fig.~\ref{fig-weak-oble-30-60} presents the spectra for weakly oblate plasma ($\xi = 0.1$) at $\theta=30^\circ$ and $\theta=60^\circ$. The main part of each figure shows the dispersion curves of the positive real modes and the inserts present the positive imaginary solutions. 

For weakly prolate and oblate systems, real $A$-, $B$- and $C$-modes exist for all wave vectors and depend only weakly on the angle. The real $A$- and $C$-modes look very much like the real isotropic transverse mode. In Figs.~\ref{fig-weak-pro-60}, \ref{fig-weak-oble-30-60} these modes are represented by the red (solid) and green (dashed) curves which almost overlay each other. The real $B$-mode looks like the real isotropic longitudinal mode and is represented by the blue (dotted) line. 

In addition to the real modes, for weakly prolate plasma there is an imaginary $C$-mode, seen in Fig.~\ref{fig-weak-pro-60}, which exists for $k < k_C$. The critical wave vector $k_C$ is maximal for $\theta=90^\circ$. When $\theta$ decreases, $k_C$ also decreases until it reaches zero at $\theta=45^\circ$ and the imaginary $C$-mode disappears. In a weakly oblate system there are two imaginary modes seen in Fig.~\ref{fig-weak-oble-30-60}a when $k < k_C < k_A$. Both $k_A$ and $k_C$ are maximal when $\theta=0^\circ$. As $\theta$ increases from $0^\circ$, $k_A$ and $k_C$ decrease. At $\theta=45^\circ$, $k_C$ goes to zero and the imaginary $C$-mode disappears. The regime of the imaginary $A$-mode shrinks to zero at $\theta=90^\circ$. 

\begin{figure}[t]
\begin{minipage}{8.5cm}
\center
\includegraphics[width=1.01\textwidth]{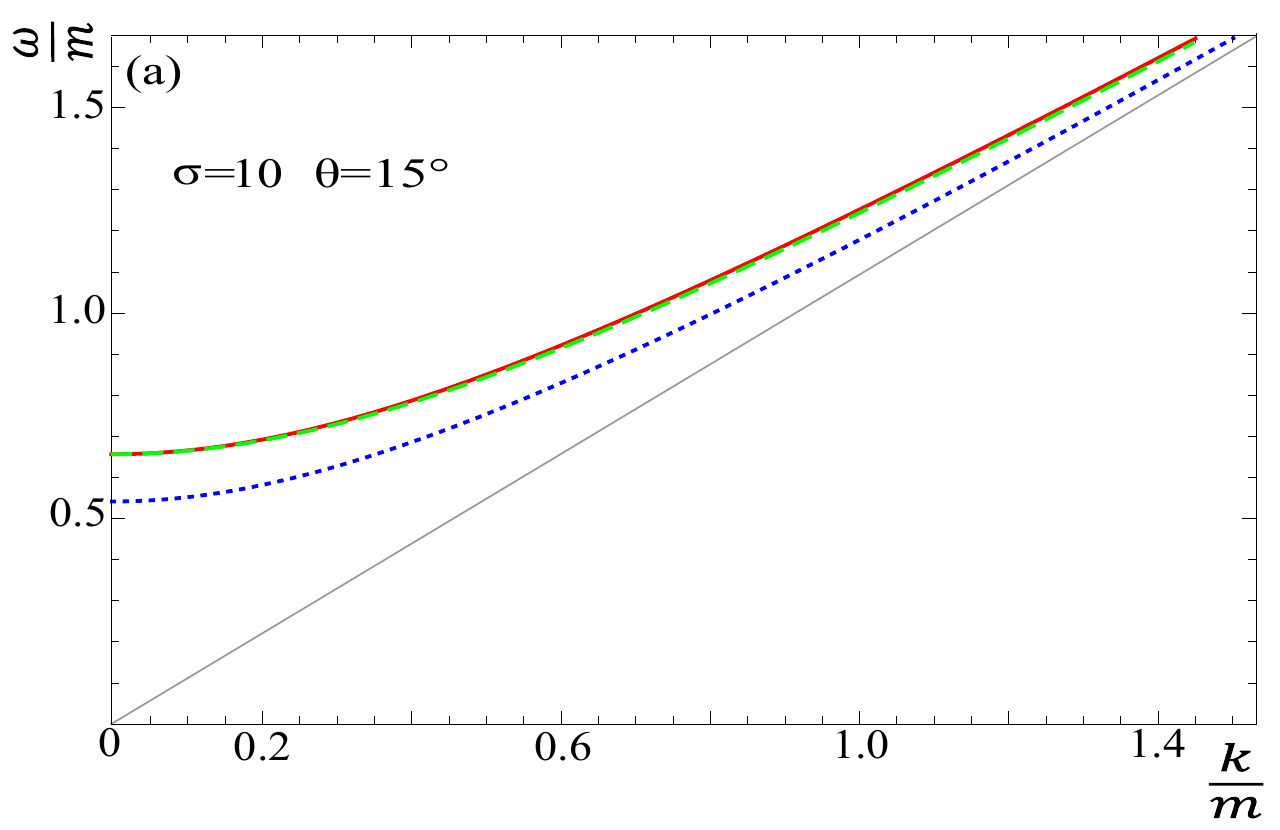}
\end{minipage}
\hspace{5mm}
\begin{minipage}{8.5cm}
\includegraphics[width=1.01\textwidth]{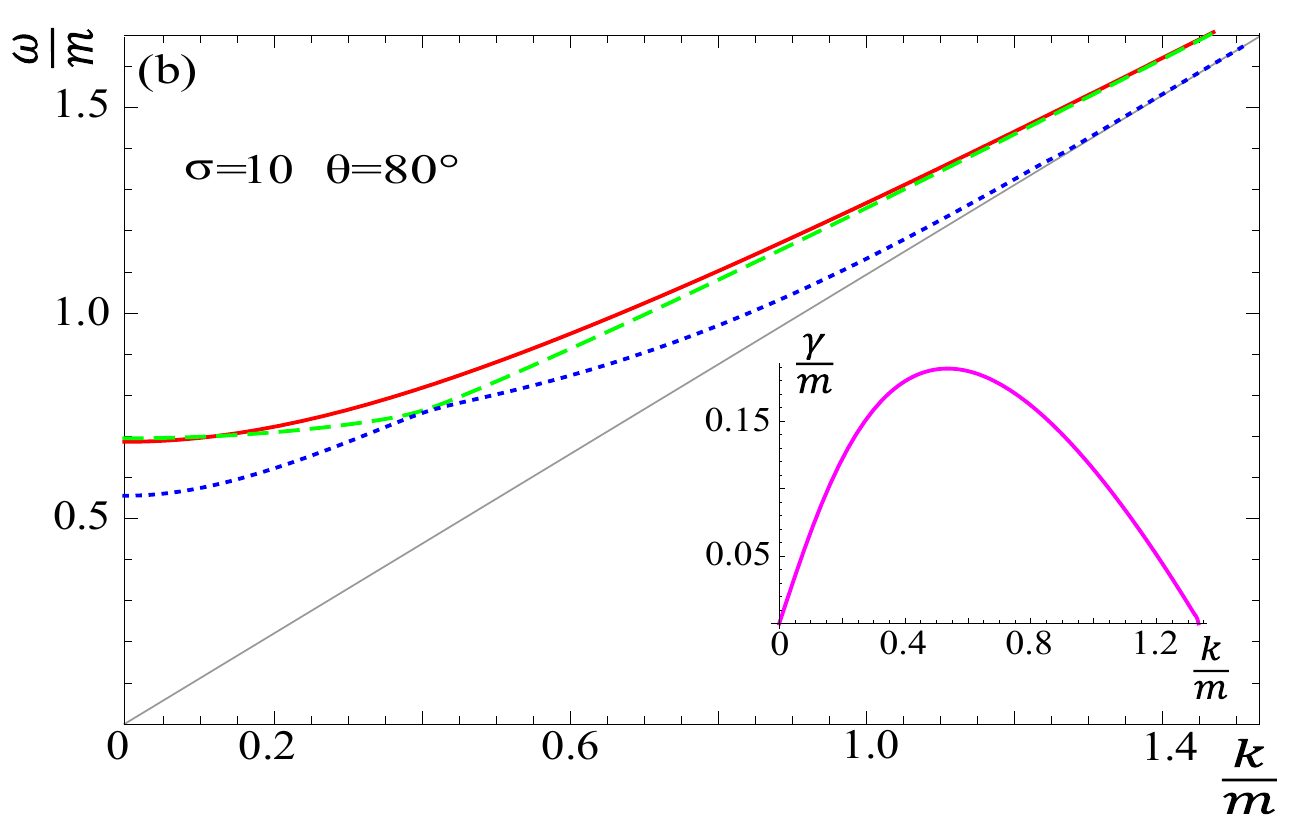}
\end{minipage}
\caption{(Color online) Dispersion curves of plasmons in prolate plasma  with $\sigma = 10$ for $\theta = 15^\circ$ (a) and $\theta = 80^\circ$ (b). }
\label{fig-prolate-80-89}
\end{figure}

In comparison with the spectra of an isotropic system, the weakly anisotropic plasma exhibits the following important differences.

\begin{itemize}

\item The transverse real mode, which is doubled in the isotropic case, is  now split into two slightly different modes, the $A$-mode and $C$-mode, which are given in Eqs. (\ref{sol-weak-aniso-T1}, \ref{sol-weak-aniso-T2}). In Figs. \ref{fig-weak-pro-60} and \ref{fig-weak-oble-30-60} the curves that correspond to these modes are represented by the red and green curves which lie almost on top of each other. 

\item In isotropic plasma longitudinal and transverse plasmons have the same plasma frequency $\omega_p = m/\!\sqrt{3}$, but in anisotropic plasma there are three different minimal frequencies for the three real modes. 

\item In isotropic plasma there are no imaginary solutions. In anisotropic plasma the number of imaginary solutions depends on the magnitude and orientation of the wave vector ${\bf k}$. In prolate plasma the number of imaginary solutions is zero or two (one pair) and in oblate plasma there are zero, two (one pair) or four (two pairs) imaginary modes.

\end{itemize}

Using the formulas (\ref{k-crit-A}, \ref{k-crit-C}), the number of modes can be written as
\ba
\label{weak-fin-A}
&&A-{\rm modes}\!:~ \left\{ \begin{array}{lll} 
2+ 2\Theta(k_{\rm A}-k) 
&& \textrm {for oblate plasma},
\\[2mm]
2
&& \textrm{for prolate plasma}, 
\end{array} \right.
\\
\label{weak-fin-B}
&&B-{\rm modes}\!:~~~~~ 2,
\\[1mm]
\label{weak-fin-C}
&&C-{\rm modes}\!:~~~~~ 2+ 2\Theta(k_{\rm C}-k),
\ea
which show that there is a maximum of 8 solutions for prolate plasma and 10 for oblate plasma. 

The analysis in this section could equally well have been done using the $\sigma$-distribution (\ref{alter-ansatz}) in the limit $|\sigma| \ll 1$. This would reproduce the results expressed by Eqs.~(\ref{cond-insta-alpha}, \ref{cond-insta-alpha-gamma}) with $\xi\to-\sigma$. Since the weakly prolate and weakly oblate systems correspond to  $\sigma>0$ and $\sigma<0$, respectively, the number of modes in Eqs.~(\ref{weak-fin-A}, \ref{weak-fin-B}, \ref{weak-fin-C}) is obviously reproduced. 

There is no anisotropy threshold for the existence of unstable modes, and even an infinitesimal anisotropy produces an instability. However, when $\xi \rightarrow 0$ (or $\sigma \rightarrow 0$) the growth rate of instability ($\gamma$) decreases and the domain of unstable modes shrinks. In this sense, the system becomes less and less unstable as it tends to isotropy. When the effect of inter-parton collisions is taken into account \cite{Schenke:2006xu}, the growth rates of unstable modes are reduced and systems of small anisotropy are effectively stabilized.

\section{Finite Anisotropy}
\label{sec-finite-aniso}

When the anisotropy parameter is not small, the coefficients $\alpha, \, \beta, \, \gamma, \, \delta$ of the decomposition (\ref{pi-decomposition}) and the solutions of the dispersion equations must be computed numerically. However, the spectrum of plasmons has the same structure as in the case of the weakly anisotropic plasma discussed in the previous section - the number of modes is the same and the behavior of the dispersion curves is very similar. 

\begin{figure}[t]
\begin{minipage}{8.5cm}
\center
\includegraphics[width=1.02\textwidth]{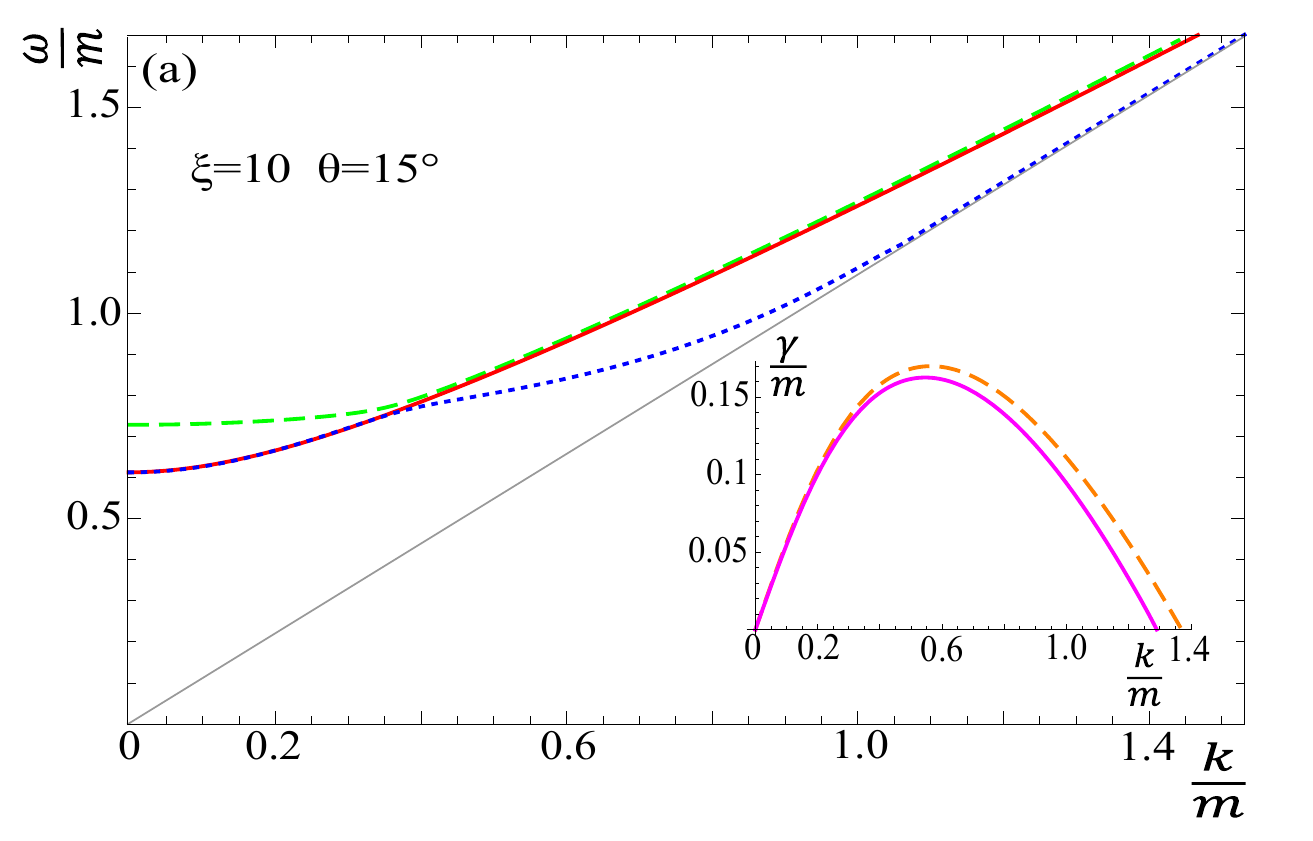}
\end{minipage}
\hspace{5mm}
\begin{minipage}{8.5cm}
\includegraphics[width=1.02\textwidth]{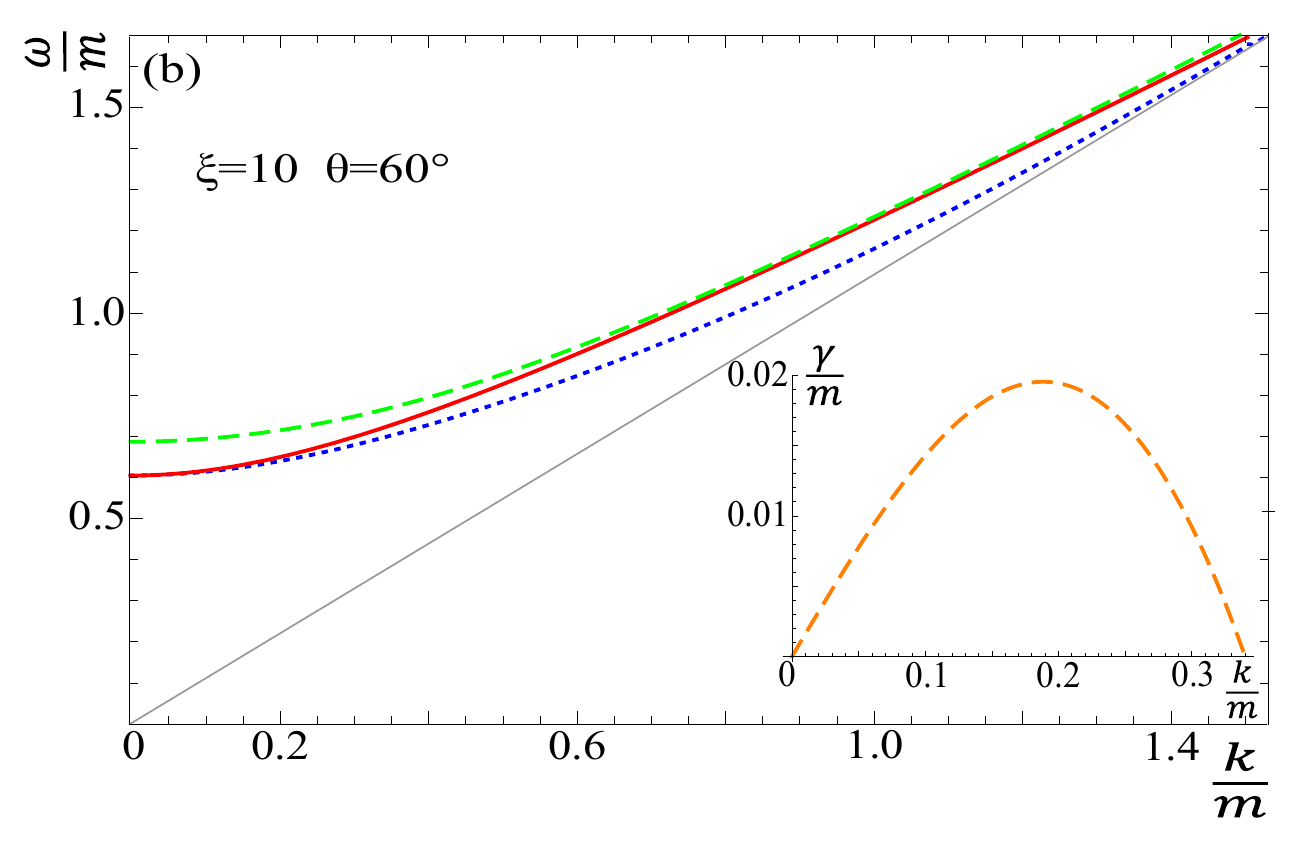}
\end{minipage}
\caption{(Color online) Dispersion curves of plasmons in oblate plasma with $\xi = 10$ for $\theta = 15^\circ$ (a) and $\theta = 60^\circ$ (b). }
\label{fig-oblate-15-60}
\end{figure}

We consider both the $\xi$-distribution (\ref{R-S-ansatz}) and the $\sigma$-distribution (\ref{alter-ansatz}), which together describe deformations of an isotropic distribution with arbitrary prolateness and oblateness. In Appendix~\ref{app-self-energy-components} we give analytic expressions for the components $\alpha, \, \beta, \, \gamma, \, \delta$ of the polarization tensor with the polar integrals unevaluated, and show some graphs of the results after the polar integrals are done. For finite $\xi$ or $\sigma$, the analytic structure of the coefficients $\alpha, \, \beta, \, \gamma, \, \delta$ is the same as in the isotropic case. For real valued $\omega$ all coefficients are complex for $\omega^2 < k^2$ and real for $\omega^2 >k^2$, and for imaginary valued $\omega$ all four coefficients are real, see Eq.~(\ref{ana-info}). 

If the anisotropy parameter is not assumed small, the coefficient $\delta$ cannot be neglected, which means that the dispersion equation for the $G$-modes (\ref{dis-eq-G}) does not factorize into equations (\ref{b-mode}, \ref{c-mode}). However it can be factorized as \cite{Romatschke:2003ms}
\be
\label{Delta-G-factor}
 \Delta^{-1}_G(\omega, {\bf k}) = \big(\omega^2 -\Omega_+^2(\omega, {\bf k})\big)\big(\omega^2 -\Omega_-^2(\omega, {\bf k})\big) = 0
\ee
where
\be
\label{Omega-pm-def}
\Omega_{\pm}^2(\omega, {\bf k})  \equiv \frac{1}{2} \Big( \alpha+\beta+\gamma +  k^2 
\pm \sqrt{(\alpha - \beta +\gamma + k)^2 + 4 k^2 n_T^2 \delta^2}\;\Big) .
\ee
The square root in Eq.~(\ref{Omega-pm-def}) is undefined if its argument is pure real and negative. When all coefficients of the polarization tensor are real, the argument of the root is positive definite. When these coefficients are complex, the root argument is also complex. Therefore, there is no case for which the argument of the root is real and negative, which means that one can find the dispersion relations by solving the equations $\omega=\pm \Omega_+(\omega, {\bf k})$ and  $\omega=\pm \Omega_-(\omega, {\bf k})$ self-consistently. 

Characteristic examples of the complete spectra of plasmons in prolate and oblate plasmas, computed with the $\xi-$ and $\sigma-$distribution, respectively,  are shown in Figs.~\ref{fig-prolate-80-89} and \ref{fig-oblate-15-60} for fixed values of $\theta$. For both prolate and oblate cases, there are six (three pairs) of real modes for all ${\bf k}$ which change slowly with $\theta$. For prolate plasmas, there is at most one pair of imaginary modes. For small angles these modes are absent. As the angle increases, the imaginary modes appear at small $k$, and extend to larger and larger $k$ as the angle increases. In oblate systems, there are at most two pairs of imaginary modes. They are both absent at $\theta=90^\circ$. When the angle decreases, the $A$-mode shows up first, and both pairs extend to larger and larger $k$ as $\theta$ continues to decrease. All of these features are the same as for the weakly anisotropic plasma discussed in Sec.~\ref{sec-weakly-aniso}.

\section{Extremely prolate plasma}
\label{sec-ex-prolate}

\begin{figure}[t]
\begin{minipage}{8.5cm}
\center
\includegraphics[width=1.04\textwidth]{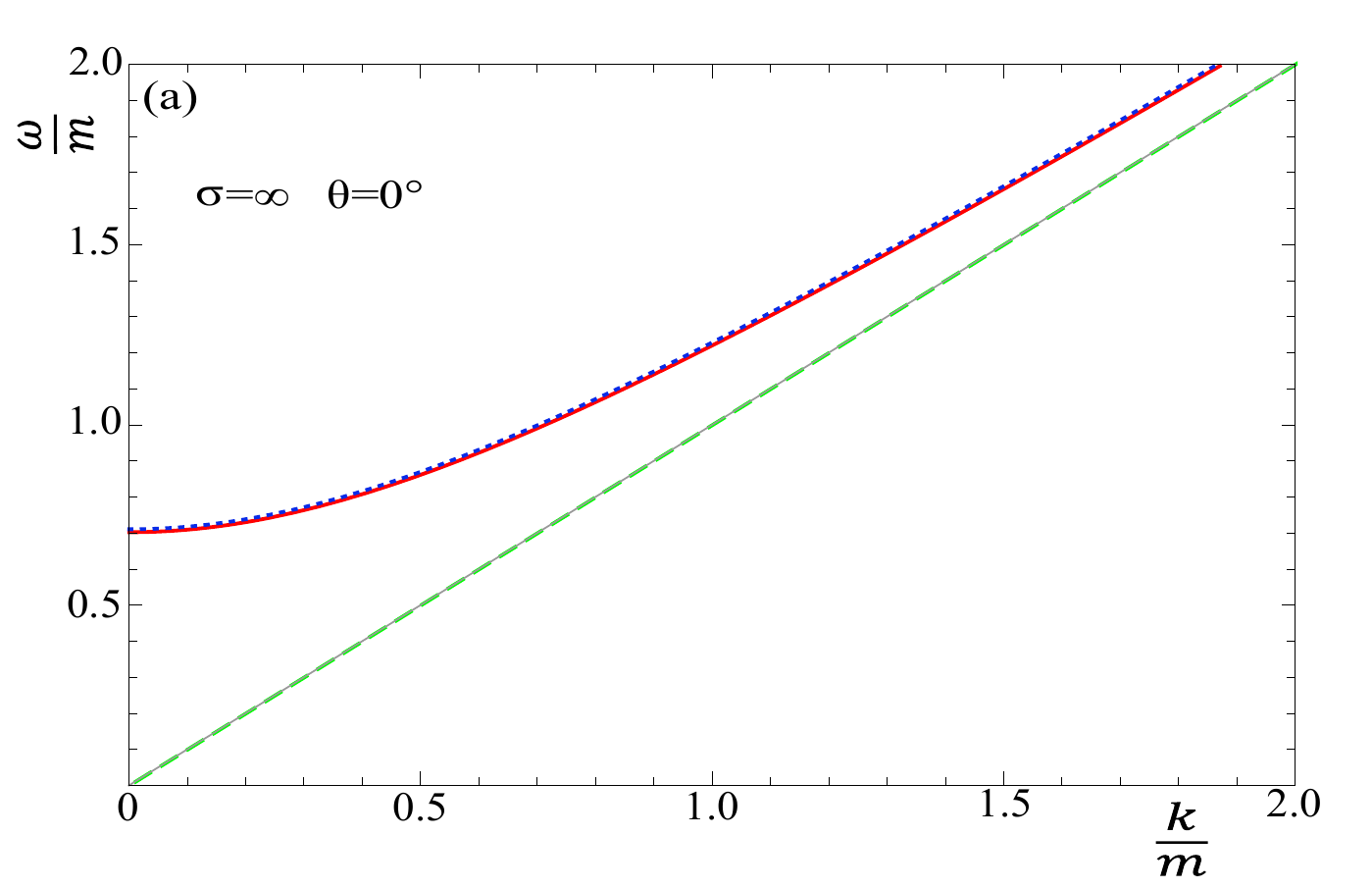}
\end{minipage}
\hspace{2mm}
\begin{minipage}{8.5cm}
\center
\includegraphics[width=1.02\textwidth]{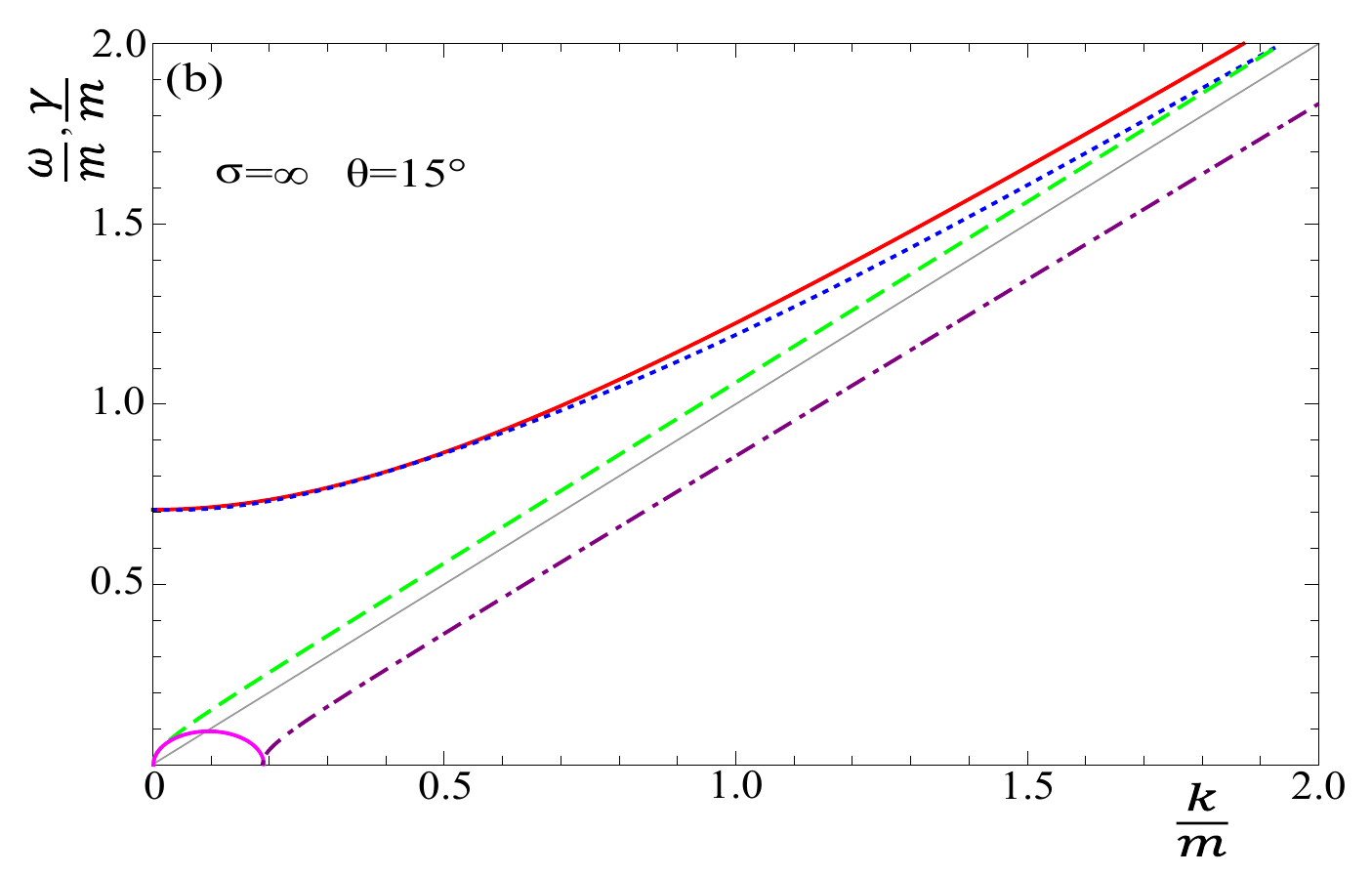}
\end{minipage}
\vspace{2mm}
\begin{minipage}{8.5cm}
\center
\includegraphics[width=1.02\textwidth]{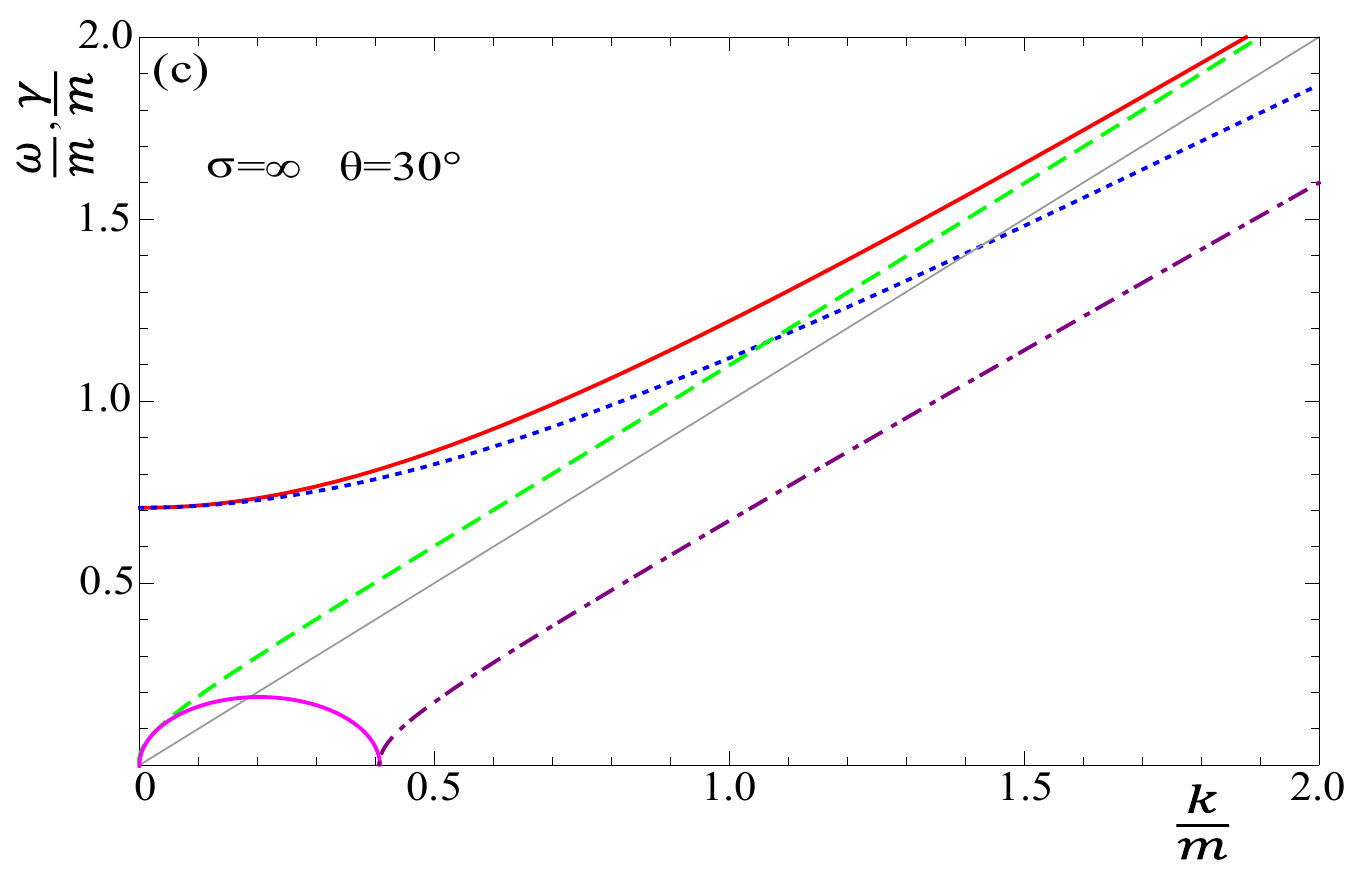}
\end{minipage}
\hspace{2mm}
\begin{minipage}{8.5cm}
\center
\includegraphics[width=1.02\textwidth]{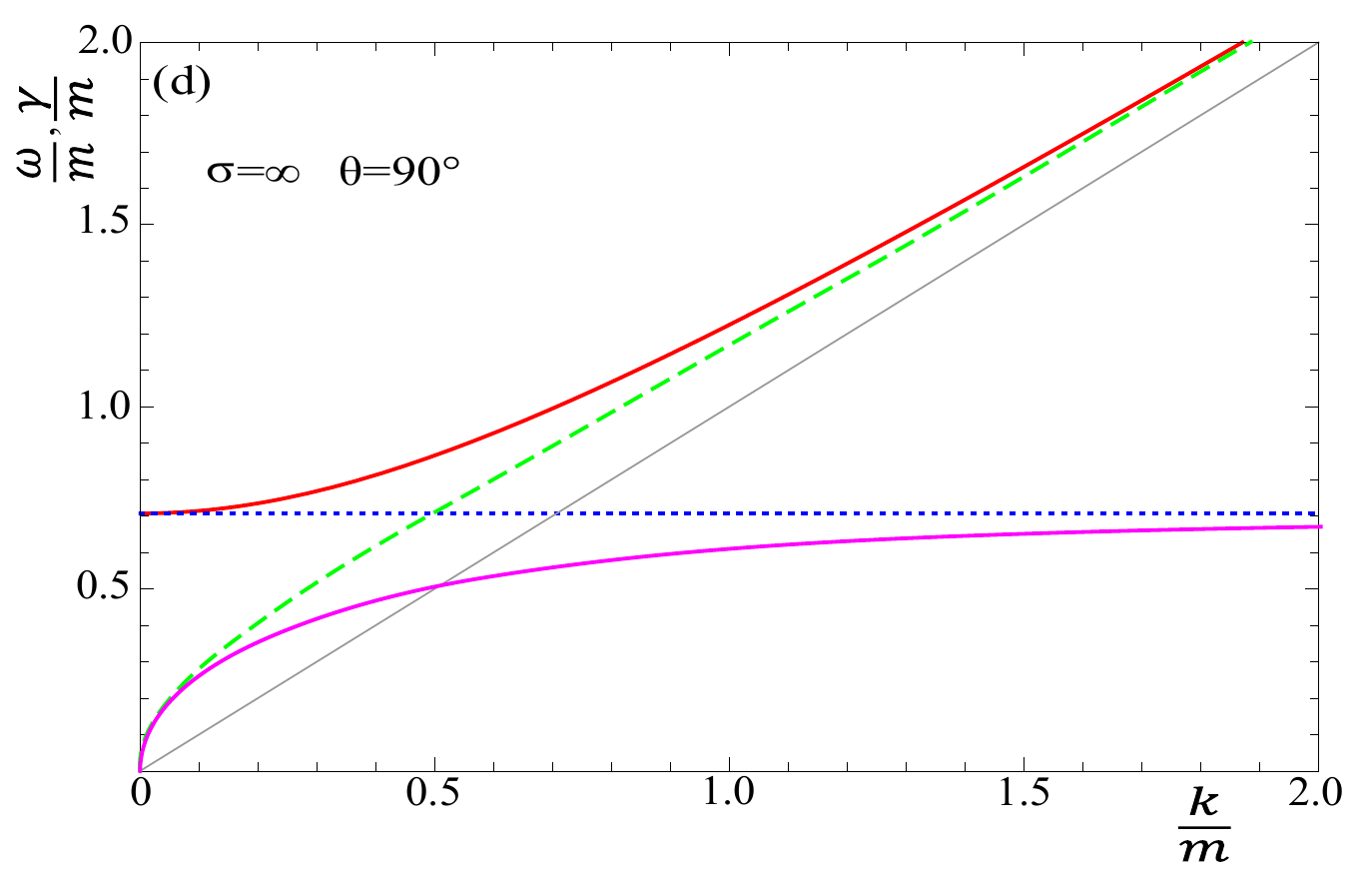}
\end{minipage}
\caption{(Color online) Dispersion curves of plasmons in extremely prolate plasma  for $\theta = 0^\circ$ (a), $\theta = 15^\circ$ (b), $\theta = 30^\circ$ (c) and $\theta = 90^\circ$ (d). }
\label{fig-ex-prolate-0-15-30-90}
\end{figure}

The extremely prolate system with the momentum distribution (\ref{extreme-prolate}) is the easiest case of all to study analytically. It was solved in \cite{Arnold:2003rq} using a different method. The coefficients $\alpha, \, \beta, \, \gamma, \, \delta$ and the inverse propagator $\Sigma$ defined by Eq.~(\ref{matrix-Sigma}) can be computed analytically without even specifying the coordinate system.  Since the velocity ${\bf v}$ of a massless parton is ${\bf v}={\bf n}$ for ${\bf p} \cdot {\bf n} > 0$ and ${\bf v} = - {\bf n}$ for ${\bf p} \cdot {\bf n} < 0$, the matrix $\Sigma$, which is pure real, is found to be
\ba
\label{Sigma-final-ex-pro}
\Sigma^{ij}(\omega,{\bf k}) = 
(\omega^2 - \frac{m^2}{2} -{\bf k}^2) \delta^{ij} +k^ik^j
-\frac{m^2 {\bf k}\cdot {\bf n}}
{2\big(\omega^2 - ({\bf k}\cdot {\bf n})^2\big)}
(k^i n^j + n^i k^j)
-\frac{m^2 \big(\omega^2 + ({\bf k}\cdot {\bf n})^2\big)
({\bf k}^2 - \omega^2)}
{2\big(\omega^2 - ({\bf k}\cdot {\bf n})^2\big)^2}
n^i n^j ,
\ea
and the coefficients $\alpha, \; \beta, \; \gamma, \; \delta$ are
\ba
\label{a-ex-pro}
\alpha (\omega,{\bf k}) &=& \frac{m^2}{2}  ,
\\[2mm]
\label{b-ex-pro}
\beta (\omega,{\bf k}) &=& \frac{m^2}{2} + \frac{m^2  ({\bf k}\cdot {\bf n})^2}
{\omega^2 - ({\bf k}\cdot {\bf n})^2}
+ \frac{m^2 \big(\omega^2 + ({\bf k}\cdot {\bf n})^2\big)
({\bf k}^2 - \omega^2)}
{2\big(\omega^2 - ({\bf k}\cdot {\bf n})^2\big)^2}
\frac{({\bf k}\cdot {\bf n})^2}{{\bf k}^2} ,
\\[2mm]
\label{c-ex-pro}
\gamma (\omega,{\bf k}) &=& 
\frac{m^2(\omega^2+(\mathbf{k}\cdot\mathbf{n})^2)({\bf k}^2-\omega^2)}
{2(\omega^2-(\mathbf{k}\cdot\mathbf{n})^2)^2}
\left(1 - \frac{(\mathbf{k}\cdot\mathbf{n})^2}{{\bf k}^2} \right) ,
\\[2mm]
\label{d-ex-pro}
\delta (\omega,{\bf k}) &=& 
\frac{m^2  ({\bf k}\cdot {\bf n})}
{2\big(\omega^2 - ({\bf k}\cdot {\bf n})^2\big)}
+ \frac{m^2 \big(\omega^2 + ({\bf k}\cdot {\bf n})^2\big)
({\bf k}^2 - \omega^2)}
{2\big(\omega^2 - ({\bf k}\cdot {\bf n})^2\big)^2}
\frac{({\bf k}\cdot {\bf n})}{{\bf k}^2} .
\ea

The dispersion equation for the $A$-modes (\ref{dis-eq-A}) has the simple solution 
\be
\label{prolate-Amode}
\omega_\alpha^2({\bf k}) = \frac{m^2}{2} + {\bf k}^2 .
\ee
Although the dispersion equation for the $G$-modes (\ref{dis-eq-G}) is rather complicated, it also has three relatively simple solutions
\ba
\label{general-solution-1}
\omega_2^2({\bf k}) &=&  \frac{m^2}{2} + ({\bf k}\cdot {\bf n})^2 ,
\\ [2mm]
\label{general-solution-2}
\omega_{\pm}^2({\bf k}) &=& 
\frac{1}{2}\Big({\bf k}^2 + ({\bf k}\cdot {\bf n})^2
\pm
\sqrt{{\bf k}^4 + ({\bf k}\cdot {\bf n})^4 
+ 2m^2 {\bf k}^2 - 2m^2 ({\bf k}\cdot {\bf n})^2
-2 {\bf k}^2({\bf k}\cdot {\bf n})^2} \; \Big) .
\ea
The modes $\omega_\alpha$, $\omega_2$ and  $\omega_+$ are real and exist for any ${\bf k}$. The solutions $\omega_\alpha$ and $\omega_+$ lie always above the light cone. The mode $\omega_2$ lies above the light cone for $k<\frac{m}{\sqrt{2}\sin\theta}$ and below for $k>\frac{m}{\sqrt{2}\sin\theta}$. The modes $\omega_+$ and $\omega_2$ cross each other at $k=\frac{m}{2\sin\theta}$. The solution $\omega_-$ can be either pure real or pure imaginary. It is imaginary for 
\be
\label{k-crit-pro}
k < k_{\rm pG} \equiv \frac{m}{\sqrt{2}} |\tan\theta| ,
\ee
and real for $k > k_{\rm pG}$. The solution  $i\gamma$, where $\gamma \equiv |\omega_-|$,  is the Weibel unstable mode, and $-i\gamma$ is its  overdamped partner. When ${\bf k} \perp {\bf n}$ or $\theta = 90^\circ$, the unstable mode exists for all values of $k$, as $k_{\rm pG}$ given by Eq.~(\ref{k-crit-pro}) goes to infinity. When ${\bf k} || {\bf n}$ or $\theta = 0^\circ$ the configuration is cylindrically symmetric and  there is no instability, since $k_{\rm pG} \to 0$. The real modes are  $\omega_\alpha^2({\bf k}) = \omega_2^2({\bf k})=m^2/2 + k^2$ and $\omega_+^2({\bf k}) = \omega_-^2({\bf k})=k^2$ in this limit.  

Some spectra of plasmons in an extremely prolate plasma are shown in Fig.~\ref{fig-ex-prolate-0-15-30-90} for different orientations of the wave vector ${\bf k}$. We use the following color scheme: red (solid) is $\omega_\alpha$, green (dashed) is $\omega_+$, blue (dotted) is $\omega_2$, pink (solid) is $\Im\omega_-$ and purple (dotted-dashed) is $\Re \omega_-$. The imaginary mode emerges at finite $\theta$ and it extends to infinite $k$ at $\theta =90^\circ$. The mode $\omega_\alpha({\bf k})$ is independent of $\theta$, and $\omega_2({\bf k})$ changes qualitatively when $\theta$ grows from $0^\circ$ to $90^\circ$. The mode $\omega_+({\bf k})$ is massless, that is $\omega_+(0) =0$, and its dispersion curve is everywhere concave, in contrast to other real dispersion curves which are usually convex. 

There is a qualitative difference between the plasmon spectra of the extremely prolate system, which is discussed here, and that of a system with prolateness characterized by the parameter $\sigma\gg 1$. In extremely prolate plasma, the mode $\omega_-$ given by the formula (\ref{general-solution-2}) exists for any wave vector ${\bf k}$: it is real for $k>k_{\rm pG}$ and imaginary for $k<k_{\rm pG}$. For a very large but finite $\sigma$, only the imaginary piece at $k<k_{\rm pG}$ is found. One could suspect that a solution has been missed in the numerical calculation, but the Nyquist analysis, which is presented in Sec.~\ref{app-ex-prolate}, proves that this is not the case. The key point is that when $\sigma \to \infty$ there is a change in the analytic properties of the left-hand-side of the $G$-mode dispersion equation (\ref{dis-eq-G}) as a function of $\omega$. The cut singularity at $\omega \in [-k,k]$ is replaced by double poles at  $\omega = \pm {\bf k} \cdot {\bf n}$ and the number of modes in extremely prolate plasma equals 8 for any ${\bf k}$. 

Another important point is that the limit of extreme prolateness is approached very slowly as $\sigma \to \infty$. To illustrate this point we consider, as an example, the coefficient $\alpha (\omega, {\bf k}=0)$ which is the plasma frequency of $A-$mode.  When computed with the $\sigma-$distribution (\ref{alter-ansatz}) we obtain
\be
\alpha_\sigma (\omega, {\bf k}=0) = \frac{m^2}{8} \frac{C_\sigma}{\sigma} 
\Big[ \Big(\sqrt{\frac{\sigma}{\sigma +1}} + \sqrt{\frac{\sigma + 1}{\sigma}}\Big)
\ln\Big(\frac{\sqrt{\sigma + 1} + \sqrt{\sigma}}{\sqrt{\sigma + 1} - \sqrt{\sigma}}\Big) - 2 \Big] ,
\ee
and for $\sigma \gg 1$ we have
\be
\label{pro-limit}
\alpha_\sigma (\omega, {\bf k}=0) \approx \frac{m^2}{2} \Big(1 - \frac{1}{\ln 4\sigma} \Big).
\ee
From this expression we find that even for $\sigma$ as large as $10^4$, the coefficient differs from its extremely prolate value at $\sigma \to \infty$ by 10\%. As will be shown in the subsequent section, when $\xi\to\infty$ the parameter $\alpha_\xi (\omega, {\bf k}=0)$ approaches the extremely oblate value much more quickly. 

\section{Extremely oblate plasma}
\label{sec-ex-oblate}

In this section we consider the second limiting case - the extremely oblate plasma with the momentum distribution given by  Eq.~(\ref{extreme-oblate}). The coefficients $\alpha, \, \beta, \, \gamma, \, \delta$, which have a much more complicated structure than for the extremely prolate plasma, equal 
\ba
\label{alpha-extremely-oblate}
\alpha(\omega,{\bf k}) &=&  \frac{m^2}{2(1-x^2)}
\Big[\hat\omega ^2-x^2+\frac{\hat\omega(1 -\hat\omega ^2)}{r_+ r_-}\Big],
\\[2mm]
\label{beta-extremely-oblate}
\beta(\omega,{\bf k})  &=& \frac{m^2\hat\omega^2}{2}
\Big[-1+\frac{\hat\omega (2 x^2+\hat\omega ^2-1)}{r_+^3 r_-^3}\Big],
\\[2mm]
\label{gamma-extremely-oblate}
\gamma(\omega,{\bf k})  &=& \frac{m^2 (\hat\omega ^2-1)}{2(1-x^2)} 
\Big[\frac{\hat\omega  \big(2 x^4+(x^2+1) \hat\omega ^2-x^2-1\big)}{r_+^3r_-^3}-x^2-1\Big],
\\[2mm]
\label{delta-extremely-oblate}
k\delta(\omega,{\bf k})  &=&\frac{m^2\hat\omega \, x }{2(1-x^2)}
\Big[\frac{-2 (x^2-1) \hat\omega ^2+x^2-\hat\omega ^4-1}{r_+^3 r_-^3} + \hat\omega\Big] ,
\ea
where $\hat\omega \equiv \omega/k$, $x \equiv \cos\theta$ and
\be
 r_+ r_- \equiv \Big(\hat\omega  + \sqrt{1-x^2} +i0^+ \Big)^{1/2} \Big(\hat\omega - \sqrt{1-x^2} +i0^+\Big)^{1/2} .
\ee
These results appeared previously in \cite{Romatschke:2004jh} with a different normalization.

The dispersion equations  (\ref{dis-eq-A}, \ref{dis-eq-G}) with the coefficients (\ref{alpha-extremely-oblate} - \ref{delta-extremely-oblate})  cannot be solved analytically. We have found numerically that there are no complex solutions, only pure real and pure imaginary ones. Using a Nyquist analysis we have verified that all solutions have been found by our numerical method. Details are given in Sec.~\ref{app-ex-oblate}. The $A$-mode dispersion equation (\ref{dis-eq-A}) has a pair of real solutions for all ${\bf k}$ and a pair of imaginary solutions if the wave vector obeys
\be
\label{k-crit-obl-A}
k < k_{\rm oA}  \equiv \frac{m}{\sqrt{2}} |\cot \theta| .
\ee
The $G$-mode dispersion equation (\ref{dis-eq-G}) has two pairs of real solutions for all ${\bf k}$ and a pair of imaginary solutions when the wave vector satisfies the condition
\be
\label{k-crit-obl-G}
k < k_{\rm oG}  \equiv \frac{m}{2}\Re \sqrt{\frac{|\cos\theta|\sqrt{\cos^2\theta + 4}+\cos^2\theta-2}{\sin^2\theta}} .
\ee
When $\cos^2\theta < 1/2$ (that is $90^\circ > \theta > 45^\circ$), the argument of the square root is negative, the real part of the root is zero, and the critical wave vector $k_{\rm oG}$ vanishes. One observes that $k_{\rm oA}$ is obtained from $k_{\rm pG}$ by changing the tangent function into a cotangent. As explained in Sec.~\ref{app-ex-oblate}, the critical values (\ref{k-crit-obl-A}, \ref{k-crit-obl-G}) are the values of $k$ for which the inverse propagators $\Delta_A^{-1}$ and $\Delta_G^{-1}$, given by Eqs.~(\ref{dis-eq-A}, \ref{dis-eq-G}), vanish at $\omega=0$.  

The total number of modes is 6, 8 or 10 exactly as in the weakly oblate case (\ref{weak-fin-A}, \ref{weak-fin-B}, \ref{weak-fin-C}). The numbers  can be written in a compact form as
\ba
\label{ob-fin-A}
A-{\rm modes}:~~ 2+ 2\Theta(k_{\text{oA}}-k),
\\[2mm]
\label{ob-fin-G}
G-{\rm modes}:~~ 4+ 2\Theta(k_{\text{oG}}-k).
\ea

\begin{figure}[t]
\begin{minipage}{8.5cm}
\center
\includegraphics[width=1.05\textwidth]{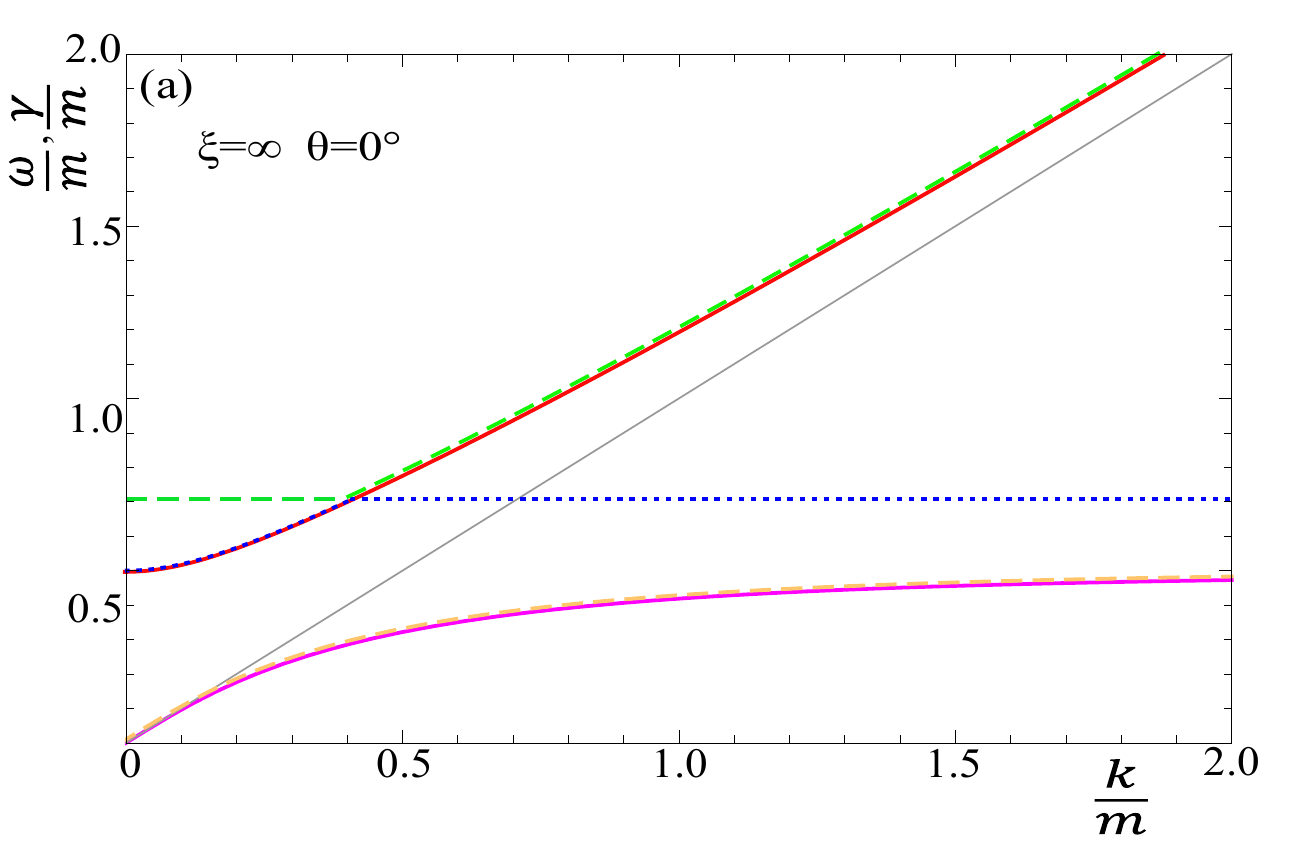}
\end{minipage}
\hspace{2mm}
\begin{minipage}{8.5cm}
\center
\includegraphics[width=1.02\textwidth]{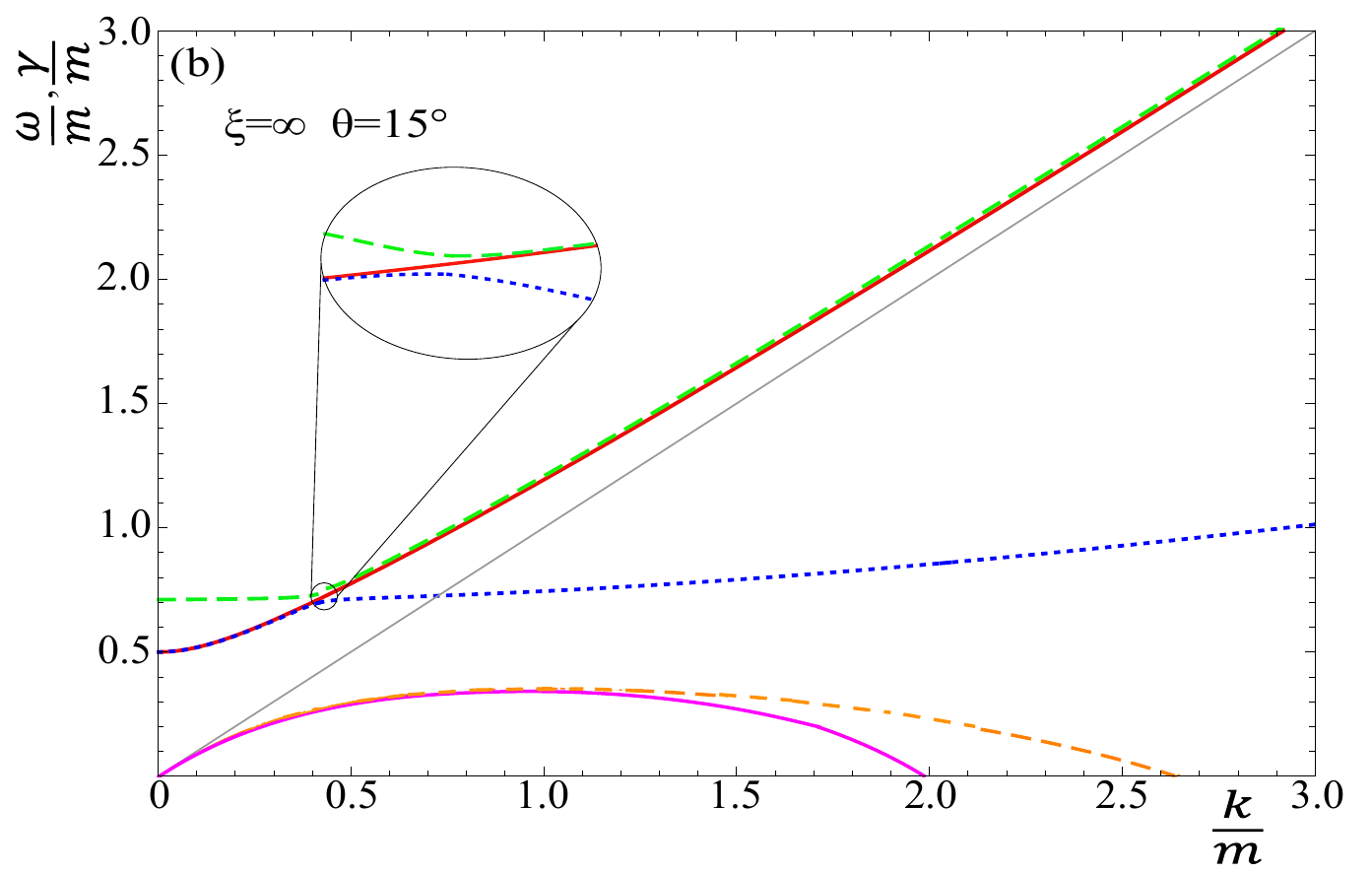} 
\end{minipage}
\vspace{2mm}
\begin{minipage}{8.5cm}
\center
\includegraphics[width=1.02\textwidth]{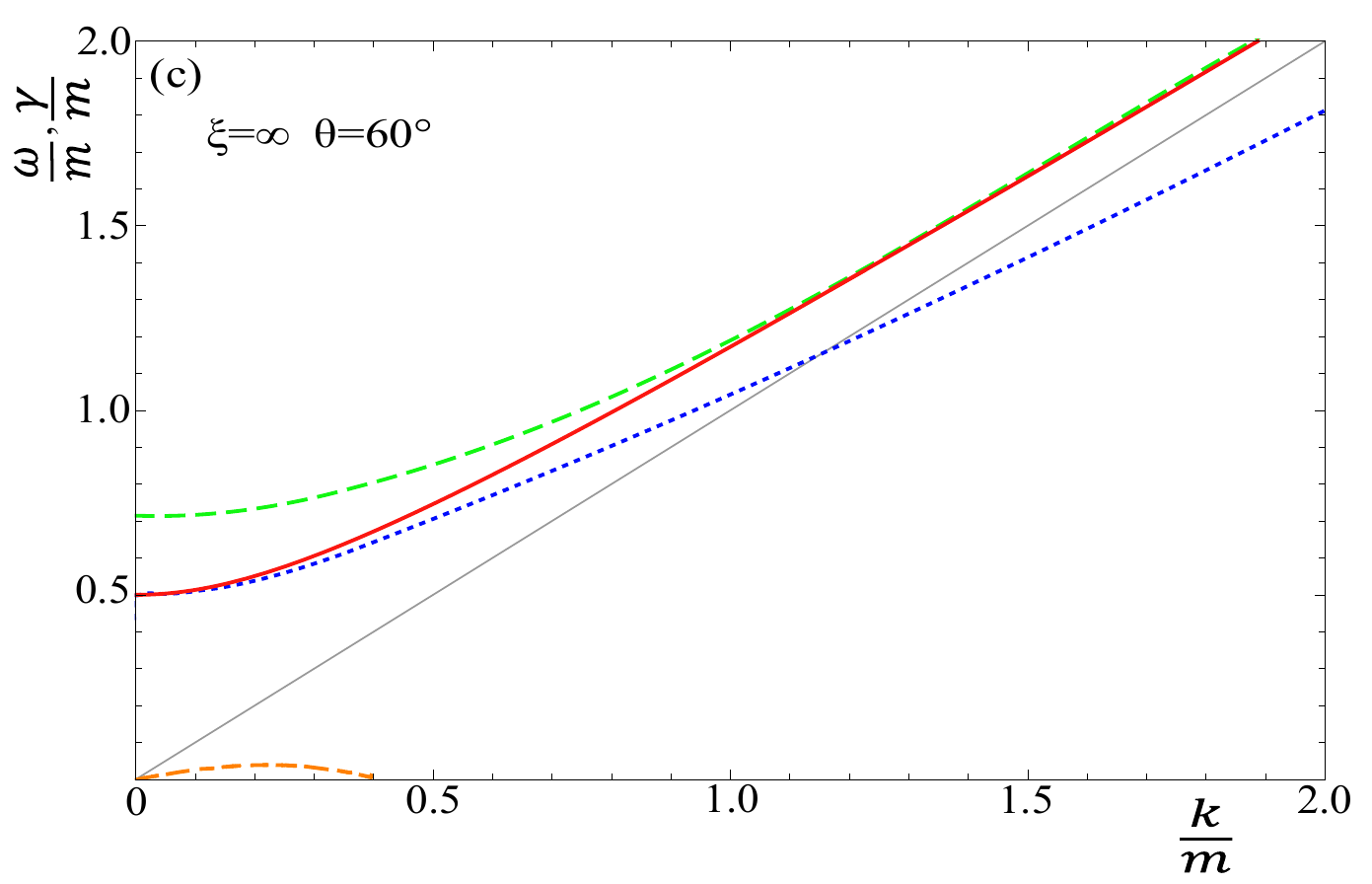}
\end{minipage}
\hspace{2mm}
\begin{minipage}{8.5cm}
\center
\includegraphics[width=1.02\textwidth]{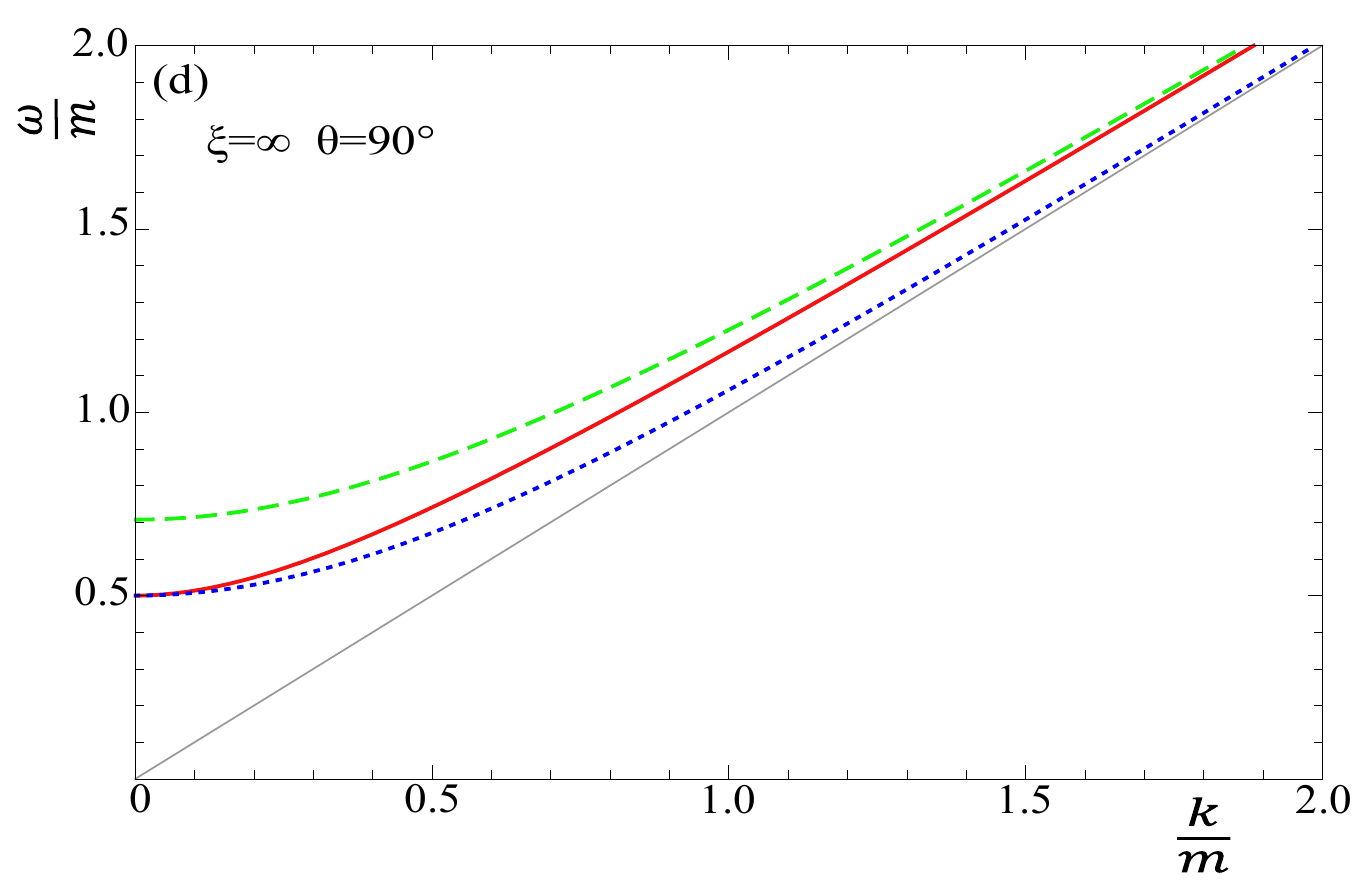}
\end{minipage}
\caption{(Color online) Dispersion curves of plasmons in extremely oblate plasma  for $\theta = 0^\circ$ (a), $\theta = 15^\circ$ (b), $\theta = 60^\circ$ (c) and $\theta = 90^\circ$ (d).}
\label{fig-extreme-oblate}
\end{figure}

In Fig.~\ref{fig-extreme-oblate} we show the dispersion curves obtained numerically from Eqs.~(\ref{dis-eq-A}, \ref{dis-eq-G}) for the angle $\theta$ equal $0^\circ,\; 15^\circ,\; 60^\circ$ and $90^\circ$.  When $\theta=0^\circ$ the real solutions $\omega_-$ and $\omega_+$ exhibit sharp corners at the same value of $k$. The $\omega_\alpha$ solution lies on top of the $\omega_-$ solution at small $k$ and on top of $\omega_+$ at large $k$. The two imaginary solutions extend through all values of $k$ and lie on top of each other, which is consistent with the observation that $k_{\rm oA}$ and $k_{\rm oG}$ both go to infinity at $\theta=0$. At $\theta=15^\circ$ we see that increasing the angle softens the corner in the real modes and causes the imaginary modes to retreat. The inset shows a blow-up of the region where the real modes approach each other. When $\theta$ has increased to $60^\circ$, the imaginary $G$-mode has dropped out, and at $90^\circ$ both imaginary modes are gone. 

The structure of the plasmon spectrum in Fig.~\ref{fig-extreme-oblate} is rather complicated. To understand it better, we consider three special limits which can be treated analytically. We start with ${\bf k}\parallel{\bf n}$ ($\theta=0^\circ$), then we discuss the situation when ${\bf k}$ is almost parallel to ${\bf n}$ ($|\sin\theta|\ll 1$), and finally we analyze the limit ${\bf k}\perp{\bf n}$ ($\theta=90^\circ$).

\subsection{Special case: ${\bf k} || {\bf n}$ }
\label{xequalspm1}

We first consider ${\bf k}=(0,0,k)$ parallel to ${\bf n}=(0,0,1)$. In this case, the vector ${\bf n}_T$, which is defined by Eq.~(\ref{nT-def}), vanishes. The decomposition using the basis $A,B,C,D$, which is introduced in Sec.~\ref{sec-decom-ABCD}, is therefore singular.  However, since there is only one independent vector in this case, one can decompose the inverse propagator or dielectric tensor using the same basis as in the isotropic case (\ref{Sigma-A-B}). One finds easily
\ba
\label{sf-parallel}
\alpha(\omega,{\bf k}) &=&  \frac{m^2}{2} - \frac{m^2( \omega^2 - k^2) }{4 \omega^2},
\\[2mm]
\label{sf-parallel-beta}
\beta(\omega,{\bf k}) &=&  \frac{m^2}{2} .
\ea
Alternatively, one can obtain these results in a straightforward way from the dielectric tensor (\ref{eij-1}) which can be easily computed. 
For the extremely oblate distribution with ${\bf k}  \perp {\bf n}$, we have ${\bf v}  \perp {\bf k}$ and the denominators of the second and third terms in the integrand in Eq.~(\ref{eij-1}) are $\omega$ and $\omega^2$, respectively. Observing further that the second term of the integral vanishes due to azimuthal symmetry, the dielectric tensor is diagonal. It is easy to show that the matrix $\Sigma$ equals
\ba
\label{sigma-k-||-n}
\Sigma (\omega, {\bf k}) = \left[
\begin{array}{ccc}
\omega^2 - k^2 - \alpha(\omega,{\bf k})  & 0 &0
\\[2mm] 
0 & 
\omega^2 - k^2 - \alpha(\omega,{\bf k}) & 0
\\[2mm]
0 & 0 & \omega^2 - \beta(\omega,{\bf k}) 
\end{array}
\right] \,.
\ea

The structure of the matrix $\Sigma$ given by Eq.~(\ref{sigma-k-||-n}) is similar to the isotropic case (\ref{sigma-iso}), except that the matrix components 11 and 33 are interchanged because the wave vector was chosen as ${\bf k}=(k,0,0)$ in Sec.~\ref{sec-iso}. The conclusion is therefore the same as for isotropic plasma: the solutions of the dispersion equation $\omega^2 - k^2 - \alpha(\omega,{\bf k})=0$ are transverse modes, which appear twice, and the solutions to $\omega^2  - \beta(\omega,{\bf k}) =0$ are longitudinal modes. Using Eqs. (\ref{sf-parallel}, \ref{sf-parallel-beta}) it is easy to find the dispersion relations which are  
\ba
\label{omega-pm-ex-oblate}
\omega^2_{\alpha} (k) &=& \frac{1}{2} \bigg( \frac{1}{4} \, m^2 + k^2 + \sqrt{ \Big( \frac{1}{4}\,m^2 + k^2 \Big)^2 + m^2 k^2} \; \bigg) 
\approx 
\left\{ \begin{array}{ccc} 
\frac{1}{4}\, m^2 +  2 k^2 \;\;\;\;\;
 & {\rm for} &  \;\;\;\;\; m^2 \gg k^2 ,
\\ [2mm]
k^2 
 & {\rm for} & \;\;\;\;\;  m^2 \ll k^2 ,
\end{array} \right.
\\[2mm]
\label{omega-pm-ex-oblate2}
\omega^2_{\alpha i} (k) &=& \frac{1}{2} \bigg( \frac{1}{4} \, m^2 + k^2 - \sqrt{ \Big( \frac{1}{4}\,m^2 + k^2 \Big)^2 + m^2 k^2} \; \bigg) 
\approx 
\left\{ \begin{array}{ccc} 
-k^2 \;\;\;\;\;
 & {\rm for} &  \;\;\;\;\; m^2 \gg k^2 ,
\\ [2mm]
- \frac{1}{4}\, m^2 
 & {\rm for} & \;\;\;\;\;  m^2 \ll k^2 ,
\end{array} \right.
\\[2mm]
\label{omega-pm-ex-oblate3}
\omega^2_\beta (k) &=& \frac{1}{2}\,m^2 .
\ea
Both $\omega_\alpha$ and $\omega_\beta$ are real solutions which exist for all $k$, and $\omega_{\alpha i} = i\gamma $ is an imaginary solution which also exists for all $k$. The maximum of the imaginary frequency is $\gamma_{\rm max} =  m/2$. 

From Eqs.~(\ref{ob-fin-A}, \ref{ob-fin-G}) the maximal number of solutions in extremely oblate plasma is 10.  
When ${\bf k} \parallel {\bf n}$ we have $\theta=0^\circ$, which means $k_{\rm oA}$ and $k_{\rm oG}$ both approach infinity. Therefore, our analysis of the special case ${\bf k} \parallel {\bf n}$ should produce the maximal number of solutions. Remembering that the transverse ($\alpha$) modes are doubled, Eqs. (\ref{omega-pm-ex-oblate}, \ref{omega-pm-ex-oblate2}, \ref{omega-pm-ex-oblate3}) correspond to 10 solutions. 

The solutions $\omega_\alpha^2$ and $\omega_\beta^2$ cross each other  at 
\be
\label{kc-ex-oblate}
k^2 = k_c^2 = \frac{m^2}{6}. 
\ee
Let us define two combinations of the real solutions:
\ba
\label{omega-2-zero-eps}
\omega_-^2 (k)  
=
\left\{ \begin{array}{ccc} 
\omega_\alpha^2 (k)   \;\;\;\;\; & {\rm for} &  \;\;\;\;\; k < k_c  ,
\\ [2mm]
\omega_\beta^2 (k)   \;\;\;\;\; & {\rm for} &  \;\;\;\;\; k > k_c  ,
\end{array} \right.
\ea
\ba
\label{omega-3-zero-eps}
\omega_+^2 (k)  
=
\left\{ \begin{array}{ccc} 
\omega_\beta^2 (k)   \;\;\;\;\; & {\rm for} &  \;\;\;\;\; k < k_c  ,
\\ [2mm]
\omega_\alpha^2 (k)   \;\;\;\;\; & {\rm for} &  \;\;\;\;\; k > k_c  .
\end{array} \right.
\ea
The dispersion curves are shown in Fig.~\ref{fig-extreme-oblate}a using the notation defined in Sec.~\ref{sec-sol-dis-eq}: the modes denoted $\omega_-$ and $\omega_+$ are represented, respectively, by the blue (dotted) and green (dashed) lines. As will be explained in the next subsection, the modes $\omega_+$ and $\omega_-$ are physical in the sense that one can obtain them by taking the limit $\theta \to 0^\circ$ of the solutions with the same names which were found at  $\theta > 0^\circ$.

\subsection{Special case: ${\bf k}$ almost  parallel to $ {\bf n}$ }

When the wave vector is not exactly along the $z$-axis but is slightly tilted, the spectrum of collective modes is changed qualitatively. To discuss this case we assume that the wave vector has a small $x$ component $k_x = k \sin \theta \approx k \theta$. The matrix $\Sigma$, which for $\theta = 0^\circ$ is given by Eq.~(\ref{sigma-k-||-n}), now contains small off-diagonal components $\sim k^2 \theta$ and is given by
\ba
\label{sigma-k-alomost-||-n}
\Sigma (\omega, {\bf k}) = \left[
\begin{array}{ccc}
-k^2 + \omega^2 - \frac{m^2}{2} + \frac{m^2( \omega^2 - k^2) }{4 \omega^2} & 0 & k^2 \theta 
\\[2mm]
0 & -k^2 + \omega^2 - \frac{m^2}{2} + \frac{m^2( \omega^2 - k^2) }{4 \omega^2} & 0
\\[2mm] 
 k^2 \theta & 0 & 
\omega^2 - \frac{m^2}{2}
\end{array}
\right] .
\ea

Computing the determinant of $\Sigma$, one finds two dispersion equations. The first reproduces the $\alpha$ modes in Eq.~(\ref{omega-pm-ex-oblate}), and the solutions are doubled as was the case for ${\bf k}$ parallel to ${\bf n}$. The second dispersion equation can be written as 
\be
\label{dis-eq-kT-small-3}
\frac{1}{\omega^2} \big(\omega^2 - \omega^2_\alpha (k) \big)\big(\omega^2 - \omega^2_{\alpha i} (k) \big)
\big(\omega^2 - \omega^2_\beta (k) \big) = k^4 \theta^2.
\ee
When $\theta=0^\circ$ we clearly recover the solutions of the previous section. Since the mode $\omega_{\alpha i}^2$ crosses neither $\omega_\alpha^2$ nor $\omega_\beta^2$, we express it as $\omega_{\alpha i}^2 = - \gamma^2$ and rewrite Eq.~(\ref{dis-eq-kT-small-3}) in the form
\be
\label{dis-eq-kT-small-4}
\big(\omega^2 - \omega^2_{\alpha} (k) \big)\big(\omega^2 - \omega^2_\beta (k) \big) = \epsilon ,
\ee
where $\epsilon \equiv \frac{\omega^2 k^4 \theta^2}{\omega^2 + \gamma^2}$. We want to look at the modes $\omega_\alpha$ and $\omega_\beta$ in the vicinity of the point where they cross. To lowest order in deviations from the solutions with $\theta=0^\circ$, we take $\epsilon$ as constant and solve the quadratic equation to obtain
\ba
\label{omega-2}
\omega^2_-  &=& \frac{1}{2} \Big(\omega^2_\alpha + \omega^2_\beta 
- \sqrt{\big(\omega^2_\alpha - \omega^2_\beta\big)^2 + 4\epsilon } \; \Big) ,
\\[2mm]
\label{omega-3}
\omega^2_+  &=& \frac{1}{2} \Big(\omega^2_\alpha + \omega^2_\beta
+ \sqrt{\big(\omega^2_\alpha - \omega^2_\beta\big)^2 + 4\epsilon } \; \Big) .
\ea
From these expressions, it is clear that the small parameter $\epsilon$ plays a role only in the vicinity of the crossing point where $\omega_\alpha = \omega_\beta$. Since both $\omega_\alpha^2$ and $\omega_\beta^2$ are positive, we have $\epsilon \ge 0$. Assuming that $(\omega^2_\alpha - \omega^2_\beta\big)^2 \gg \epsilon$, we expand the square roots in the formulas (\ref{omega-2}, \ref{omega-3}) to obtain
\ba
\omega_-^2 (k)  
=
\left\{ \begin{array}{ccc} 
\omega_\alpha^2 (k) - \frac{\epsilon}{| \omega^2_\alpha - \omega^2_\beta |}  \;\;\;\;\; & {\rm for} &  \;\;\;\;\; k < k_c  ,
\\ [2mm]
\omega_\beta^2 (k)  - \frac{\epsilon}{| \omega^2_\alpha - \omega^2_\beta |}  \;\;\;\;\; & {\rm for} &  \;\;\;\;\; k > k_c  ,
\end{array} \right.
\ea
\ba
\omega_+^2 (k)  
=
\left\{ \begin{array}{ccc} 
\omega_\beta^2 (k) + \frac{\epsilon}{| \omega^2_\alpha - \omega^2_\beta |}  \;\;\;\;\; & {\rm for} &  \;\;\;\;\; k < k_c  ,
\\ [2mm]
\omega_\alpha^2 (k)  + \frac{\epsilon}{| \omega^2_\alpha - \omega^2_\beta |}  \;\;\;\;\; & {\rm for} &  \;\;\;\;\; k > k_c  .
\end{array} \right. 
\ea
This result shows that the modes $\omega_-^2$ and $\omega_+^2$ approach each other at $k=k_c$ but do not cross. This is referred to as {\it mode coupling}, which is a general phenomenon that is explained in \S 64  of  \cite{Landau-Lifshitz-1981}. One can also show that the double imaginary mode $\omega_{\alpha i}$ splits into two different modes when $\theta$ is finite. 

The complete spectrum is presented in Fig.~\ref{fig-extreme-oblate}b for $\theta=15^\circ$.  As shown in the inset, the $\omega_+$ and $\omega_-$ modes  approach each other at $k = k_c$ but do not cross. The number of modes is the same as for the extremely oblate distribution with arbitrary values of $\theta$. 

\subsection{Special case: ${\bf k} \perp {\bf n}$ }

When ${\bf k} \perp {\bf n}$, the wave vector can be written as ${\bf k} = (k\cos\phi,k\sin\phi,0)$ and therefore the system can be treated as  effectively two-dimensional and {\em isotropic} in the $x\!-\!y$ plane. From Eqs. (\ref{k-crit-obl-A}, \ref{k-crit-obl-G}) we see that both of the critical wave vectors $k_{\text{oA}}$ and $k_{\text{oG}}$ go to zero in the limit $\theta \to 90^\circ$ and therefore the two imaginary modes disappear, as expected for an isotropic system. There should be two real solutions (one pair) from the $A$-mode dispersion equation (\ref{dis-eq-A}) and four real solutions (two pairs) from the $G$-mode equation (\ref{dis-eq-G}).

When $x \equiv \cos\theta=0$  the coefficients (\ref{alpha-extremely-oblate}, \ref{beta-extremely-oblate}, \ref{gamma-extremely-oblate}, \ref{delta-extremely-oblate}) simplify to
\ba
\label{a-kL0}
\alpha(\omega, {\bf k}) &=& 
\frac{m^2}{2} \frac{\omega^2}{k^2} \bigg( 1 -  \frac{\sqrt{\omega^2 - k^2}}{\omega} \bigg),
\\ [4mm]
\label{b-kL0}
 \beta(\omega, {\bf k}) &=&
\frac{m^2}{2} \frac{\omega^2}{k^2} \bigg(\frac{\omega}{\sqrt{\omega^2 - k^2}} - 1 \bigg) ,
\\ [4mm]
\label{c-kL0}
\gamma(\omega, {\bf k}) &=&
 \frac{m^2}{2} \frac{\omega^2 - k^2}{ k^2}
\bigg(\frac{\omega}{\sqrt{\omega^2 - k^2}} -1 \bigg) ,
\\ [4mm]
\label{d-kL0}
\delta(\omega, {\bf k}) &=& 0 ,
\ea
where $\omega \in \mathbb{R}$ and $\omega^2 > k^2$. 

Since $\delta(\omega, {\bf k}) = 0$, the second dispersion equation factors into two equations, as in the case of the weakly anisotropic plasma discussed in Sec.~\ref{sec-weakly-aniso}, and we solve the dispersion equations for $A$-modes, $B$-modes, and $C$-modes  (\ref{dis-eq-A}, \ref{b-mode}, \ref{c-mode}). The $A$-mode dispersion equation (\ref{dis-eq-A}) has the form
\be
(\omega^2 - k^2)k^2 +  \frac{m^2}{2} \Big(  \omega  \sqrt{\omega^2 - k^2}  - \omega^2 \Big) = 0 ,
\ee
which is quadratic in $\omega^2$ and can be solved analytically. The solution is 
\ba
\omega_\alpha^2 (k) = \frac{m^4 + 4 m^2k^2 -8 k^4 + m^3 \sqrt{ m^2 + 8 k^2}}{8(m^2-k^2)}  
\approx 
\left\{ \begin{array}{ccc} 
\frac{1}{4}\, m^2 +  \frac{5}{4}\, k^2 \;\;\;\;\;
 & {\rm for} &  \;\;\;\;\; m^2 \gg k^2 ,
\\ [2mm]
k^2 
 & {\rm for} & \;\;\;\;\;  m^2 \ll k^2 .
\end{array} \right.
\ea
The $B$-mode dispersion equation (\ref{b-mode}) simplifies to
\be
k^2 + \frac{m^2}{2} \bigg(1 -  \frac{\omega}{\sqrt{\omega^2 - k^2}} \bigg) = 0
\ee
and the solution gives the longitudinal mode
\ba
\omega_\beta^2 (k) = \frac{\big(\frac{m^2}{2} + k^2 \big)^2}{m^2 + k^2} \approx 
\left\{ \begin{array}{ccc} 
\frac{1}{4}\, m^2 +  \frac{3}{4}\, k^2 \;\;\;\;\;
 & {\rm for} &  \;\;\;\;\; m^2 \gg k^2 ,
\\ [2mm]
k^2 
 & {\rm for} & \;\;\;\;\;  m^2 \ll k^2 .
\end{array} \right.
\ea
Finally, the $C$-mode dispersion equation (\ref{c-mode}) becomes 
\be
\omega^2 - k^2 - \frac{m^2}{2} =  0 ,
\ee
which produces the solution
\be
\omega_{\alpha\gamma}^2 (k) = \frac{1}{2}\, m^2 + k^2 .
\ee
The $B$-mode and $C$-mode solutions are the limits $\theta \to 90^\circ$ of those found for arbitrary angles by solving numerically the $G$-mode dispersion equation (\ref{dis-eq-G}). The solution $\omega_{\alpha\gamma}$ is the larger of the two real $G$-modes (which we call $\omega_+$) and $\omega_\beta$ is the smaller $G$-mode (called $\omega_-$) which stays above the light cone for all $k$ when $\theta=90^\circ$. The dispersion curves for ${\bf k} \perp {\bf n}$ are shown in Fig.~\ref{fig-extreme-oblate}d. 

\begin{figure}[t]
\center
\includegraphics*[width=0.35\textwidth]{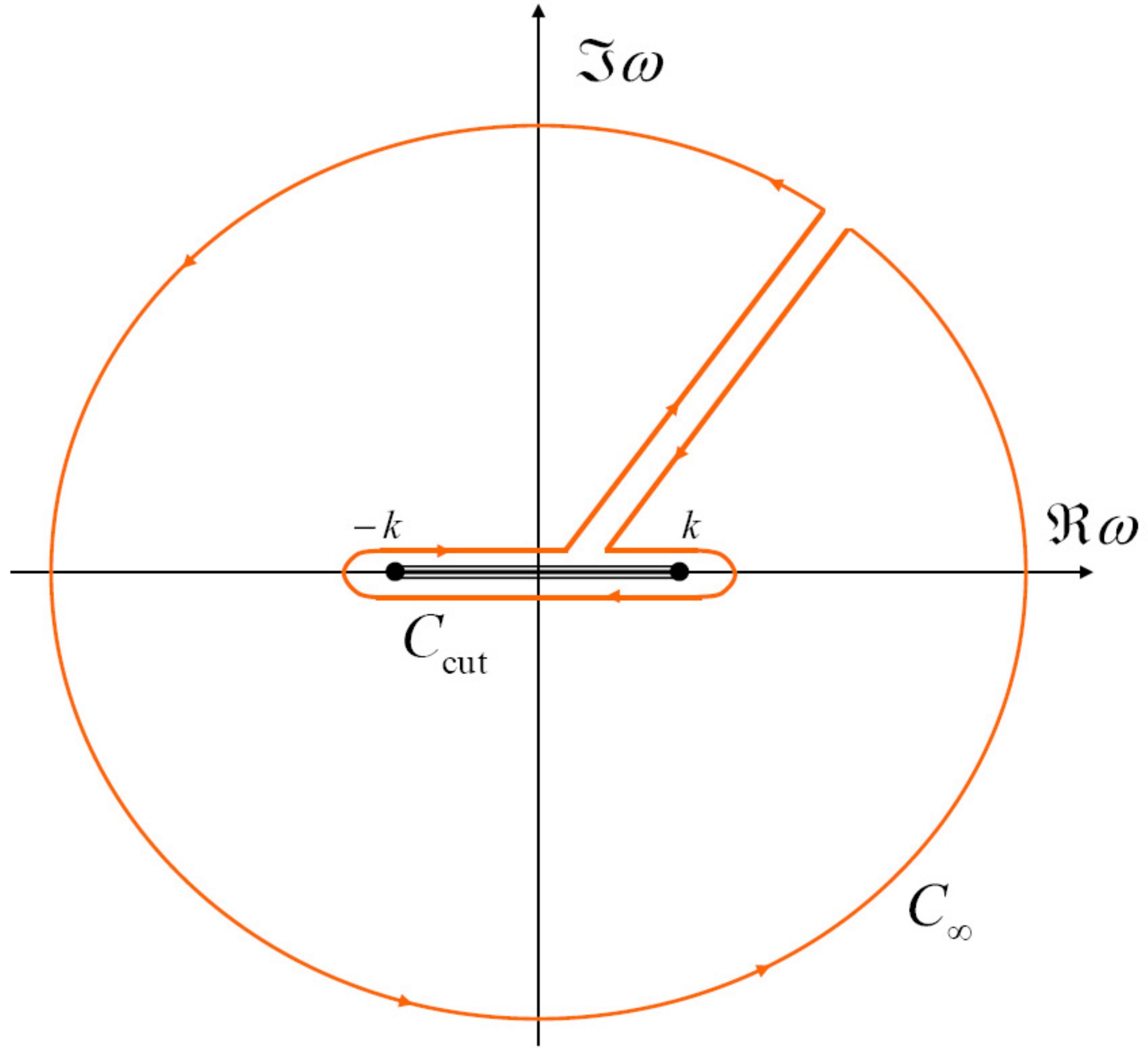}
\caption{(Color online) The contour $C$ in the plane of complex $\omega$ which is used to compute the number of solutions of some dispersion equations.} 
\label{fig-Nyquist-1}
\end{figure}

The spectrum of the extremely oblate system coincides with that of large but finite $\xi$. We also note that the limit $\xi \to \infty$ is approached much more quickly than the limit $\sigma \to \infty$, which was discussed at the end of Sec.~\ref{sec-ex-prolate}. In order to compare the two limits, we consider the same example: the coefficient $\alpha (\omega, {\bf k}=0)$ which is the mass of $A-$mode. Using the $\xi-$distribution (\ref{R-S-ansatz}) we find
\be
\alpha_\xi (\omega, {\bf k}=0) = \frac{m^2}{4} \frac{C_\xi}{\sqrt{\xi}} 
\Big[ \Big(1 - \frac{1}{\xi} \Big) {\rm Arctan}\sqrt{\xi} + \frac{1}{\sqrt{\xi}}  \Big] ,
\ee
and for $\xi \gg 1$ we have
\be
\alpha_\xi (\omega, {\bf k}=0) \approx \frac{m^2}{4} \Big(1 + \frac{2}{\pi \sqrt{\xi}} \Big).
\ee
From this expression we find that for $\xi = 10^4$ the coefficient $\alpha_\xi (\omega, {\bf k}=0)$ differs from the extremely oblate limit ($\xi \to \infty$) by only 0.6\%. In comparison, when $\sigma=10^4$, $\alpha_\sigma (\omega, {\bf k}=0)$ differs from the extremely prolate limit by 10\% (see Eq. (\ref{pro-limit})).

\section{Nyquist analysis}
\label{sec-Nyquist}

A Nyquist analysis allows one to determine the number of solutions of a given equation without solving the equation. Knowing the number of solutions is very important for our analysis of plasmons, because in all cases except that of the extremely prolate distribution, it is not possible to obtain exact analytic solutions of the dispersion equations. In some cases we have used analytic approximations, and in others we have found solutions numerically. When an approximation is used, there is a danger to find solutions that are artifacts of the approximation. When numerical methods are used, a solution that is outside the range of the search can be missed. 

To explain the idea of a Nyquist analysis, we discuss a generic equation of the form
\be
\label{general-eq}
f(\omega) = 0
\ee
and we define the function 
\be
\label{F-def}
F(\omega) \equiv \frac{f^\prime(\omega)}{f(\omega)} = \frac{d}{d\omega}{\rm ln}f(\omega) .
\ee
We consider the contour integral 
\be
\label{Nyq-int-1}
\oint_C \frac{d\omega}{2\pi i} F(\omega) ,
\ee
where the contour is a positively (counterclockwise) oriented closed loop, which is chosen so that $F(\omega)$ is analytic inside the loop except at isolated points. The integral is equal to the sum of the residues. It is straightforward to show that the residue of $F(\omega)$ at a zero of $f(\omega)$ of order $l$ is $l$, and the residue of $F(\omega)$ at a pole of $f(\omega)$ of order $l$ is $-l$. Thus, we have
\be
\label{Nyq-int-2}
\oint_C \frac{d\omega}{2\pi i} F(\omega) = n_Z - n_P ,
\ee
where $n_Z$ and $n_P$ are the numbers of zeros and poles of $f(\omega)$ inside the contour $C$, taking into account the fact that each zero and pole of order $l$ is counted $l$ times. Our aim is to determine $n_Z$.

\begin{figure}[t]
\begin{minipage}{8.5cm}
\center
\includegraphics[width=1.03\textwidth]{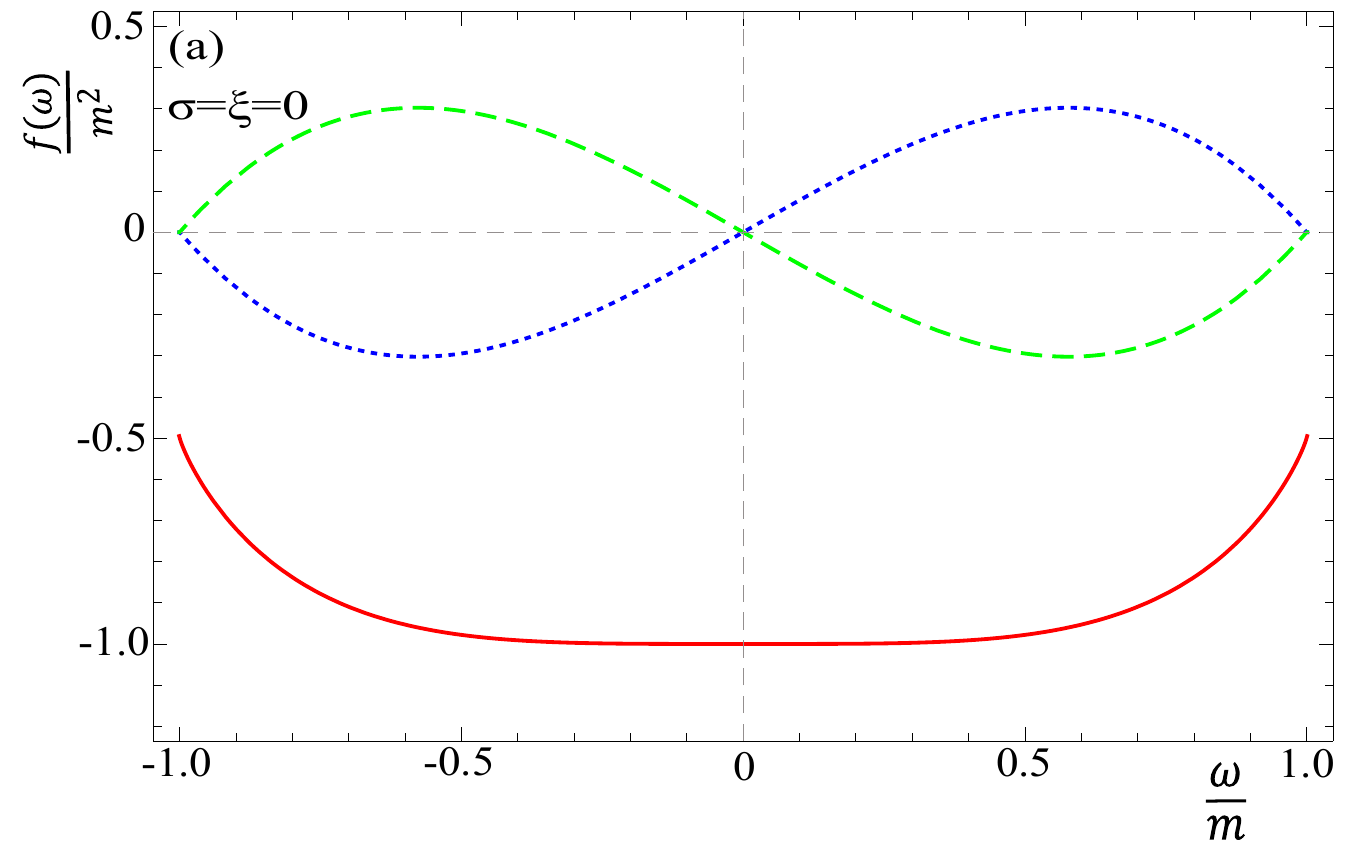}
\end{minipage}
\hspace{4mm}
\begin{minipage}{8.5cm}
\center
\includegraphics[width=1.01\textwidth]{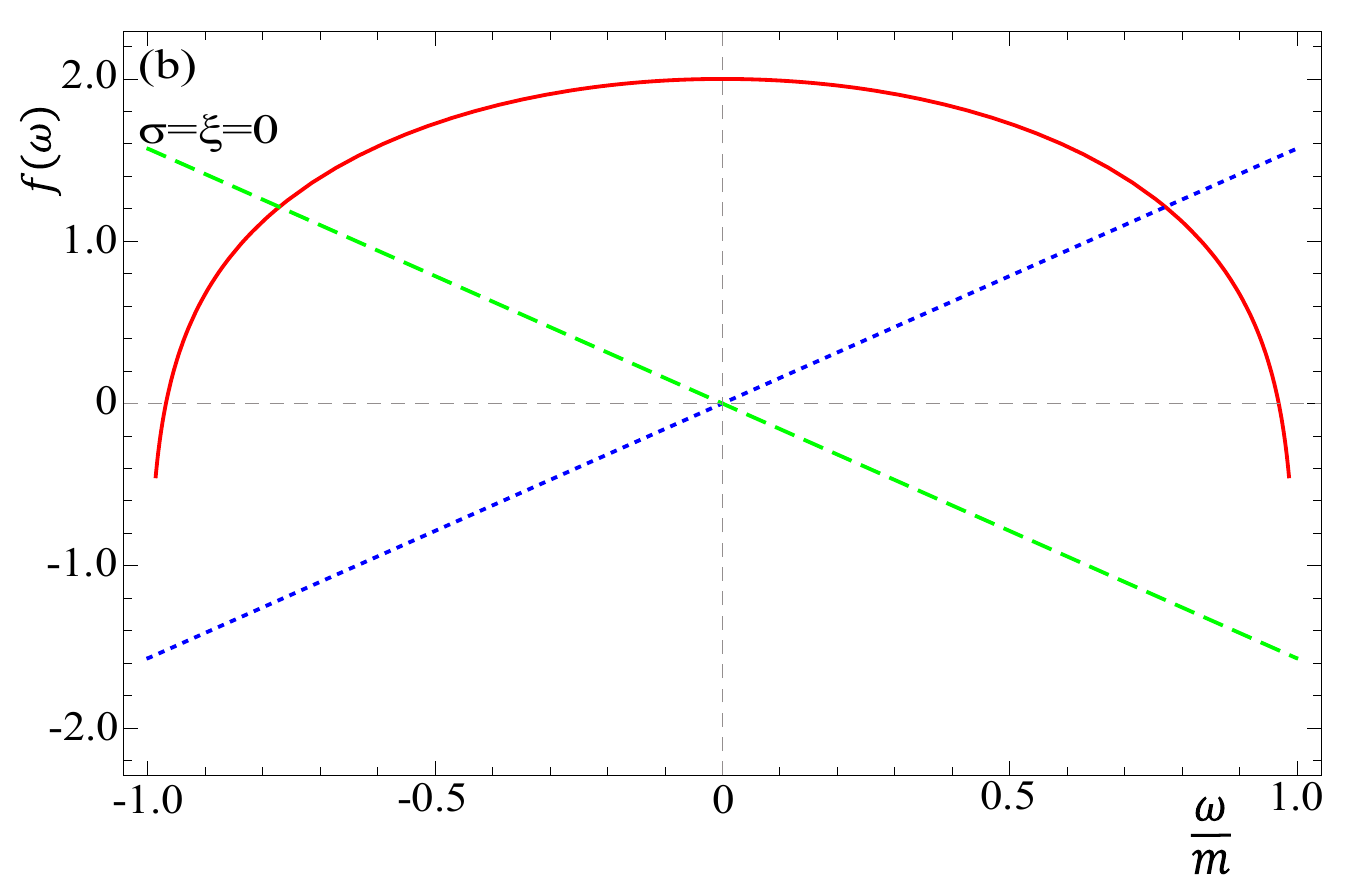}
\end{minipage}
\caption{(Color online) The real and imaginary parts of $f(\omega)$ for $A$-modes (a) and $B$-modes (b) in isotropic plasma for $k/m=1$ and $\omega$ along the cut. }
\label{iso-f-plots}
\end{figure}

The first step in the Nyquist analysis is to choose the contour $C$. If $f(\omega)$ has only isolated singular points, then $C$ can be chosen as the big circle which includes the entire plane of complex $\omega$. If $f(\omega)$ has cuts, then the contour must be chosen to exclude these cuts. For example, for isotropic plasma, which will be discussed in Sec.~\ref{app-iso}, $f(\omega)$ has a cut for $\omega \in [-k,k]$ and consequently the contour $C$ is chosen as in Fig.~\ref{fig-Nyquist-1}. For all of the momentum distributions we have considered in this paper (except the extremely prolate case), the dispersion equation has a cut along the real axis.  The contours we will use in the Nyquist analysis of all dispersion equations are all similar to those in Fig.~\ref{fig-Nyquist-1}, but the length of the cut depends on the particular distribution. The integrals along the lines connecting the circular contour $C_\infty$ to $C_{\rm cut}$ always compensate each other and therefore the contour integral (\ref{Nyq-int-2}) equals
\be
\label{Nyq-int-LM}
\oint_{C_\infty} \frac{d\omega}{2\pi i} F(\omega) +
\oint_{C_{\rm cut}} \frac{d\omega}{2\pi i} F(\omega) = n_Z - n_P.
\ee

The contribution from the big circle is easy to calculate by writing $\omega = |\omega|e^{i \phi}$ and taking $|\omega|\to\infty$. Using $d\omega = i \omega d\phi$, we have 
\be
\label{Nyq-int-infty}
\oint_{C_\infty} \frac{d\omega}{2\pi i} F(\omega) = \lim_{|\omega| \rightarrow \infty} \omega F(\omega) \equiv n_\infty .
\ee

The integral along the cut can be calculated using the fact that $F(\omega)$, defined by Eq.~(\ref{F-def}), is the logarithmic derivative of $f(\omega)$. Consequently 
\be
\label{Nyq-int-cut}
\oint_{C_{\rm cut}} \frac{d\omega}{2\pi i} F(\omega)  
= \frac{1}{2\pi i}\oint_{C_{\rm cut}}  \frac{d}{d\omega}{\rm ln}f(\omega)
= \frac{1}{2\pi i} \Big( {\rm ln} f(\omega_e) - {\rm ln} f(\omega_s) \Big) \equiv n_W ,
\ee
where $\omega_s$ is the (arbitrarily chosen) starting point of the contour which encloses the cut, and $\omega_e$ is the end point. The points $\omega_s$ and $\omega_e$ have the same modulus, but their phases differ by $2\pi$. The value of the right-hand-side of Eq.~(\ref{Nyq-int-cut}) can be found by mapping the closed contour $C_{\rm cut}$ in the plane of complex $\omega$ onto a path in the plane of complex $f(\omega)$.  Since the logarithm of $f$ has a cut, which runs along the real axis from $f = -\infty$ to $f = 0$, the value of the integral (\ref{Nyq-int-cut}) is a {\it winding number} (denoted $n_W$) which equals the number of times that the curve in the plane of complex $f$, which starts at  $f(\omega_s)$ and ends at $f(\omega_e)$, travels counterclockwise around the point $f=0$.  

Combining the results (\ref{Nyq-int-infty}, \ref{Nyq-int-cut}), we rewrite Eq.~(\ref{Nyq-int-LM}) as 
\be
\label{ny-all}
n_Z = n_P + n_\infty + n_W ,
\ee
which tells us that the number of zeros of the function $f(\omega)$ inside the contour $C$ equals the number of poles of $f(\omega)$ inside this contour, plus $n_\infty$ given by the limit (\ref{Nyq-int-infty}), plus the winding number (\ref{Nyq-int-cut}).  In the following subsections we show how to evaluate the numbers $n_P$, $n_\infty$ and $n_W$ for the dispersion equations we have solved in this paper. In each case, the only difficult piece is the calculation of $n_W$, for which we will need to determine the signs of the real and imaginary parts of the function $f(\omega)$ along the contour $C_{\rm cut}$. 

To illustrate the procedure, we will produce graphs of $f(\omega)$ using the following conventions. We plot $f(\omega)$ as a function of $\omega$ along the cut for various choices of the angle $\theta$. The real part of $f(\omega)$ is the same whether $\omega$ has a positive or negative infinitesimal imaginary part, and is represented as a red (solid) line. The imaginary part with positive infinitesimal imaginary part (values of $\omega$ along the top of the cut) is a blue (dotted) line, and for negative infinitesimal imaginary part (on the bottom of the cut) it is a green (dashed) line. 

\begin{figure}[t]
\begin{minipage}{6.5cm}
\center
\includegraphics[width=0.9\textwidth]{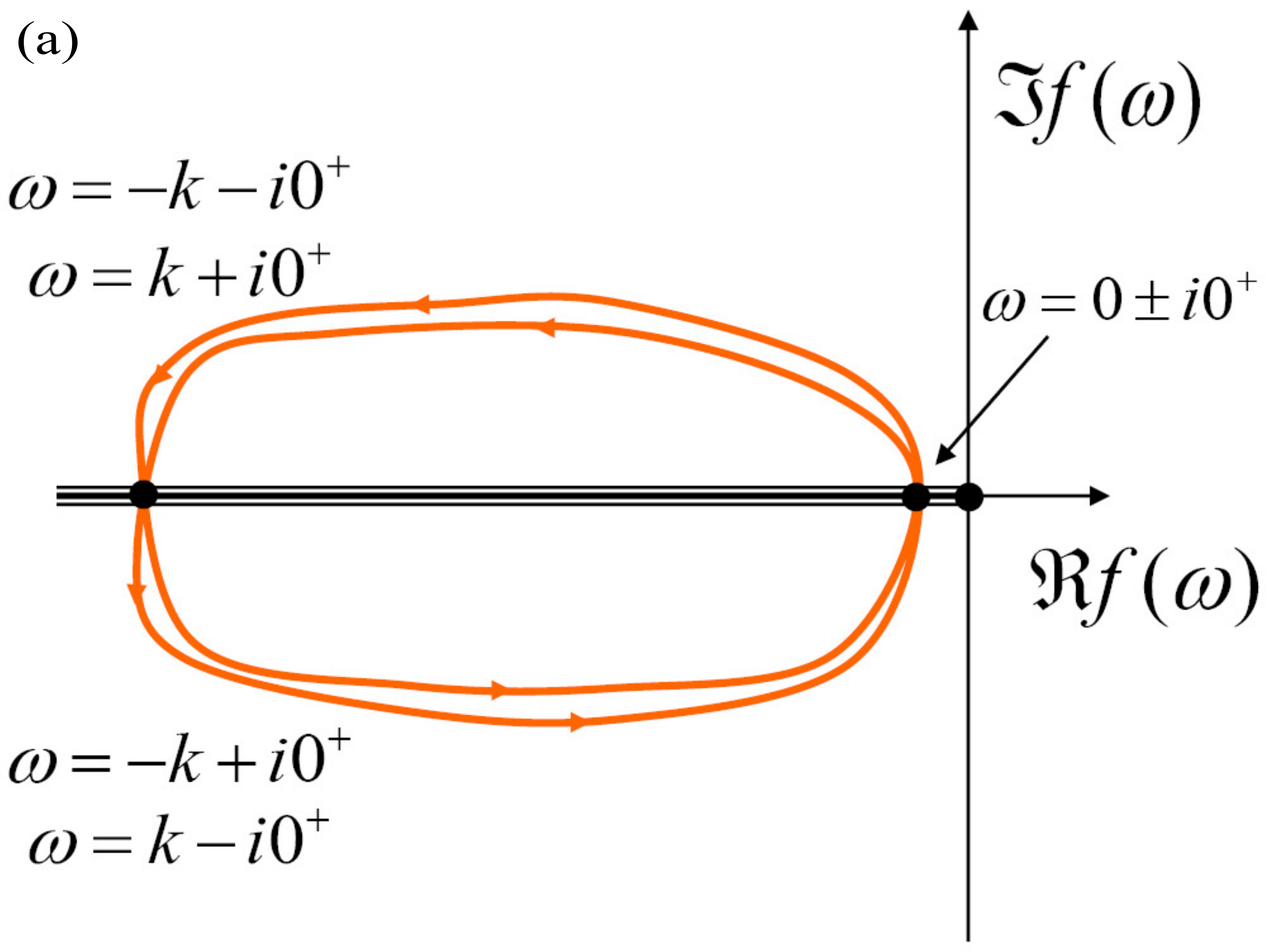}
\end{minipage}
\hspace{5mm}
\begin{minipage}{10.5cm}
\center
\includegraphics[width=0.72\textwidth]{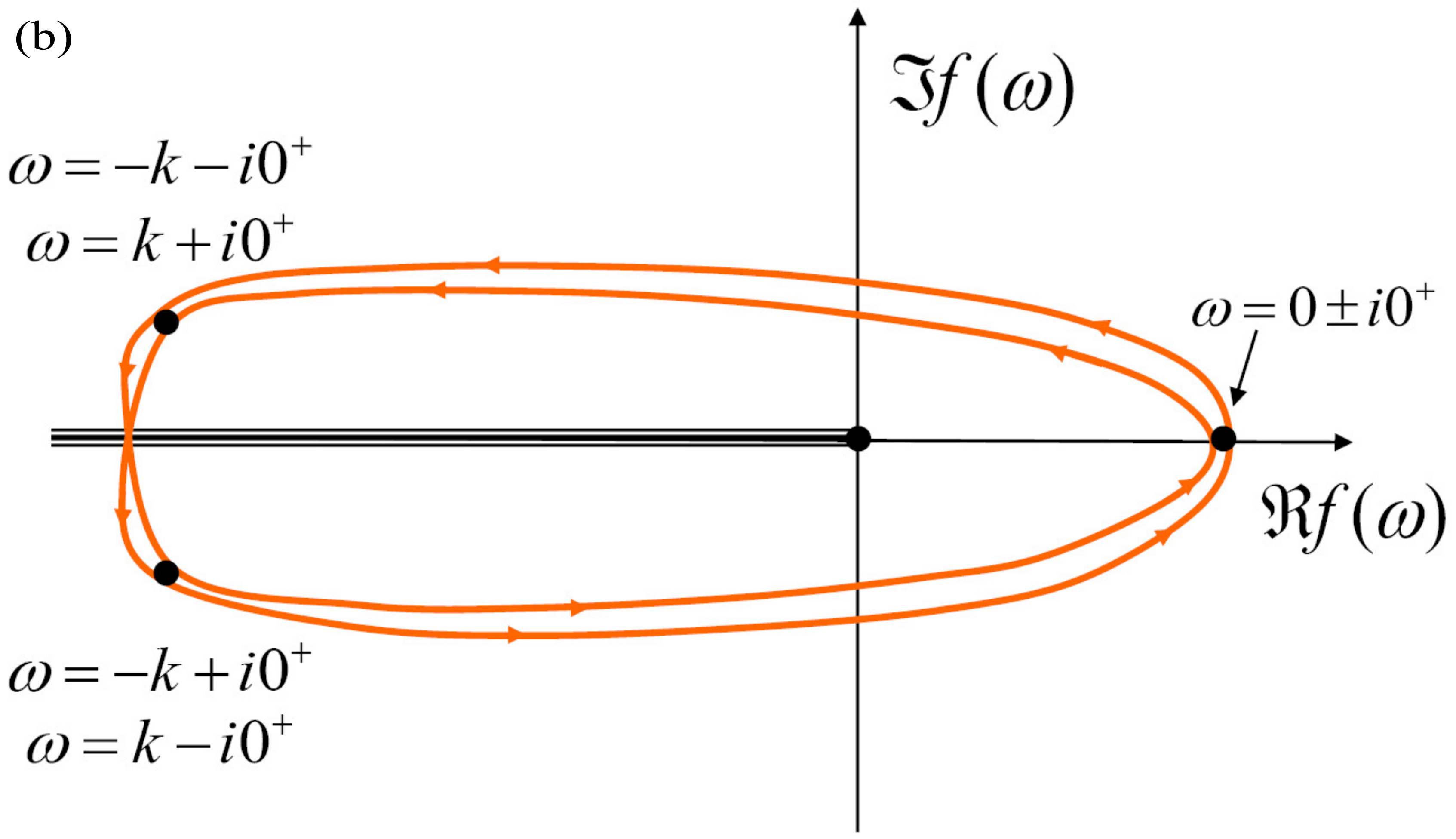}
\end{minipage}
\caption{(Color online) The mapping of the contour $C_{\rm cut}$ onto the path in complex $f(\omega)$ for isotropic plasma. The panel (a) shows the mapping for the $A$-mode dispersion equation and the panel (b) is for the $B$-mode dispersion equation.}
\label{fig-Nyquist-2}
\end{figure}

\subsection{Isotropic plasma}
\label{app-iso}

We start with the familiar case of isotropic plasma. It is instructive to see how the Nyquist analysis works in this case for which the answer is known. The function $f(\omega)$ for (transverse) $A$-modes and (longitudinal) $B$-modes is given by equations (\ref{dis-eq-iso-T}) and (\ref{dis-eq-iso-L}), respectively. The coefficients $\alpha_{\rm iso}(\omega,{\bf k})$ and $\beta_{\rm iso}(\omega,{\bf k})$ (given by Eqs.~(\ref{alpha-iso}, \ref{beta-iso})) have cuts for $\omega \in [-k,k]$, which means that the contour $C$ should be chosen as in Fig.~\ref{fig-Nyquist-1}. 
In both cases, $f(\omega)$ does not have any poles inside the contour, and therefore $n_P=0$. Using Eq.~(\ref{Nyq-int-infty}) it is straightforward to show that $n_\infty = 2$ for $A$-modes and $n_\infty = 0$ for $B$-modes. 

The winding number $n_W$ is calculated from Eq.~(\ref{Nyq-int-cut}) by mapping the contour $C_{\rm cut}$ in the plane of complex $\omega$ onto the plane of complex $f(\omega)$. In Fig.~\ref{iso-f-plots} we show the real and imaginary parts of $f(\omega)$ as functions of $\omega$, considering only values of $\omega$ which lie infinitesimally above and below the cut. The structure of the graph reflects the well known symmetry properties of the dielectric functions (\ref{alpha-beta-eT-eL}) 
\be
\Re \varepsilon_{L,T}(- \omega, {\bf k}) = \Re \varepsilon_{L,T}(\omega, {\bf k}) ,
~~~~~~~~
\Im \varepsilon_{L,T}(- \omega, {\bf k}) = -\Im \varepsilon_{L,T}(\omega, {\bf k}) .
\ee

We move around the cut shown in Fig.~\ref{fig-Nyquist-1} in the counter-clockwise direction, using the data presented in Fig.~\ref{iso-f-plots}. We describe the process of mapping for $A$-modes. Let us start, for example, at the top left corner of the cut where $\omega=-k+i0^+$. The red (solid) line in Fig.~\ref{iso-f-plots}a tells us that the real part of the dispersion equation at this value of $\omega$ is negative, and the blue (dotted) curve tells us that the imaginary part is also negative. Combining these results, $f(\omega= - k+ i0^+)$ is found to be in the third quadrant of its complex plane, as shown in the bottom left corner of Fig.~\ref{fig-Nyquist-2}a. Continuing in the same fashion, one produces the map shown in Fig.~\ref{fig-Nyquist-2}a where the point $f=0$ is circumnavigated zero times. Equivalently, the cut of the function ${\rm ln}f$ is crossed twice in each direction, so that ${\rm ln}f$ remains on the same Riemann sheet, and consequently the integral (\ref{Nyq-int-cut}) vanishes. The conclusion is that the winding number $n_W$ equals zero. The equivalent mapping for $B$-modes is shown in Fig.~\ref{fig-Nyquist-2}b. In this case, the circular trajectory is shifted to the right so that the origin is enclosed within the loop. The mapping circumnavigates the origin twice, and the winding number is therefore $n_W=2$. 

Combining the results derived above, Eq.~(\ref{ny-all}) gives
\ba
A-{\rm modes}:~~n_Z=0+2+0=2,\nonumber\\
B-{\rm modes}:~~n_Z=0+0+2=2,
\ea
which agrees with the known result that the dispersion equations for (transverse) $A$-modes and (longitudinal) $B$-modes both have two solutions (one pair of positive and negative real solutions). 

We note that for $B$-modes, the non-zero winding number is obtained because the circular trajectory in Fig.~\ref{fig-Nyquist-2}b is shifted to the right so that the origin is inside the loop, or equivalently, because the real part of $f(\omega=0)$ is positive. In the following sections we will see that for anisotropic plasmas (except in the case of the extremely prolate distribution) there are cases for which $f(\omega=0)>0$, $n_W\ne 0$, and extra solutions to the dispersion equations appear, when a specific condition on the  wave vector $\bf k$ is satisfied. These extra solutions are imaginary modes, which do not exist in isotropic plasmas. We comment that the Nyquist analysis itself can only tell us the number of solutions of a given equation but gives no information about the nature (real, imaginary or complex) of these solutions.

\subsection{Weakly anisotropic plasma}
\label{app-weakly-aniso}

In the case of a weakly anisotropic plasma, there are three dispersion equations (\ref{dis-eq-A}, \ref{b-mode}, \ref{c-mode}) whose solutions give $A$-, $B$- and $C$-modes. The components  $\alpha, \, \beta, \, \gamma$ of the polarization tensor are given in Sec.~\ref{sec-weakly-aniso}. We first note that the analytic structure of $f(\omega)$ in all three cases is the same as for isotropic plasma, and therefore we can use the contour shown in Fig.~\ref{fig-Nyquist-1}. For all three equations there are no poles inside the contour, which means $n_P=0$.

We start by considering Eq.~(\ref{b-mode}) which produces $B$-modes. It is easy to show that $n_\infty = 0$, as for the isotropic case. The mapping of the cut into the complex plane of $f(\omega)$ also looks like the mapping for the isotropic case (Fig.~\ref{fig-Nyquist-2}b), and therefore $n_W=2$, also as for the isotropic case. The conclusion is that there are 2 solutions - the same as for longitudinal modes in isotropic plasmas. 

Now we consider $A$-modes and $C$-modes.  Equation ~(\ref{Nyq-int-infty}) gives $n_\infty = 2$ in both cases, and therefore equation (\ref{ny-all}) tells us that the number of solutions is $n_Z=2 +n_W$. For isotropic $A$-modes we showed in the previous section that $f(\omega=0) \le 0$ for any ${\bf k}$, $n_W=0$, and the number of solutions is always 2. We will show below that for anisotropic plasmas with arbitrarily small $\xi$, for both $A$-modes and $C$-modes, there are wave vectors for which $f(\omega=0)>0$, $n_W > 0$, and additional solutions appear. 

\begin{figure}[t]
\begin{minipage}{8.5cm}
\center
\includegraphics[width=0.97\textwidth]{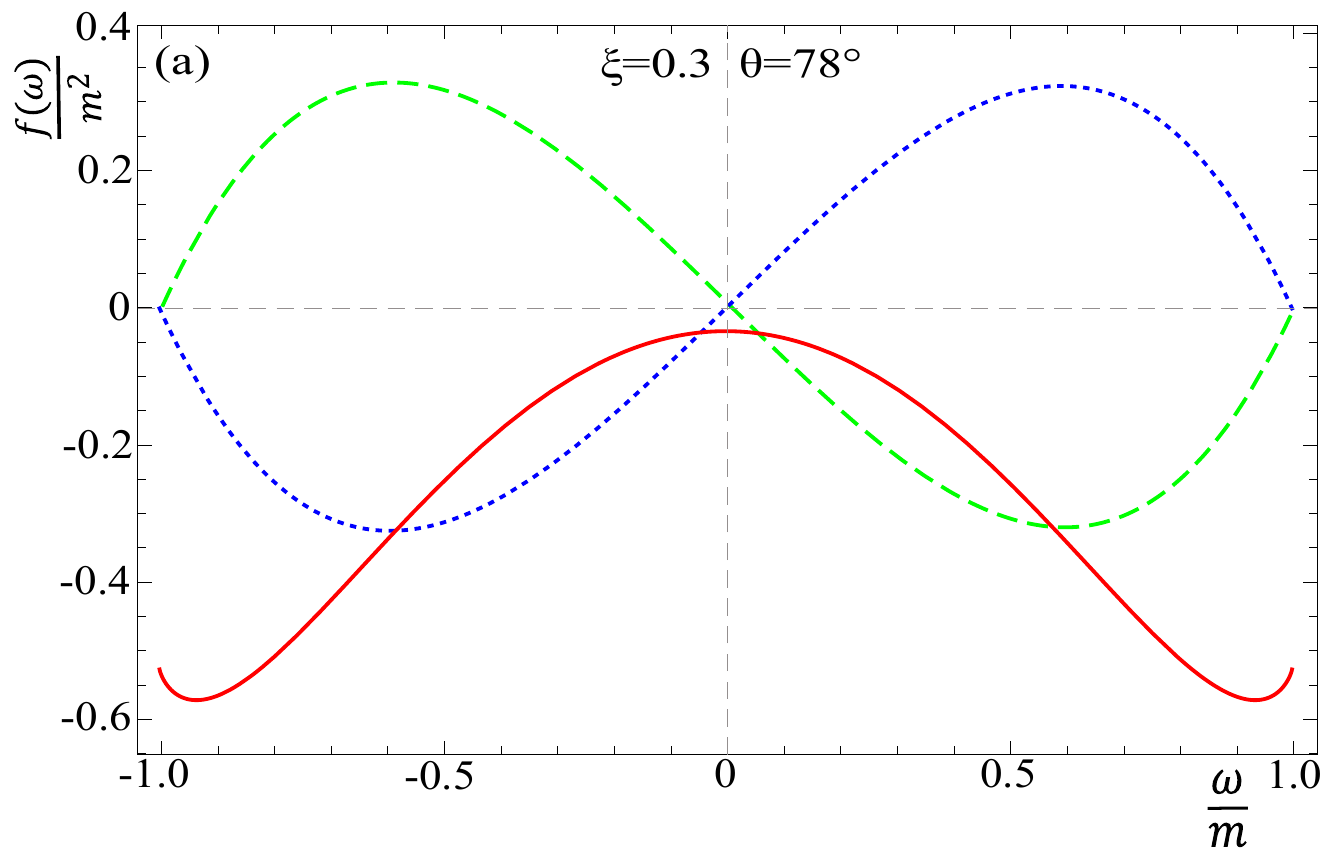}
\end{minipage}
\hspace{5mm}
\begin{minipage}{8.5cm}
\center
\includegraphics[width=1.0\textwidth]{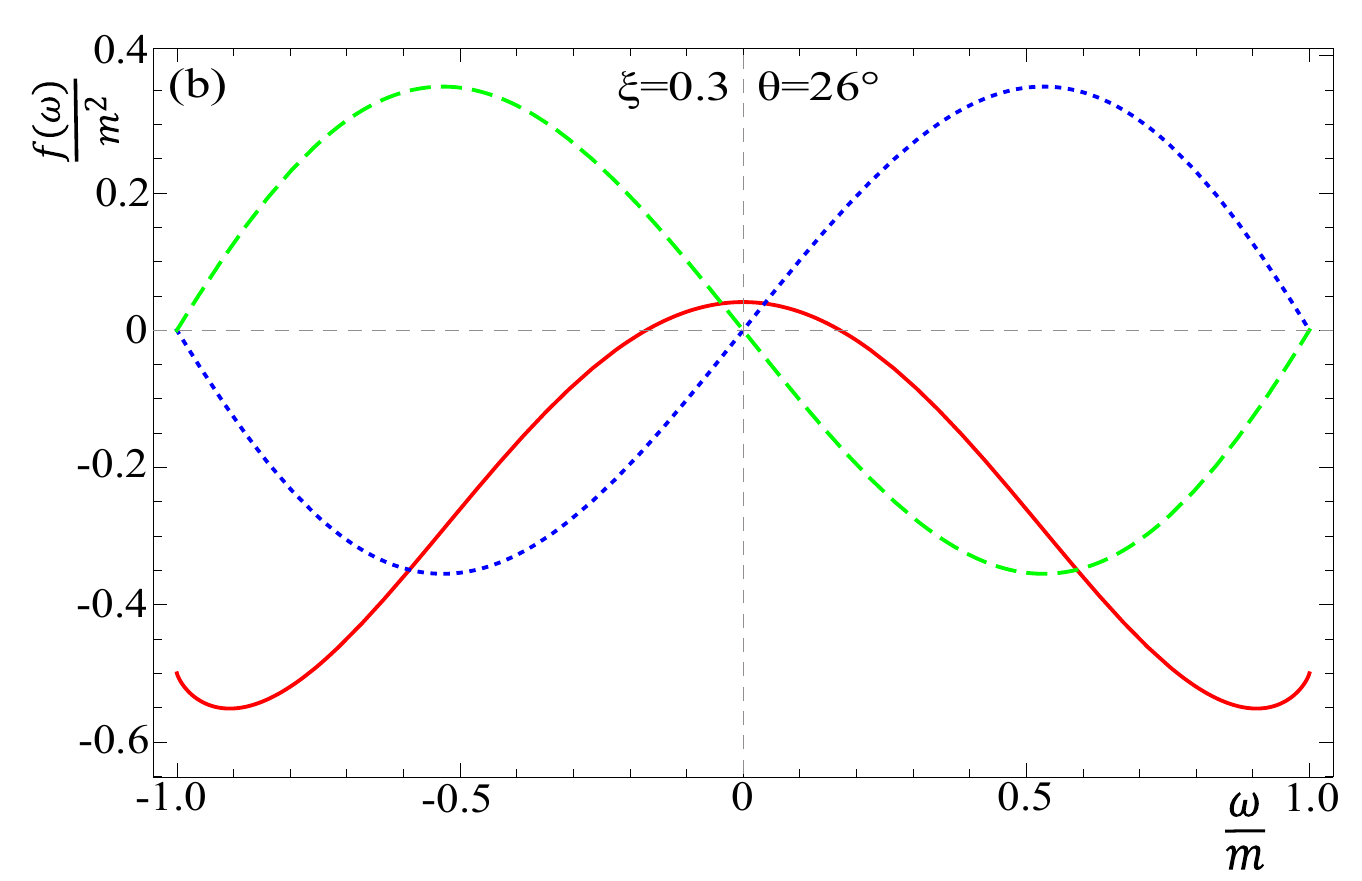}
\end{minipage}
\caption{(Color online) The real and imaginary parts of $f(\omega)$ for $A$-modes in weakly anisotropic plasma with $\xi=0.3$ for $\omega$ along the cut and $k/m=0.2$. In the panel (a) $\theta=78^\circ$ and in the panel (b) $\theta =26^\circ$.}
\label{fig-weakA}
\end{figure}

For the $A$-mode and $C$-mode dispersion equations, $f(\omega=0)$ can be either negative or positive depending on the length and orientation of the wave vector ${\bf k}$. Two examples are shown in Fig.~\ref{fig-weakA} for $A$-modes. When $f(\omega=0) \le 0$, as in the panel (a) of the figure, the mapping of the contour $C_{\rm cut}$ in the plane of complex $\omega$ onto the plane of complex $f(\omega)$ looks qualitatively like in Fig.~\ref{fig-Nyquist-2}a and we have $n_W =0$. When $f(\omega=0) > 0$, as shown in Fig.~\ref{fig-weakA}b, the mapping looks like Fig.~\ref{fig-Nyquist-2}b, and $n_W =2$. The corresponding analysis of $C$-modes is qualitatively similar.

We can find analytically the condition that distinguishes the case of $n_W=0$ from that of $n_W=2$.  Using Eqs. (\ref{alpha-iso}, \ref{alpha-final}, \ref{gamma-final}), one obtains
\ba 
&& A-{\rm modes}:~~f(\omega=0) = - k^2 + \xi \, \frac{m^2}{3} \,\cos^2\theta ,
\\[2mm]
&& C-{\rm modes}:~~f(\omega=0) =  - k^2 - \xi \, \frac{m^2}{3} \,\big( 1 - 2 \cos^2\theta \big).
\ea
From these expressions we can derive a critical value of the magnitude of the wave vector (as a function of $\theta$) at which $f(\omega=0)$ becomes positive and the winding number changes from 0 to 2. These critical values are denoted $k_A$ and $k_C$ and given in Eqs. (\ref{k-crit-A}, \ref{k-crit-C}). 
The number of solutions of the $A$-mode dispersion equation is either 2 or 4, depending on whether $k$ is smaller or greater than $k_A$, and the $C$-mode dispersion equation has either 2 or 4 solutions, depending on whether $k$ is smaller or greater than $k_C$. These results agree with Eqs. (\ref{weak-fin-A}, \ref{weak-fin-C}).

\subsection{Extremely prolate plasma}
\label{app-ex-prolate}

A Nyquist analysis is not necessary for the extremely prolate system, because one can find exact analytic solutions (\ref{prolate-Amode}, \ref{general-solution-1}, \ref{general-solution-2}) to the dispersion equations (\ref{dis-eq-A}, \ref{dis-eq-G}), and therefore there is no possibility that solutions have been missed.  The extremely prolate distribution is interesting for a different reason however. There is a qualitative difference between the numerical solutions found using the $\sigma$-distribution (\ref{alter-ansatz}) with very large values of $\sigma$ and the solutions obtained from the extremely prolate distribution, which corresponds to $\sigma \to \infty$. The analogous statement is not true for the $\xi$-distribution. In order to understand this point, we have done a Nyquist analysis of the extremely prolate case. 

The analysis of the $A$-mode dispersion equation (\ref{dis-eq-A}) is completely trivial. The function $f(\omega)$ has no poles and no cuts, and therefore we choose the contour as a big circle that includes the whole complex plane. Equation~(\ref{dis-eq-A}) has two solutions, $n_Z = n_\infty = 2$. In the case of the $G$-mode dispersion equation (\ref{dis-eq-G}), there are no cuts, the contour is chosen as the same big circle, and the winding number is zero. The function $f(\omega)$ has two double poles at $\omega = \pm {\bf k}\cdot {\bf n}$ and thus $n_P = 2 \times 2 = 4$. From the definition (\ref{Nyq-int-infty}) one obtains $n_\infty = 2$.  Combining these results, Eq.~(\ref{ny-all}) gives $n_Z = 6$. Adding the $A$-modes and $G$-modes together we reproduce the result from Sec.~\ref{sec-ex-prolate}, that the extremely prolate system has a total of 8 solutions at all wave vectors, in contrast to the system with very large prolate anisotropy, which has 8 solutions only for certain wave vectors. 

\begin{figure}[t]
\begin{minipage}{8.5cm}
\center
\includegraphics[width=1.08\textwidth]{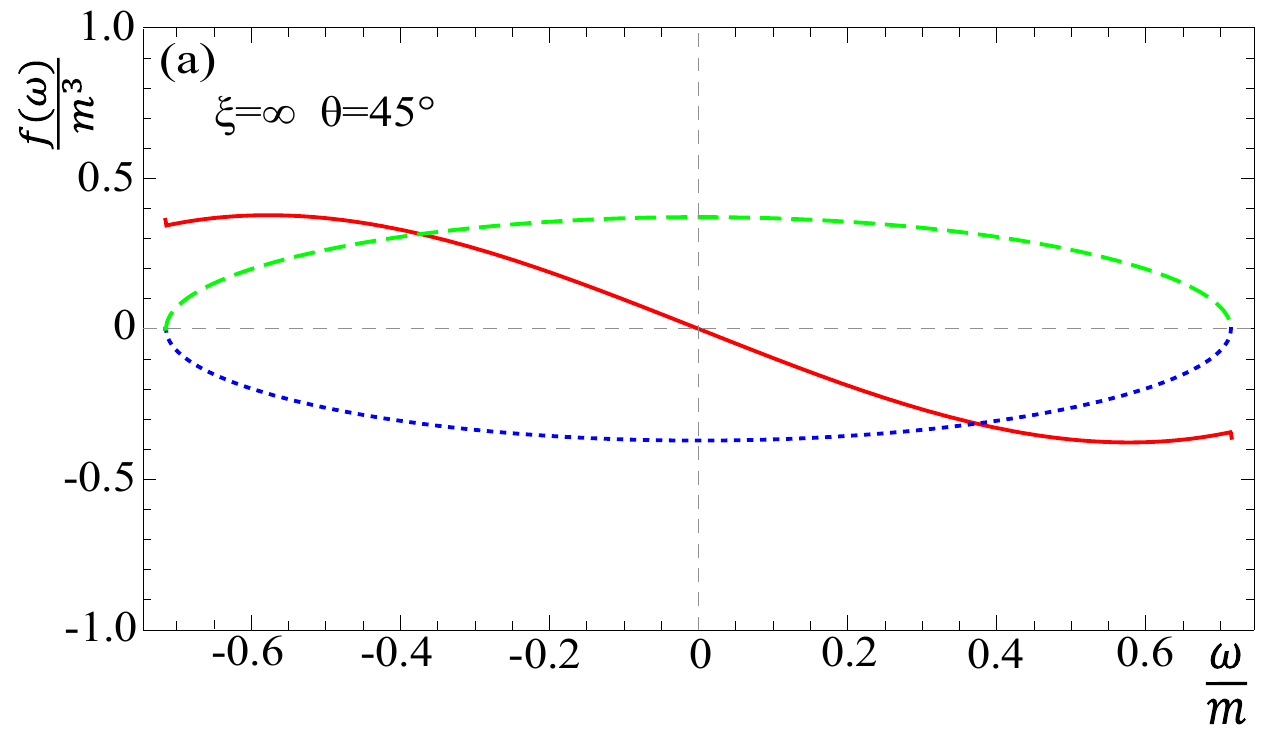}
\end{minipage}
\hspace{5mm}
\begin{minipage}{8.5cm}
\center
\includegraphics[width=0.94\textwidth]{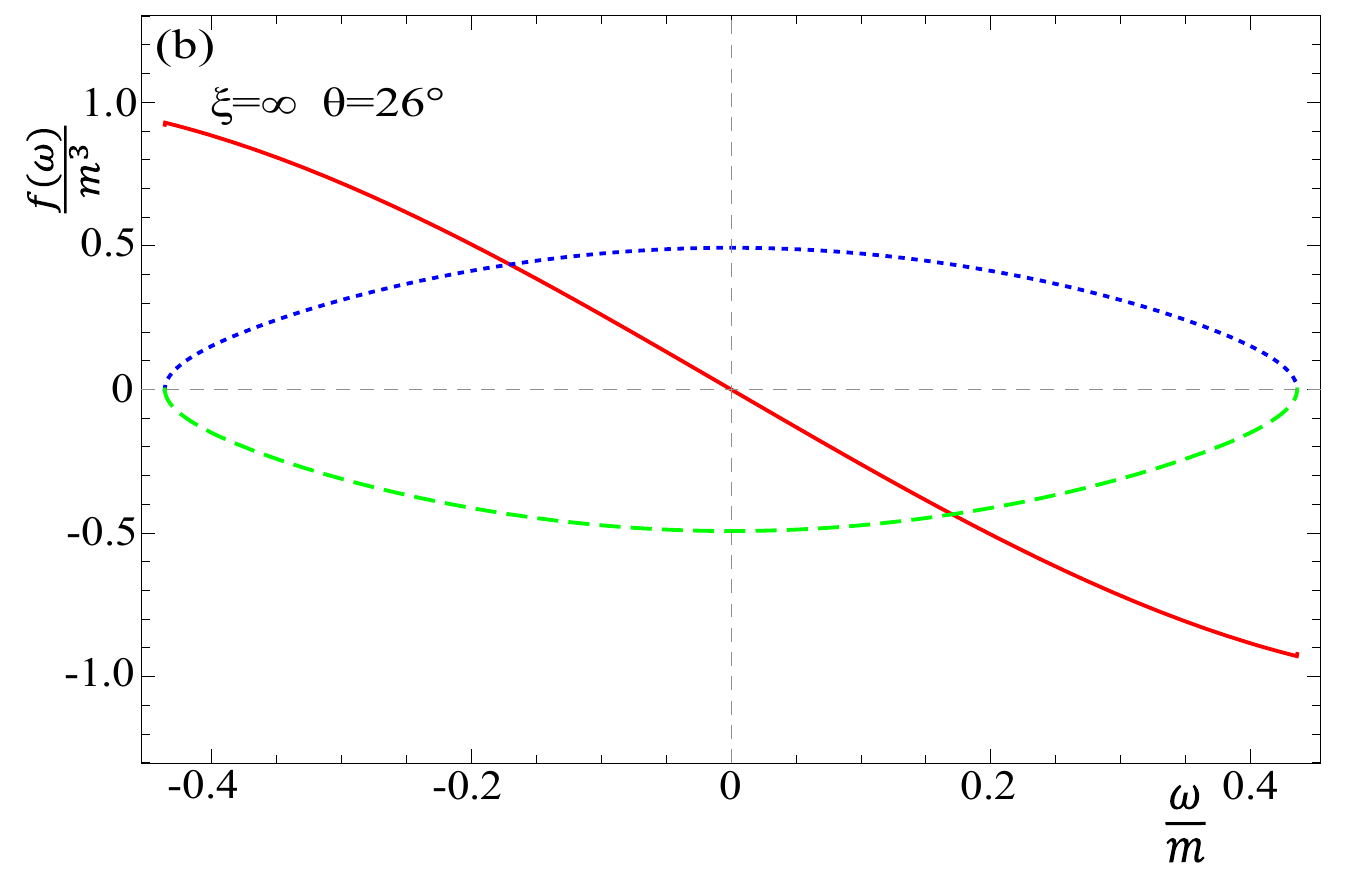}
\end{minipage}
\caption{(Color online) The real and imaginary parts of $f(\omega)$ for $A$-modes in extremely oblate plasma for $\omega$ along the cut and $k/m=1$. In the panel (a) we use $\theta =45^\circ$, which corresponds to $k > k_{\rm oA}$ and gives $n_W=-1$, and in the panel (b) $\theta =26^\circ$, which gives $k < k_{\rm oA}$ and $n_W=1$.}
\label{oblateA-wind-m1}
\end{figure}

The important point is that the analytic properties of the left-hand-side of the $G$-mode dispersion equation (\ref{dis-eq-G}) as a function of $\omega$ change when $\sigma \to \infty$. In this limit, the cut singularity at $\omega \in [-k,k]$ changes into double poles at  $\omega = \pm {\bf k} \cdot {\bf n}$. For $\sigma$ very large but finite, there is a contribution to $n_Z$ from $n_W =2$ for $k<k_{\rm pG}$ (see Eq. (\ref{k-crit-pro})). When $\sigma \to \infty$, the cut, and therefore also the winding number, disappears, but $n_Z$ acquires a contribution from $n_P$ which exists for all $k$. Therefore the limit $\sigma \to\infty$  produces the same number of solutions as the $k<k_{\rm pG}$ region of the large $\sigma$ distribution, but not the $k>k_{\rm pG}$ region. There are always 8 solutions of the dispersion equations in case of extremely prolate plasma, and there are 8 or 6 solutions for large $\sigma$ depending whether or not the condition $k<k_{\rm pG}$ is satisfied.  

\subsection{Extremely oblate plasma}
\label{app-ex-oblate}

The $A$-modes are obtained from the dispersion equation (\ref{dis-eq-A}) with the coefficient $\alpha(\omega,{\bf k})$ given by Eq.~(\ref{alpha-extremely-oblate}). The function $\Delta^{-1}_A(\omega,{\bf k}) \equiv \omega^2 - {\bf k}^2 - \alpha(\omega,{\bf k})$ equals
\be
\label{f-orig}
\Delta^{-1}_A(\omega,{\bf k})  = \omega^2 - k^2 - \frac{m^2}{2(1 - \cos^2\theta)k^2}
\bigg[\omega^2 - k^2\cos^2\theta - \frac{\omega (\omega^2 - k^2)}{k\sqrt{\omega + k\sqrt{1-\cos^2\theta}}\;\sqrt{\omega - k\sqrt{1-\cos^2\theta}}} \bigg] .
\ee
This function has a cut due to the square root, and we choose the contour $C$ as in Fig.~\ref{fig-Nyquist-1}, but with the cut extending from  $-k\sqrt{1-\cos^2\theta}$ to $k\sqrt{1-\cos^2\theta}$ instead of $[-k, k]$, as in the isotropic and weakly anisotropic cases. The mapping of the path along the cut to the plane of complex $f(\omega)$ is discontinuous, because of the fact that $\Im f(\omega)$  is infinite at $\omega = \pm k\sqrt{1-\cos^2\theta}$. To avoid this problem, we analyze the function
\be
\label{f-tilde}
f(\omega)  \equiv \sqrt{\omega + k\sqrt{1-\cos^2\theta}}\;\sqrt{\omega - k\sqrt{1-\cos^2\theta}} \; \Delta^{-1}_A(\omega,{\bf k}) ,
\ee
which has the same cut and the same number of zeros as the original function of the $A$-mode dispersion equation (\ref{f-orig}). The function  $f(\omega)$  has no singularity inside the contour and thus $n_P =0$. Equation~(\ref{Nyq-int-infty}) gives $n_\infty=3$ and thus the number of zeros of $f(\omega)$ is $n_Z=3+n_W$. To map the path around the cut to the plane of complex $f(\omega)$, we use Fig.~\ref{oblateA-wind-m1} where the real and imaginary parts of $f(\omega)$ along the cut are shown for two different values of $\cos\theta$ for $k/m=1$. For the case shown in Fig.~\ref{oblateA-wind-m1}a, where $\Im f(\omega = i0^+) < 0$, the point $f=0$ is encircled once in the negative (clockwise) direction which gives $n_W = - 1$. In the case corresponding to Fig.~\ref{oblateA-wind-m1}b, where $\Im f(\omega = i0^+) > 0$, the point $f=0$ is encircled once in the positive (counterclockwise) direction and $n_W = 1$. The winding number is therefore determined by the sign of $\Im f(\omega = i0^+)$. The number of solutions thus equals $2+2\theta(\Im f(\omega = i0^+)) = 2+2\Theta(k_{\rm oA}-k)$ which gives Eq. (\ref{ob-fin-A}).

\begin{figure}[t]
\begin{minipage}{8.5cm}
\center
\includegraphics[width=1.0\textwidth]{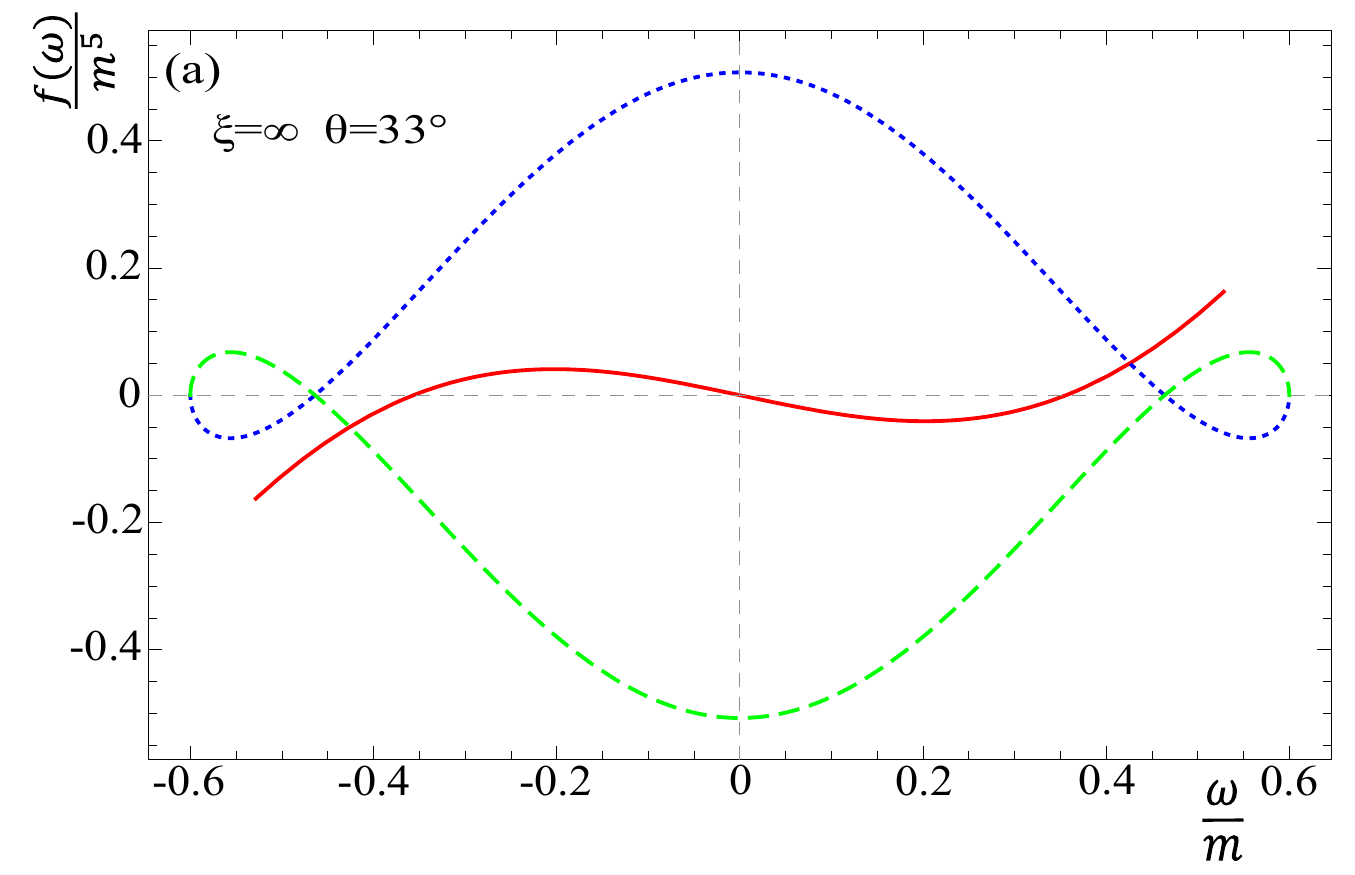}
\end{minipage}
\hspace{3mm}
\begin{minipage}{8.5cm}
\center
\includegraphics[width=1.02\textwidth]{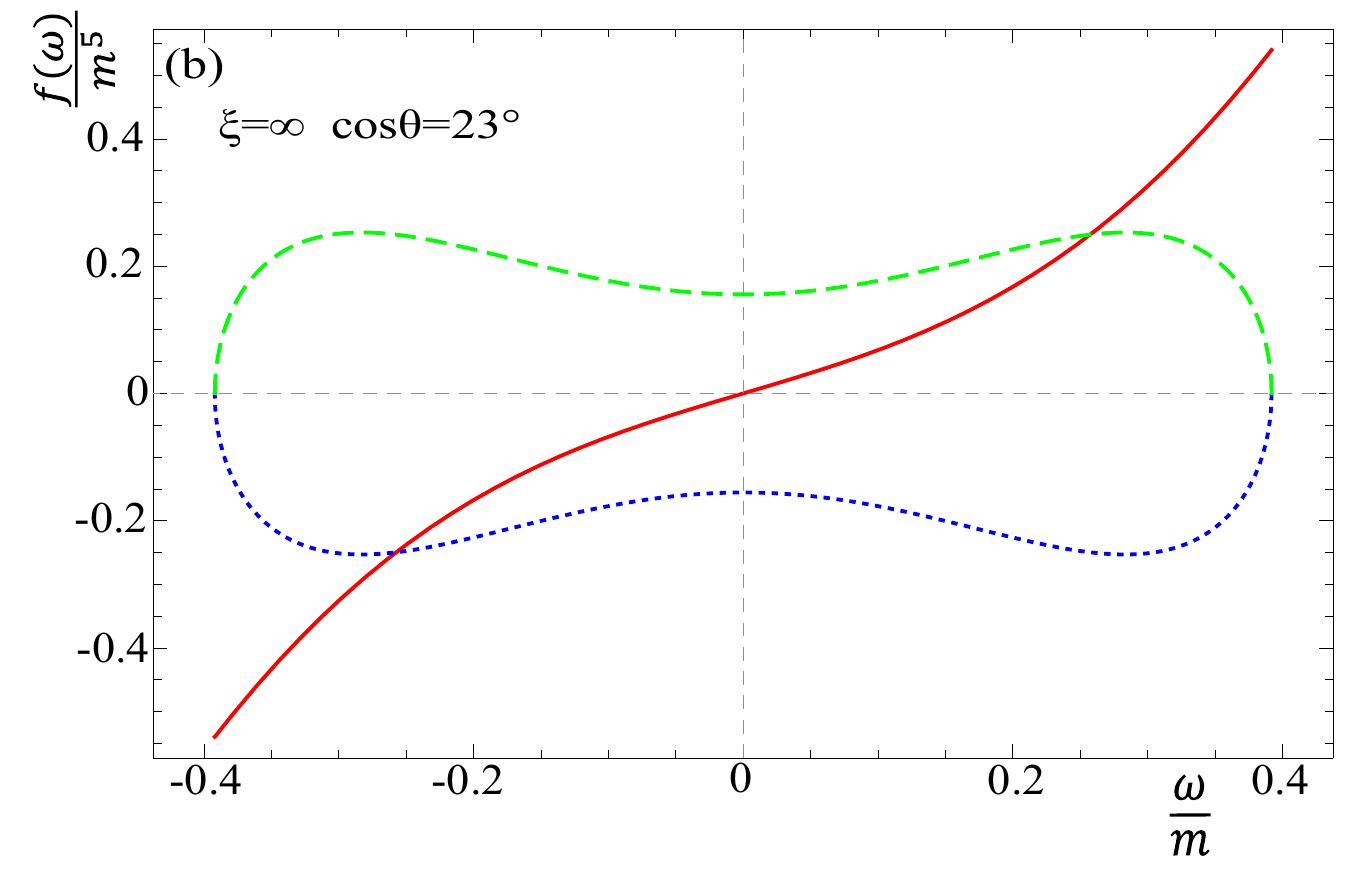}
\end{minipage}
\caption{(Color online) The real and imaginary parts of $f(\omega)$ for $G$-modes in extremely oblate plasma for $\omega$ along the cut and $k/m=1$. In the panel (a) we use $\theta =33^\circ$, which corresponds to $k > k_{\rm oG}$ and gives $n_W=-1$, and in the panel (b) $\theta =23^\circ$, which gives $k < k_{\rm oG}$ and $n_W=1$.}
\label{oblateG-wind-m1}
\end{figure}

The $G$-modes are obtained from the dispersion equation (\ref{dis-eq-G}) with the coefficients $\alpha, \, \beta, \, \gamma, \, \delta$ given by Eqs.~(\ref{alpha-extremely-oblate}-\ref{delta-extremely-oblate}). In this case it is easier to work with the function defined as
\ba
\label{f-G-ex-oblate}
f(\omega) = \frac{k^2}{\omega^2} 
\Big(\omega + k\sqrt{1 - \cos\theta^2}\,\Big)^{3/2} \Big(\omega - k \sqrt{1 - \cos\theta^2}\, \Big)^{3/2}
\Delta^{-1}_G( \omega, {\bf k}) .
\ea
Since this function has a cut for $\omega \in \big[-k\sqrt{1-\cos^2\theta}, k\sqrt{1-\cos^2\theta}\,\big]$, we choose the same contour as for the analysis of the $A$-modes described above. The function $f(\omega)$ has no singularities inside the contour and therefore $n_P =0$. Eq.~(\ref{Nyq-int-infty}) gives $n_\infty=5$,  and consequently, the number of zeros of $f(\omega)$ inside the contour equals $n_Z = 5+n_W$. The winding number is calculated from the graphs shown in Fig.~\ref{oblateG-wind-m1}. In the panel (a) $n_W = -1$ and in the panel (b) $n_W =1$. The number of solutions of the dispersion equation (\ref{dis-eq-G}) is therefore either 4 or 6, depending on the sign of $\Im f(\omega = i0^+)$. The result can be written as in Eq.~(\ref{ob-fin-G}) with the critical value of wave vector defined by Eq.~(\ref{k-crit-obl-G}).

\section{Summary and final remarks}
\label{section:conclusions}

We have performed a systematic analysis of plasmons - gluon collective modes - in quark-gluon plasma where the momentum distribution is obtained by either squeezing or stretching the isotropic distribution. We have considered distributions with all degrees of deformation along the beam axis, from the extremely prolate distribution, which is infinitely elongated along the beam, to the extremely oblate distribution, which is infinitely squeezed with respect to the beam axis. In every case we have calculated the dispersion curves for the full spectrum, in some cases numerically and in some cases analytically. The number of modes for each system is  summarized in Table~\ref{tab-modes}.

All solutions are either pure real or pure imaginary, and they always appear as pairs of partners with opposite sign. In all systems under consideration, except the isotropic plasma, there are unstable modes (positive pure imaginary solutions). Imaginary solutions exist only for certain wave vectors. We have used a Nyquist analysis to verify that all solutions have been found, and to determine analytically the critical values of wave vectors at which the imaginary solutions appear. The critical values of the wave vectors are given by Eqs. (\ref{k-crit-A}, \ref{k-crit-C}, \ref{k-crit-pro}, \ref{k-crit-obl-A}, \ref{k-crit-obl-G}) and collected here for the reader's convenience:
\ban
k_{\rm A} \equiv \Re\sqrt{\frac{\xi}{3}}\, m|\cos\theta| , 
~~~~~~~~~~~~~~~~
k_{\rm pG} \equiv \frac{m}{\sqrt{2}} |\tan\theta| ,
~~~~~~~~~~~~~~~~
k_{\rm oA}  \equiv \frac{m}{\sqrt{2}} |\cot\theta| ,
\\[2mm]
\label{k-crit-all}
k_{\rm C} \equiv m \Re\sqrt{\frac{\xi}{3}\,\big(2 \cos^2\theta - 1 \big)},
~~~~~~~~~~~
k_{\rm oG}  \equiv \frac{m}{2}\Re \sqrt{\frac{|\cos\theta|\sqrt{\cos^2\theta + 4}+\cos^2\theta-2}{\sin^2\theta}} .
\ean

\begin{figure}[t]
\center
\includegraphics[width=10cm]{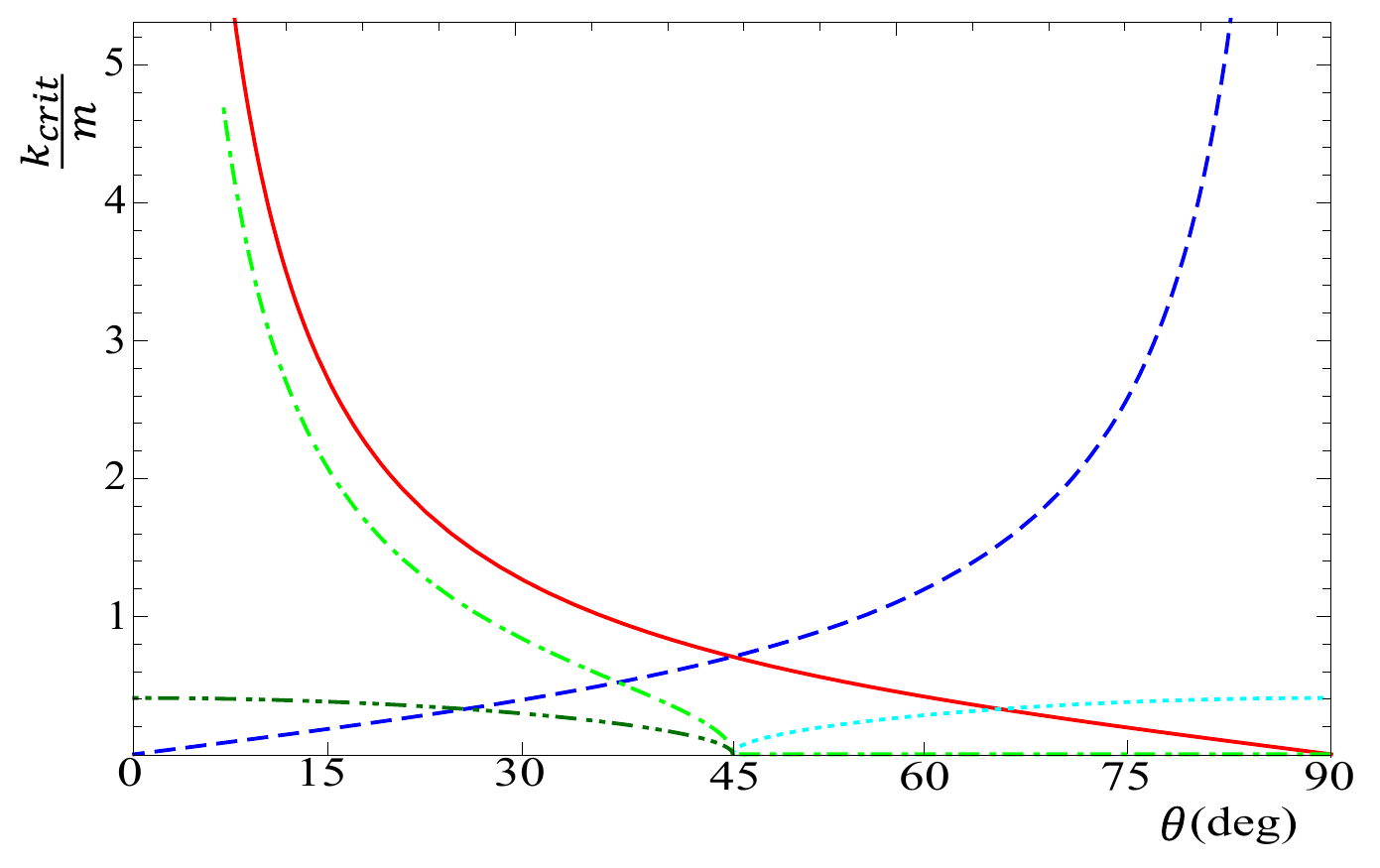}
\vspace{-2mm}
\caption{(Color online) The critical wave vectors as a function of $\theta$. The red (solid) line represents $k_{\rm oA}$,  blue (dashed) curve is $k_{\rm pG}$, cyan (dotted)  is $k_{\rm C}$ with $\sigma = -\xi= 0.5$, green (dot-dashed) is $k_{\rm oG}$, and finally the dark green (dot-dot-dashed) line is $k_{\rm C}$ with $\xi= - \sigma =0.5$.}
\label{k-crit-plot}
\end{figure}

It is interesting to look at these critical wave vectors as functions of $\theta$. In Fig.~\ref{k-crit-plot} we show the behavior of the critical vectors $k_{\rm A}, \, k_{\rm oA}, \, k_{\rm C}, \, k_{\rm oG}$ and  $k_{\rm pG}$ as a function of angle ($0^\circ <\theta<90^\circ$) for different values of the anisotropy parameters. The prolate critical wave vector has a maximum at $\theta=90^\circ$, which goes to infinity in the extreme prolate case, and the oblate critical vectors have maximum at $\theta=0^\circ$ which approaches infinity in the extreme oblate case. The prolate plasma system produces the strongest instability when the wave vector ${\bf k}$ is exactly transverse to the anisotropy vector ${\bf n}$. For the oblate system the situation is exactly reversed, and the strongest instability occurs when the wave vector and anisotropy vector are parallel to each other. 

\begin{figure}[t]
\center
\includegraphics[width=13cm]{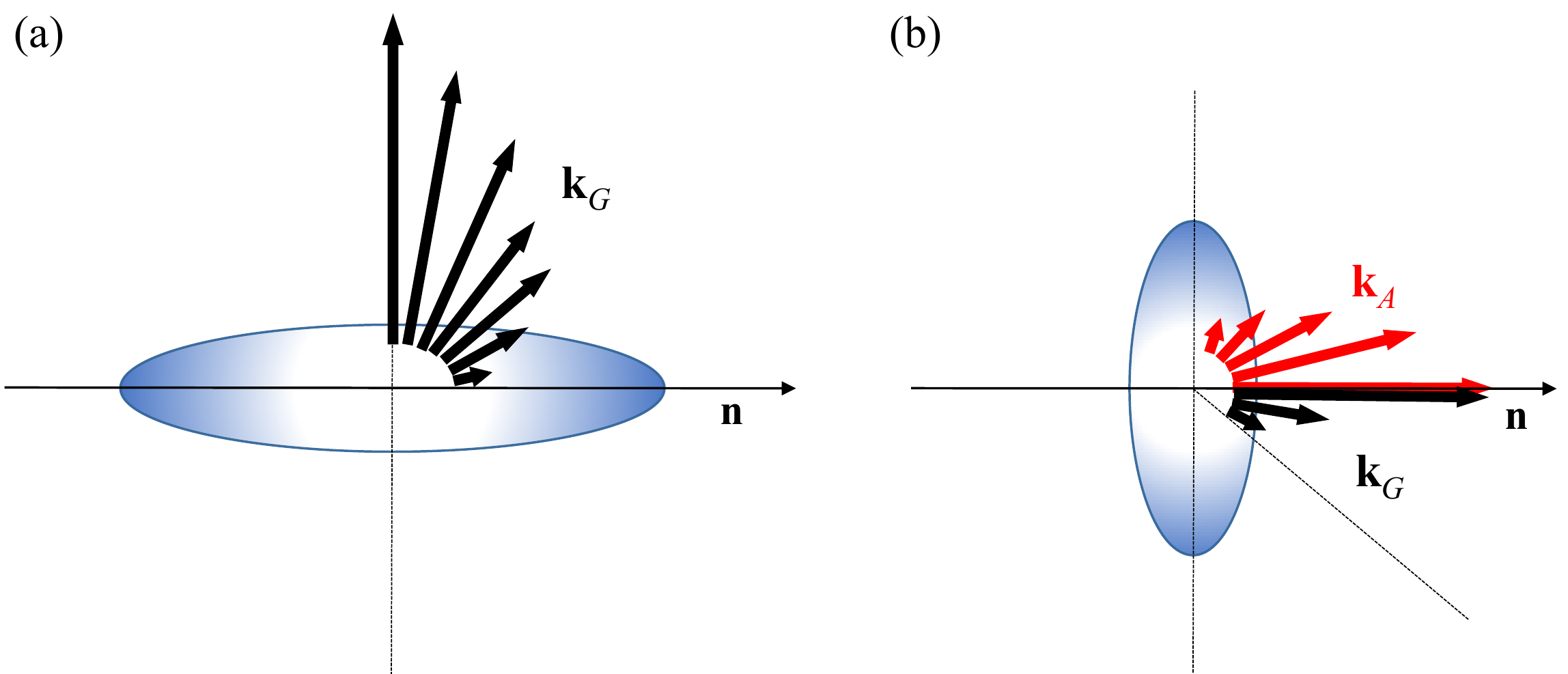}
\caption{(Color online) The largest wave vectors of unstable modes at different orientations. In the prolate plasma (a) there is one unstable $G$-mode, which exists for $0^\circ <\theta \le 90^\circ$, and in oblate plasma (b) there are two unstable $A$- and $G$-modes which exist for $0^\circ \le \theta < 90^\circ$ and $0^\circ \le \theta < 45^\circ$, respectively.}
\label{fig-configurations}
\end{figure}

The orientation of the largest wave vectors of unstable modes is schematically shown in Fig.~\ref{fig-configurations}. Since the $A$-modes are transverse, the chromoelectric field of the maximally unstable $A$-mode in oblate plasma is perpendicular to ${\bf n}$. The $G$-modes are, in general, neither transverse nor longitudinal. However, the maximally unstable $G$-modes (with ${\bf k} || {\bf n}$ in oblate plasma and ${\bf k} \perp {\bf n}$ in prolate plasma) are transverse. Therefore, for $G$-modes in prolate systems, and both $A$-modes and $G$-modes in oblate plasmas, the chromoelectric field of the maximally unstable modes is aligned with the direction where the momentum of plasma particles is maximal. This observation is important because the dynamics of unstable systems are dominated by the fastest growing unstable modes. 

There are unstable modes even for arbitrarily small anisotropy - there is no threshold value of the anisotropy parameter for which instabilities develop. However, the growth rate of instability decreases, and the domain of wave vectors for which unstable modes exist shrinks, as the anisotropy parameter decreases.  Our work does not take into account inter-parton collisions which, as discussed in \cite{Schenke:2006xu}, reduce the growth rates and domains of wave vectors of the unstable modes and effectively stabilize  systems of sufficiently small anisotropy.

\begin{table}[b]
\caption{\label{table-modes} Number of modes }
\begin{ruledtabular}
\label{tab-modes}
\begin{tabular}{ccccc}
 Momentum &  Number  & Number   & Total number & Maximal number
\\
distribution &  of real modes & of imaginary modes & of modes & of modes 
\\ \hline \hline 
extremely prolate & $6 + 2\Theta(k-k_{\rm p})$ & $2\Theta(k_{\rm p}-k)$  &  8 &  8
\\[2mm]
weakly prolate   & 6 & $2\Theta(k_{\rm C}-k)$ & $6 + 2\Theta(k_{\rm C}-k)$ & 8
\\[2mm]
isotropic & 6 & 0 & 6 & 6
\\[2mm]
weakly oblate    & 6 & $2\Theta(k_{\rm A}-k) + 2\Theta(k_{\rm C}-k)$  & $6 + 2\Theta(k_{\rm A}-k) + 2\Theta(k_{\rm C}-k)$ & 10
\\[2mm]
extremely oblate  & 6 & $2\Theta(k_{\rm oA}-k) + 2\Theta(k_{\rm oG}-k)$  & $6 + 2\Theta(k_{\rm oA}-k) + 2\Theta(k_{\rm oG}-k)$ & 10
\end{tabular}
\end{ruledtabular}
\end{table}

A systematic analysis of collective modes in anisotropic plasmas is an important part of a study of physical quantities in such systems. The results in this paper have been used extensively in our analysis of parton energy loss in unstable QGP, a preliminary account of which is published as \cite{Carrington:2012hv,Carrington:2013tz,Carrington:2014yra}. We hope that the analysis we have presented in this paper will facilitate computations of other characteristics of anisotropic quark-gluon plasma. 

\section*{Acknowledgments}

We are grateful to Michael Strickland for discussions and critical reading of the manuscript. This work was partially supported by the Polish National Science Centre under grant 2011/03/B/ST2/00110 and the Natural Sciences and Engineering Research Council of Canada. 

\appendix

\section{Classical dispersion equation}
\label{app-dis-eq}

For the completeness of our discussion, we show here how the dispersion equation of plasma waves is derived in classical electrodynamics. One starts with  sourceless Maxwell equations in a medium. After Fourier transformation to the space of frequencies $\omega$ and wave vectors ${\bf k}$ they give 
\ba
k^i D^i(\omega, {\bf k}) &=& 0, \;\;\;\;\;\;\;\;\;\;\;
\epsilon^{ijk} k^j E^k(\omega, {\bf k}) = \omega B^j (\omega,{\bf k}) ,
\\
k^i B^i(\omega, {\bf k}) &=& 0, \;\;\;\;\;\;\;\;\;\;\;
\epsilon^{ijk} k^j B^k(\omega, {\bf k}) = - \omega D^j (\omega,{\bf k}) , 
\ea
where ${\bf E}(\omega, {\bf k})$, ${\bf D}(\omega, {\bf k})$, ${\bf B}(\omega, {\bf k})$ are the electric field, electric induction and magnetic field, and  $\epsilon^{ijk}$ is the completely antisymmetric tensor. The electric induction is expressed through the electric field by means of the electric permeability tensor $\varepsilon^{ij} (\omega,{\bf k})$ as
\be
\label{D-vs-E}
D^i (\omega,{\bf k}) = \varepsilon^{ij} (\omega,{\bf k}) \, E^j (\omega,{\bf k}) .
\ee
Substituting the dielectric induction in the form (\ref{D-vs-E}) into Maxwell equations, one finds the following equation of motion of the electric  field
\be
\label{E-EOM}
\big[
- {\bf k}^2 \delta^{ij} + k^ik^j 
+ \omega^2 \varepsilon^{ij}(\omega,{\bf k}) \big] E^j (\omega,{\bf k})  = 0 .
\ee
Solutions of the homogeneous equation (\ref{E-EOM}) exist, provided the determinant of the matrix (\ref{matrix-Sigma}) vanishes. This is the dispersion equation (\ref{general-dis-eq-det}). 

\begin{figure}[t]
\begin{minipage}{8.5cm}
\center
\includegraphics[width=1.02\textwidth]{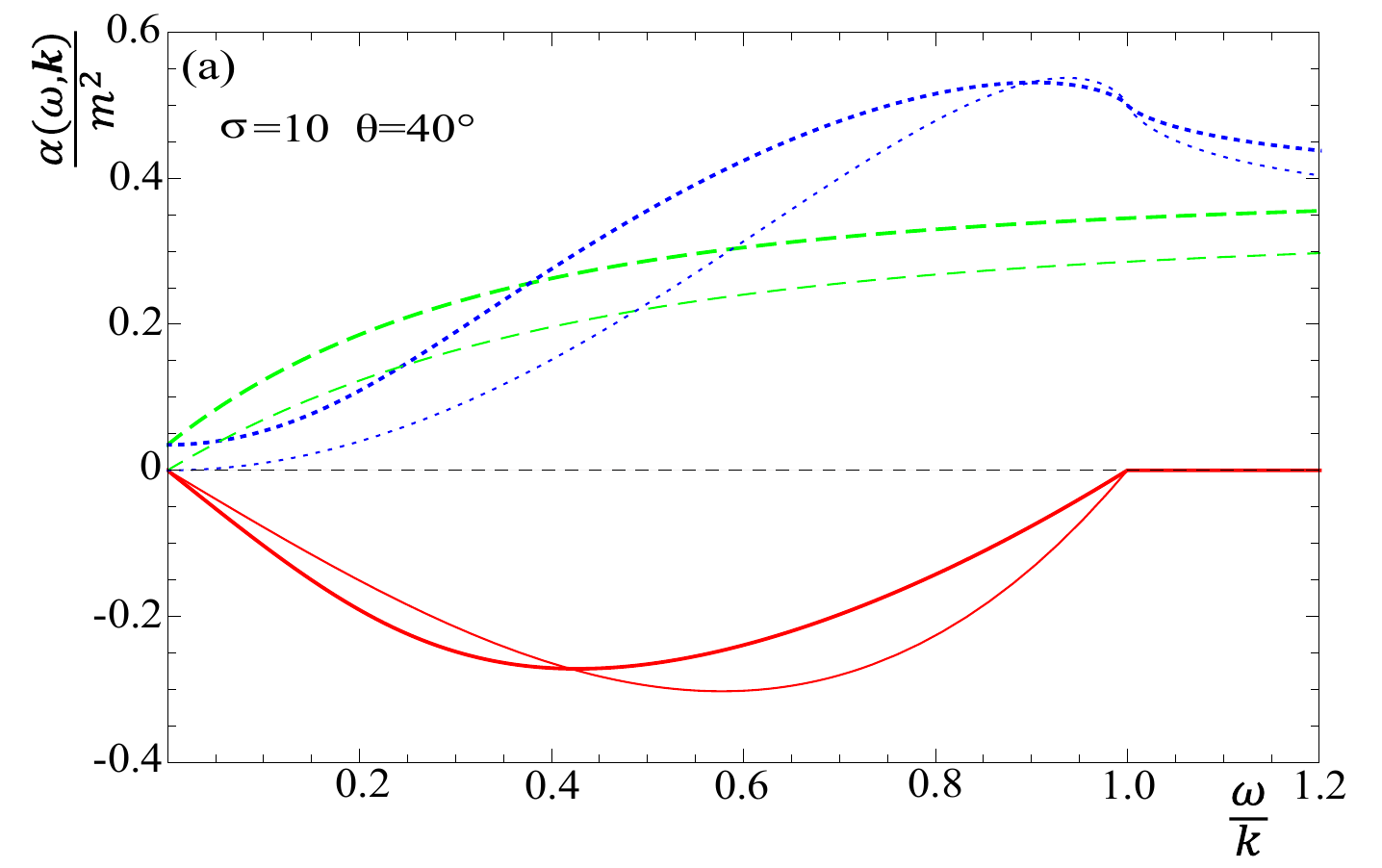}
\end{minipage}
\hspace{1mm}
\begin{minipage}{8.5cm}
\center
\includegraphics[width=1.03\textwidth]{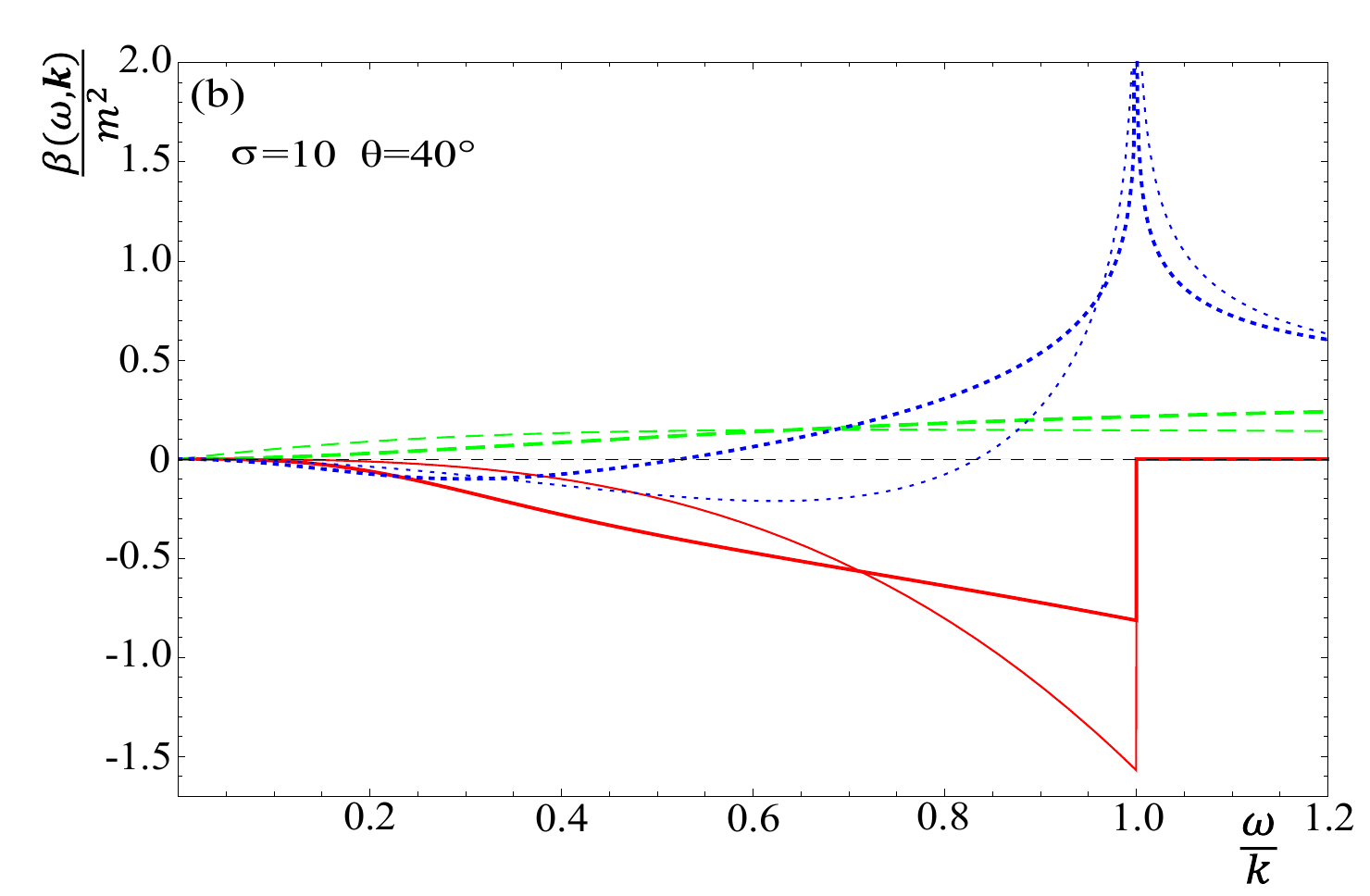}
\end{minipage}
\begin{minipage}{8.5cm}
\center
\includegraphics[width=1.0\textwidth]{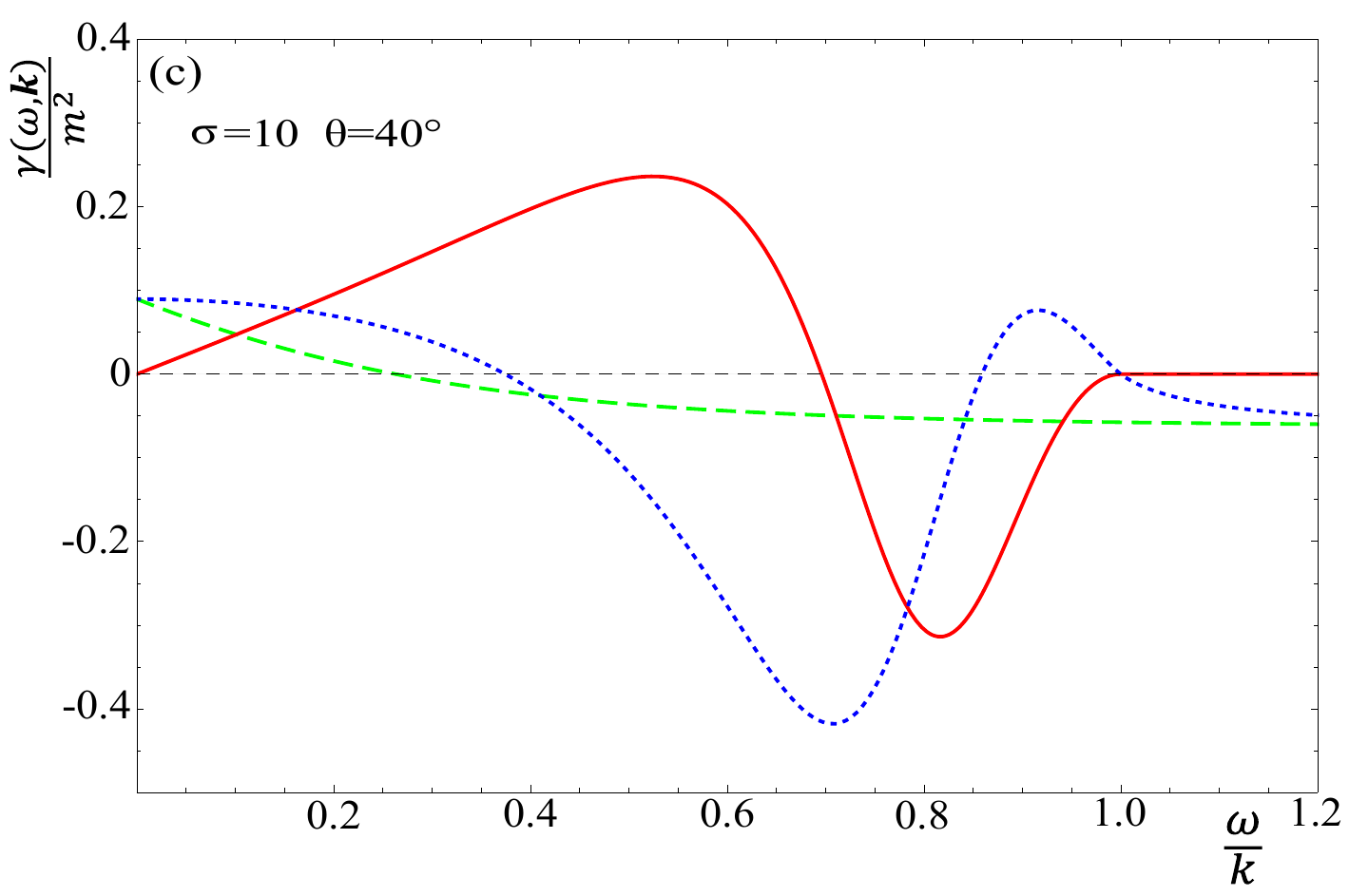}
\end{minipage}
\hspace{4mm}
\begin{minipage}{8.5cm}
\center
\includegraphics[width=1.0\textwidth]{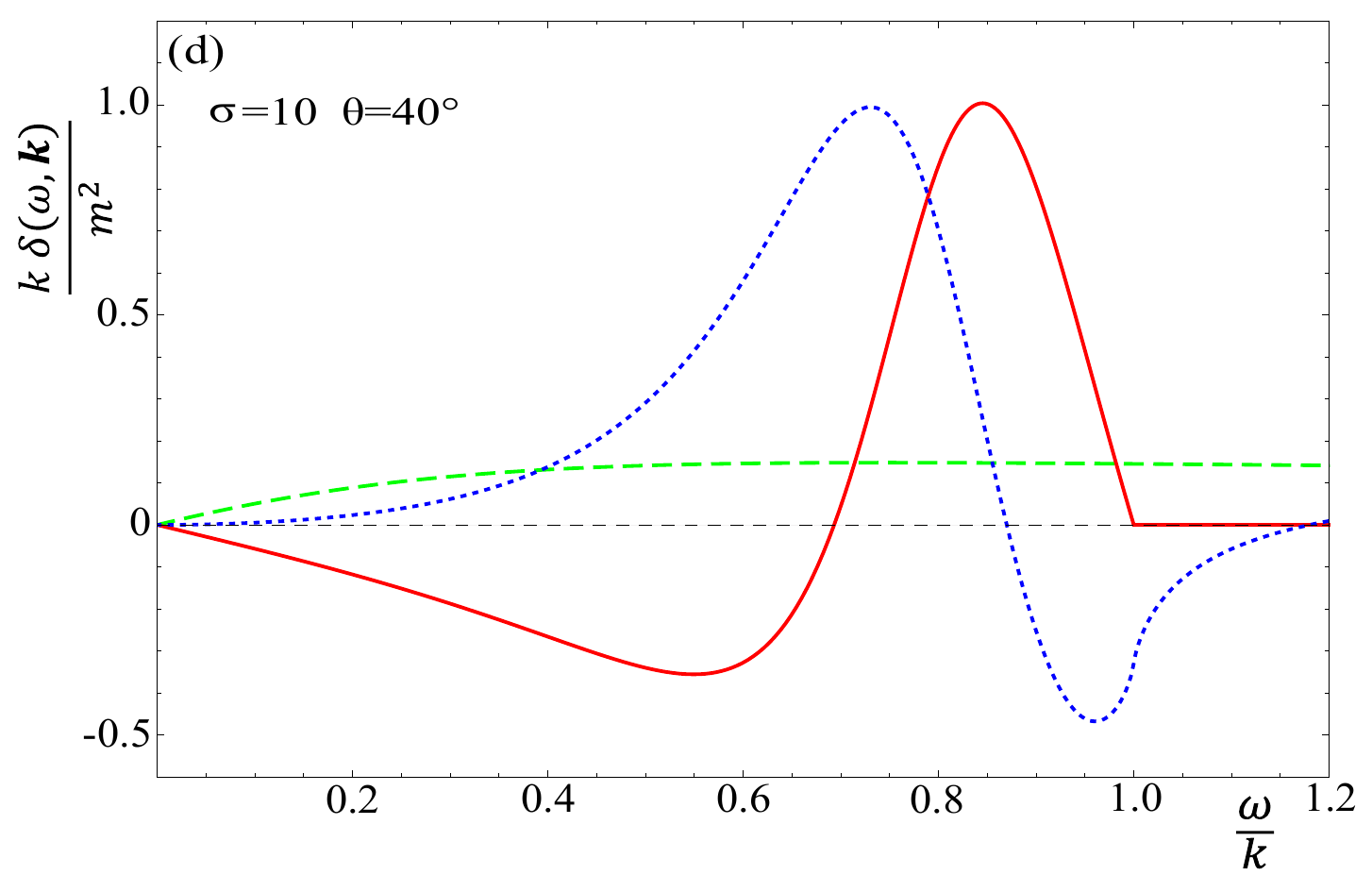} 
\end{minipage}
\caption{(Color online) The components $\alpha_\sigma, \,\beta_\sigma,\, \gamma_\sigma, \, k\,\delta_\sigma$ of the polarization tensor obtained from the $\sigma$-distribution with $\sigma=10$  and $\theta=40^\circ$. For each component, the blue (dotted) line is the real part with real $\omega$, the green (dashed) line is the imaginary part with real $\omega$, and the red (solid) line is the pure real result when $\omega$ is imaginary. The lighter lines in the panels presenting the coefficients $\alpha_\sigma$ and $\beta_\sigma$ correspond to the isotropic distribution.}
\label{structure-sigma}
\end{figure}

\section{Components of polarization tensor}
\label{app-self-energy-components}

\begin{figure}[t]
\begin{minipage}{8.5cm}
\center
\includegraphics[width=1.0\textwidth]{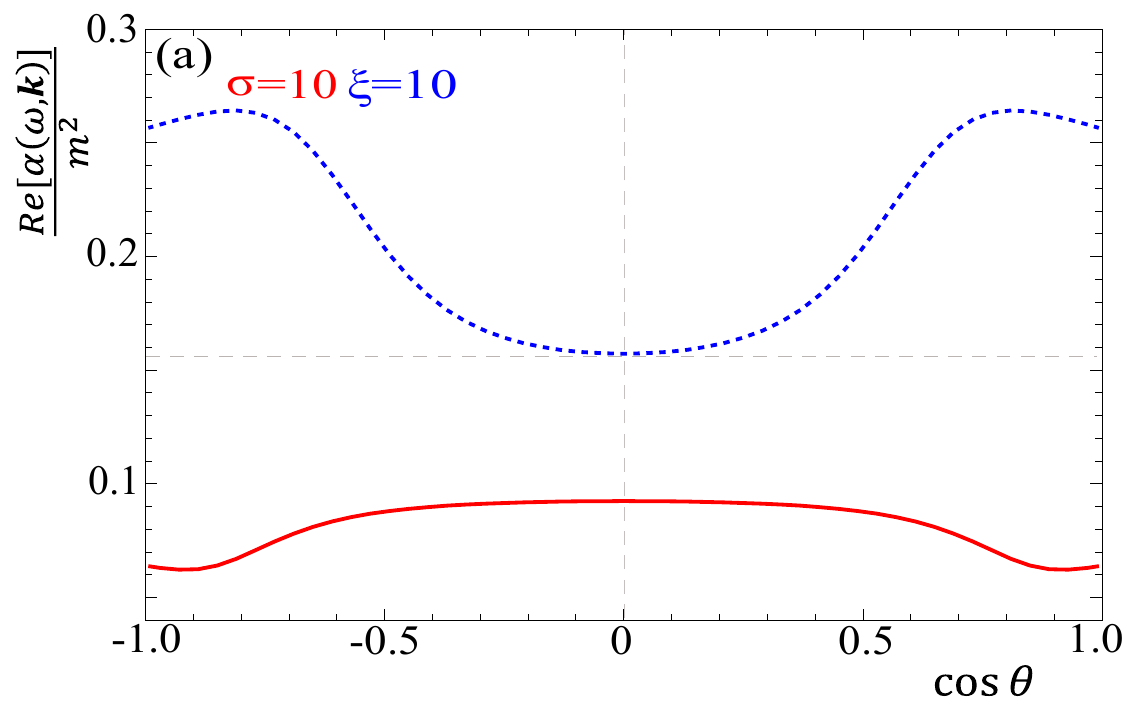}
\end{minipage}
\hspace{1mm}
\begin{minipage}{8.5cm}
\center
\includegraphics[width=1.02\textwidth]{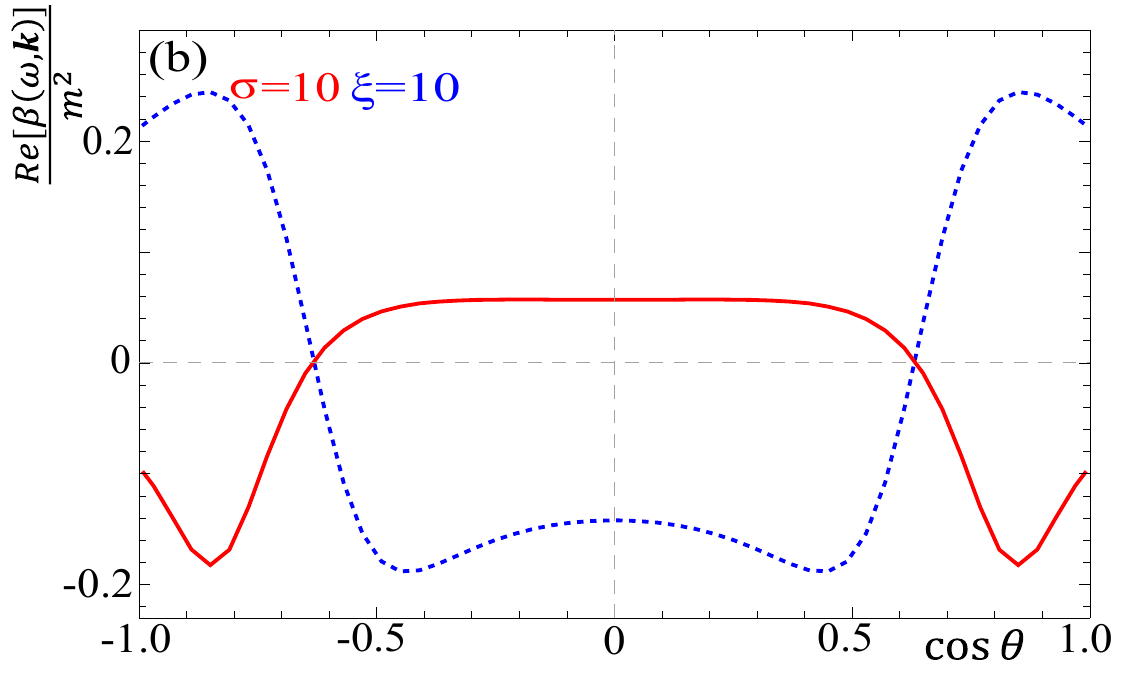}
\end{minipage}
\vspace{1mm}
\begin{minipage}{8.5cm}
\center
\includegraphics[width=1.0\textwidth]{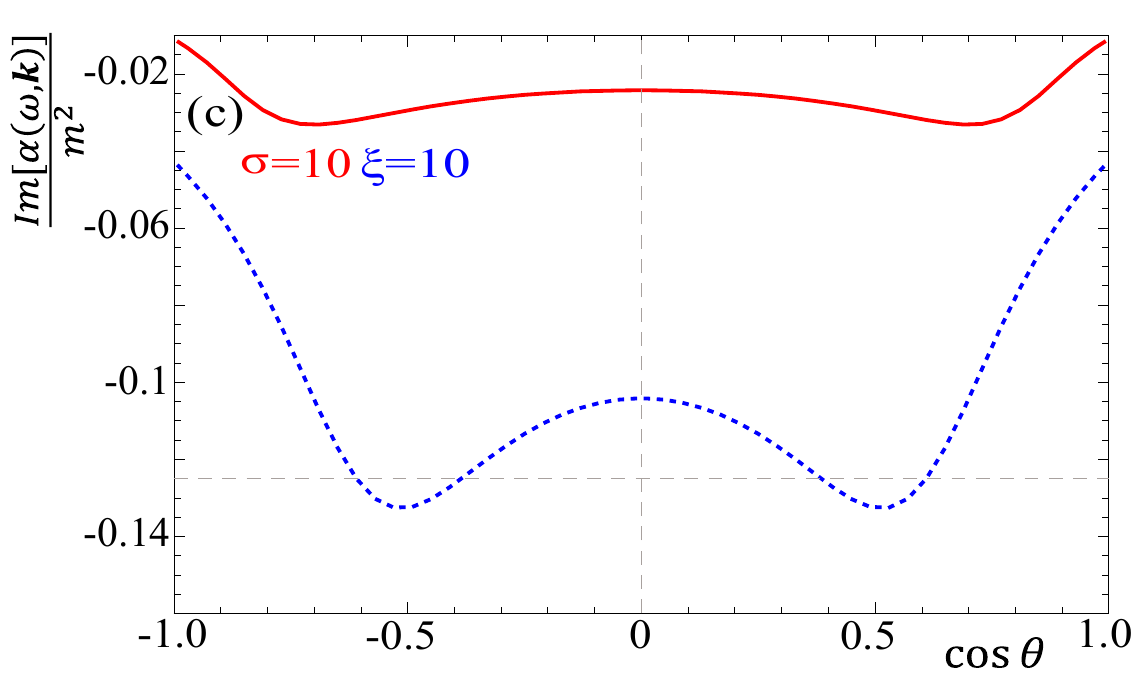}
\end{minipage}
\hspace{-3mm}
\begin{minipage}{8.5cm}
\center
\includegraphics[width=1.03\textwidth]{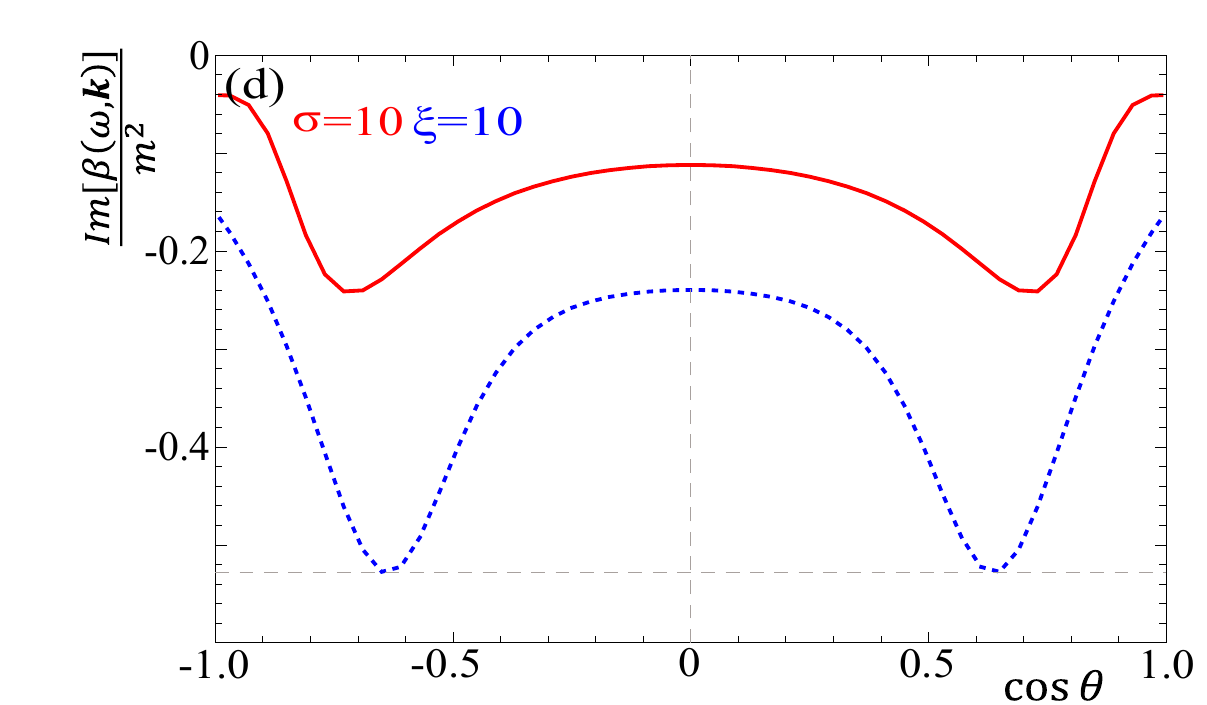}
\end{minipage}
\vspace{1mm}
\begin{minipage}{8.5cm}
\center
\includegraphics[width=1.0\textwidth]{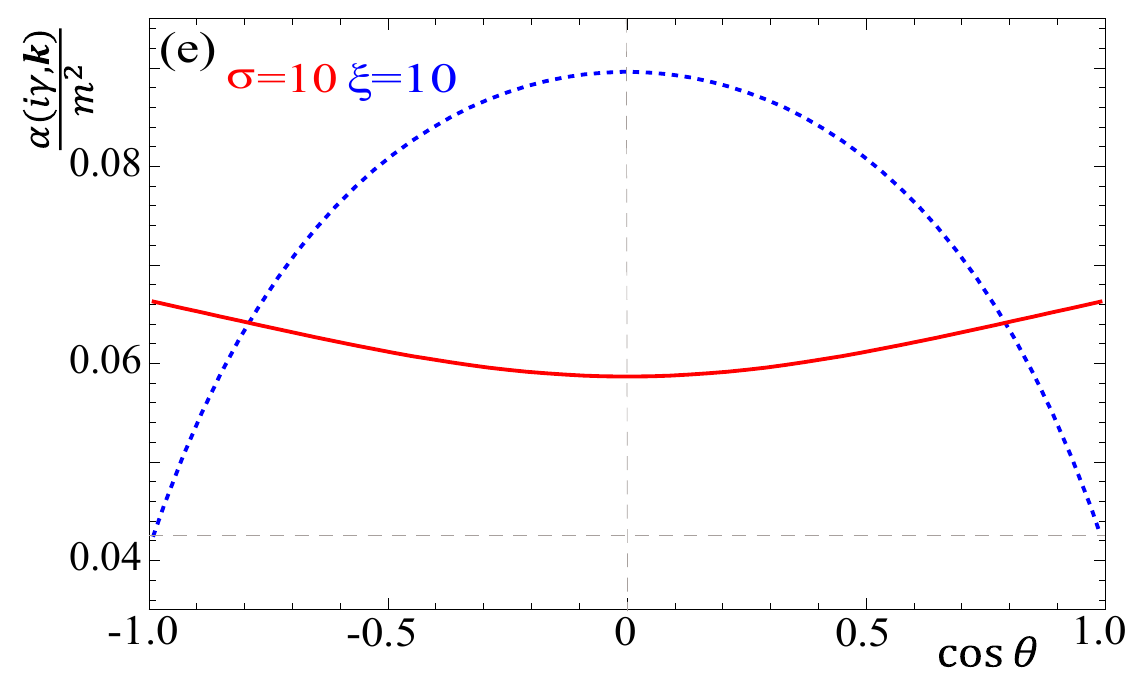}
\end{minipage}
\hspace{1mm}
\begin{minipage}{8.5cm}
\center
\includegraphics[width=1.0\textwidth]{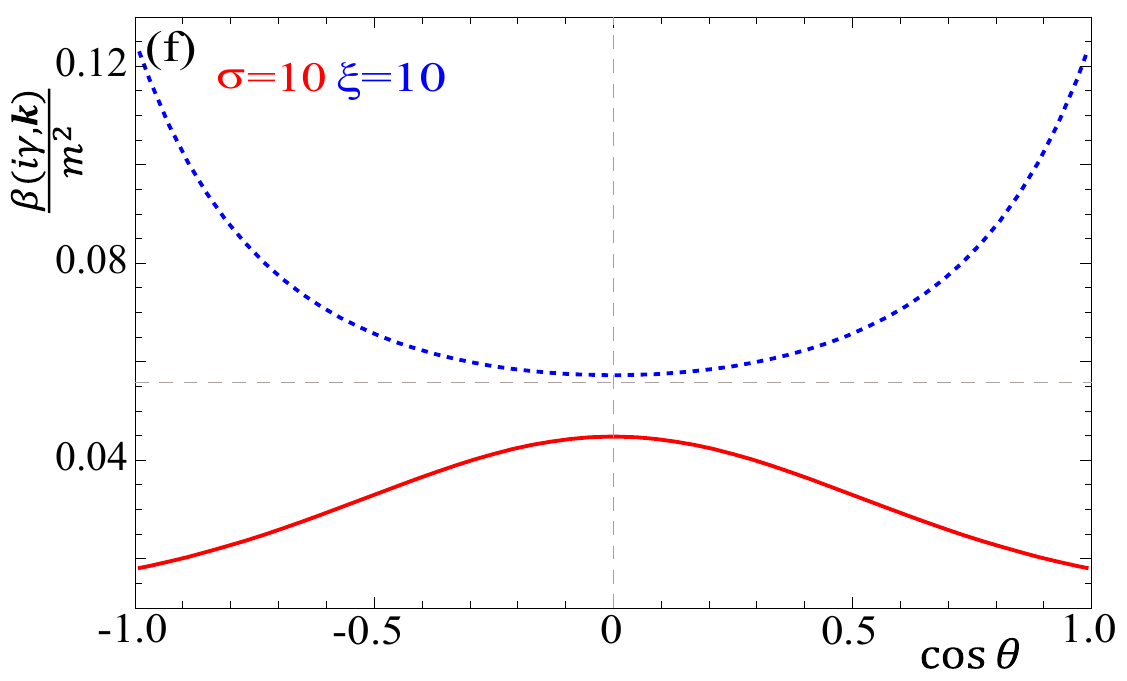} 
\end{minipage}
\caption{(Color online) The components $\alpha$ and $\beta$ of the polarization tensor with $\sigma=10$ (red (solid) line) and $\xi$=10 (blue (dotted) line)  versus $x \equiv\cos\theta$ with $\omega/k=0.8$ or $\omega/k=0.8\,i$. The top (middle) panels present the real (imaginary) parts of $\alpha$ and $\beta$ for real $\omega$. The bottom panels show $\alpha$ and $\beta$ for imaginary $\omega$. }
\label{structure-mirror}
\end{figure}

We present here the components $\alpha, \, \beta, \, \gamma, \, \delta$ of the polarization tensor computed with the $\xi$-distribution (\ref{R-S-ansatz}) and $\sigma$-distribution (\ref{alter-ansatz}). The azimuthal and polar integrals in the formulas (\ref{alpha-gen}, \ref{beta-gen}, \ref{gamma-gen}, \ref{delta-gen}) can be evaluated analytically, but the results are lengthy, and not very enlightening. We give here only the expressions with the polar integral unevaluated using the following notation: $\hat\omega \equiv \omega/k$, $x \equiv \cos\theta$ and $y \equiv \cos\theta^\prime$. We also define
\ba
\label{Rpm}
R_\pm \equiv \sqrt{\hat\omega+i0^+ - x y \pm \sqrt{1-x^2}\sqrt{1-y^2}}.
\ea
The results for the $\xi$-distribution (\ref{R-S-ansatz}), which appeared previously in \cite{Romatschke:2003ms} with a different normalization, are
\ba
\frac{\alpha_\xi (\omega,{\bf k})}{C_\xi m^2} &=& \int_{-1}^1\frac{dy (\xi  x y+\hat\omega )}{\left(x^2-1\right) \left(\xi  y^2+1\right)^2}
\bigg[\frac{x^2-2 x y \hat\omega +y^2+\hat\omega ^2-1}{R_+ R_-}+x y-\hat\omega \bigg] ,
\\[2mm]
\frac{\beta_\xi (\omega,{\bf k})}{C_\xi m^2} &=&  \int_{-1}^1\frac{dy \,\hat\omega }{\left(\xi  y^2+1\right)^2}
\bigg[\frac{\hat\omega  (\xi  x y+\hat\omega )}{R_+ R_-}+(\xi +1)    (-x) y-\hat\omega \bigg] ,
\\[2mm] \nn
\frac{\gamma_\xi (\omega,{\bf k})}{C_\xi m^2} &=&  \int_{-1}^1 \frac{dy}{\left(x^2-1\right) \left(\xi  y^2+1\right)^2}
\bigg[-\frac{\left(\left(x^2+1\right) \hat\omega ^2+x^2-4 x y \hat\omega +2
   y^2-1\right) (\xi  x y+\hat\omega )}{R_+ R_- }
\\ && ~~~~~~~~~~~~~~~~~~~~~~~~~~~~~~~~~
+ \xi  \left(x^2+1\right) y (x \hat\omega
   -y)+\hat\omega  \left(x \left(x^2-3\right) y+\left(x^2+1\right) \hat\omega  \right) \bigg],
\\[2mm]
\frac{k\,\delta_\xi (\omega,{\bf k})}{C_\xi m^2} &=&  \int_{-1}^1 \frac{dy}{\left(x^2-1\right) \left(\xi  y^2+1\right)^2}
\bigg[\frac{\hat\omega  (x \hat\omega -y) (\xi  x y+\hat\omega )}{R_+ R_-}-\hat\omega 
   \left(y \left((\xi +1) x^2-1\right)+x \hat\omega \right)\bigg] .
\ea
For the $\sigma$-distribution (\ref{alter-ansatz}) we obtain
\ba
\frac{\alpha_\sigma (\omega,{\bf k})}{C_\sigma m^2}&=& \int_{-1}^1 \frac{dy ((\sigma +1) \hat\omega -\sigma  x y)}
   {\left(x^2-1\right) \left(\sigma \left(y^2-1\right)-1\right)^2} 
\bigg[\frac{x^2-2 x y \hat\omega +y^2+\hat\omega ^2-1}{R_+ R_-}+x y-\hat\omega \bigg] ,
\\[2mm]
\frac{\beta_\sigma (\omega,{\bf k})}{C_\sigma m^2} &=&\int_{-1}^1 \frac{dy \, \hat\omega}{\left(\sigma  \left(y^2-1\right)-1\right)^2}
\bigg[ \frac{\hat\omega  ((\sigma +1) \hat\omega -\sigma x y)}{R_+ R_-} -(\sigma +1) \hat\omega - x y \bigg] ,
\\[2mm] \nn
\frac{\gamma_\sigma (\omega,{\bf k})}{C_\sigma m^2} &=& \int_{-1}^1 \frac{dy}{\left(x^2-1\right) \left(\sigma \left(y^2-1\right)-1\right)^2}
\bigg[
\frac{\left(\left(x^2+1\right)   \hat\omega ^2+x^2-4 x y \hat\omega +2 y^2-1\right) (\sigma  x y-(\sigma +1) \hat\omega )}{R_+ R_-} 
\\&&~~~~~~~~~~~~~~~~~~~~~~~~~~~~~~~~~~
+ \left((\sigma +1) \left(x^2+1\right) \hat\omega ^2+\sigma  \left(x^2+1\right) y^2+x \hat\omega  y \left(-4 \sigma
   +x^2-3\right)\right) \bigg] ,
\\[2mm]
\frac{k\, \delta_\sigma (\omega,{\bf k})}{C_\sigma m^2} &=& \int_{-1}^1 \frac{dy \, \hat\omega }{\left(x^2-1\right) \left(\sigma  \left(y^2-1\right)-1\right)^2}
\bigg[\frac{(x \hat\omega -y) ((\sigma +1) \hat\omega -\sigma  x
   y)}{R_+ R_-}-y \left(-\sigma +x^2-1\right)-(\sigma +1) x \hat\omega \bigg] .
\ea

As explained in Sec.~\ref{sec-coefficients} and summarized in Eq.~(\ref{ana-info}), all four coefficients are real when $\omega$ is real and $\omega^2 >k^2$. When $\omega$ is real but $\omega^2 < k^2$, the coefficients become complex. For imaginary valued $\omega$, all four coefficients are real. In Fig.~\ref{structure-sigma} we show the components of the polarization tensor $\alpha_\sigma, \,\beta_\sigma,\, \gamma_\sigma, \, k\,\delta_\sigma$ as functions of $\omega/k$ for $\sigma=10$ and $\theta=40^\circ$. In each part of the figure, we show the real and imaginary parts of the component for real $\omega$ and the pure real components for imaginary $\omega$. For reference, the curves for $\alpha$ and $\beta$ in isotropic plasma are also shown. 

The polarization tensors of the $\xi$- and $\sigma$-distribution exhibit a kind of symmetrical structure when they are plotted as functions of $x \equiv \cos\theta$. One finds the following pattern. The real part of $X_\xi$, where $X$ is either $\alpha, \,\beta$ or $\gamma$, has a maximum when the real part of $X_\sigma$ has a minimum and {\it vice versa}. The imaginary parts of $X_\xi$ and $X_\sigma$ are maximal and minimal, respectively, for the same values of $x$. In the case of $\delta$ the positions of the maxima (and minima) of real parts coincide, and the maximum of the imaginary part of $\delta_\xi$ coincides with the minimum of the imaginary part of $\delta_\sigma$ and {\it vice versa}. These results are shown in Fig.~\ref{structure-mirror} for the coefficients $\alpha$ and $\beta$. 


\end{document}